\documentclass[prl,amsmath,amssymb,superscriptaddress,twocolumn,longbibliography,aps]{revtex4-2}
\usepackage{graphicx}
\usepackage{subfigure}
\usepackage{adjustbox}
\usepackage{bm}
\usepackage{array}
\usepackage{color}
\usepackage{braket}
\usepackage{standalone}
\usepackage{multirow}
\usepackage{tikz}
\usepackage{mathrsfs}
\usepackage{dsfont}
\usepackage{amsmath}
\usepackage[colorlinks,bookmarks=true,citecolor=blue,linkcolor=red,urlcolor=blue]{hyperref}
\usepackage[capitalize]{cleveref}
\usepackage{changes}

\usepackage{xcolor}
\definecolor{orange}{rgb}{1,0.5,0}
\definecolor{darkblue}{rgb}{0.,0.,0.4}
\definecolor{darkred}{rgb}{0.5,0.,0.}

\newcommand{\mycomment}[1]{}

\begin{document}

\title{
Critical Majorana fermion at a topological quantum Hall bilayer transition 
}

\author{Cristian Voinea}
\affiliation{School of Physics and Astronomy, University of Leeds, Leeds LS2 9JT, United Kingdom}

\author{Wei Zhu}
\email{zhuwei@westlake.edu.cn}
\affiliation{Institute of Natural Sciences, Westlake Institute for Advanced Study, Hangzhou 310024, China}
\affiliation{Department of Physics, School of Science, Westlake University, Hangzhou 310030, China}

\author{Nicolas Regnault}
\affiliation{Center for Computational Quantum Physics, Flatiron Institute, 162 5th Avenue, New York, NY 10010, USA}
\affiliation{Department of Physics, Princeton University, Princeton, New Jersey 08544, USA}
\affiliation{Laboratoire de Physique de l'Ecole normale sup\'{e}rieure, ENS, Universit\'{e} PSL, CNRS, Sorbonne Universit\'{e}, Universit\'{e} Paris-Diderot, Sorbonne Paris Cit\'{e}, 75005 Paris, France}

\author{Zlatko Papi\'c}
\email{z.papic@leeds.ac.uk}
\affiliation{School of Physics and Astronomy, University of Leeds, Leeds LS2 9JT, United Kingdom}

\date{\today}

\begin{abstract}

Quantum Hall bilayers are a uniquely tunable platform that can realize continuous transitions between distinct topological phases of matter. One prominent example is the transition between the Halperin state and the Moore--Read Pfaffian, long predicted to host a critical theory of Majorana fermions but so far not verified in unbiased microscopic simulations. Using the fuzzy sphere regularization,  we identify the low-energy spectrum at this transition with the 3D gauged Majorana conformal field theory. We show that the transition is driven by the closing of the neutral fermion gap, and we directly extract the operator content in both integer and half-integer spin sectors.  Our results resolve the long-standing question of the nature of a topological phase transition in a setting relevant to quantum Hall experiments, while also providing a realization of emergent fermionic fields on the fuzzy sphere, previously limited to bosonic fields.
\end{abstract}

\maketitle

{\bf \em Introduction.---}The fractional quantum Hall (FQH) effect has long served as a fertile ground for discovering new phases of matter that defy the Landau paradigm~\cite{Tsui82,Laughlin83}. It brought into the spotlight many-body phases with exotic quasiparticles called anyons~\cite{Leinaas77,Wilczek82}, some of which could prove useful for fault-tolerant quantum computing~\cite{Nayak08,pachos2012introduction}. A prominent example is the even-denominator state at filling factor $\nu=5/2$~\cite{Willett87}, believed to be described by the Moore--Read wave function~\cite{Moore91}. The elementary excitations of the Moore--Read state are anyons with charge $e/4$ and  non-Abelian braiding statistics~\cite{Nayak96,Read00}. Due to the existence of anyons, it is generally challenging to study topological phase transitions between different FQH states. For example, despite theoretical evidence~\cite{Morf98,Rezayi00}, experiments have yet to unambiguously determine the nature of the $\nu=5/2$ state~\cite{Banerjee2018, Zibrov16, Li17, Zibrov17, Falson18, Dutta21, Huang23, Xiaomeng2022, Farahi2023, Willett2023, Hu2025}.

\begin{figure}[tb]
    \centering
    \includegraphics[width=0.95\linewidth]{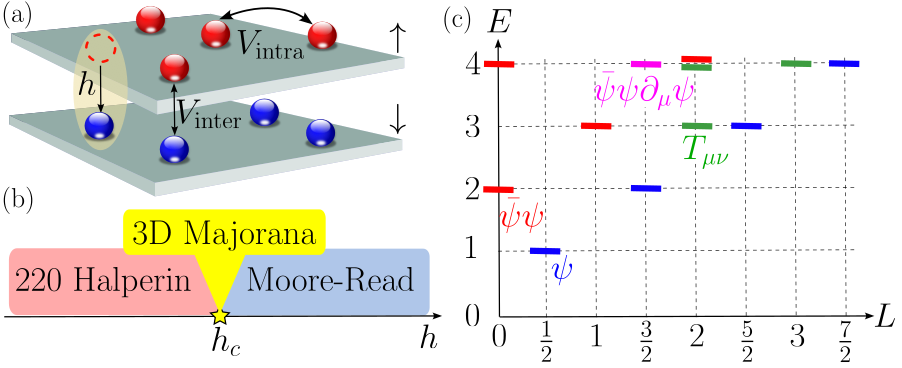}
    \caption{(a) Quantum Hall bilayer system with the layers $\uparrow$, $\downarrow$. The particles interact via intralayer ($V_\mathrm{intra}$) and interlayer ($V_\mathrm{inter}$) interactions, and they can tunnel between the layers with an amplitude $h$. (b) At a critical tunneling $h_c$, the bosonic particles at filling $\nu=1$ undergo a transition between the 220 Halperin state and the Moore--Read state. (c) At criticality, the energy spectrum on the fuzzy sphere, resolved as a function of angular momentum, matches that of the free 3D Majorana fermion conformal field theory, with the characteristic towers of primary fields (labelled) and their descendants.
    }
    \label{fig:summary}
\end{figure}

Multicomponent quantum Hall systems, such as bilayers and wide quantum wells, offer a complementary platform for exploring a broader class of FQH states and transitions between them~\cite{Girvin07,Eisenstein14}. These systems are highly tunable, as the effective interactions and tunneling can be controlled via electrostatic gates and by varying the magnetic length, see Fig.~\ref{fig:summary}(a). Such tuning can induce quantum phase transitions between FQH phases, including the one at filling $\nu=1/2$, observed in numerous experiments~\cite{Suen92, Eisenstein92, Suen94b, Shabani13, Singh24, Singh25}. The transition is believed to occur between a two-component Halperin state~\cite{Halperin83} and a single-component Moore--Read state~\cite{He93, Nomura04, Papic10, Peterson10a, Peterson10b, Liu16, Zhu16, Cabra2001, CAPPELLI2001, Repellin2015, Crepel2019}, Fig.~\ref{fig:summary}(b). Both of these gapped phases represent weakly-paired states of composite fermions~\cite{Jain89}. The tunneling-driven critical point between them  is predicted to be governed by a massless Majorana fermion~\cite{Wen00, Read00, Barkeshli10Bilayer,Barkeshli11Bilayer} -- the long-sought particle that is its own antiparticle~\cite{Elliott15,YazdaniMajoranaReview}. While the $\nu=1/2$ transition was analyzed at the level of mean-field theory in Ref.~\cite{Read00} more than two decades ago, its identification in a microscopic model has remained elusive.

In this paper, we employ the recently developed fuzzy-sphere regularization~\cite{Zhu23} to provide the first microscopic verification of the 3D gauged Majorana conformal field theory (CFT) emerging in the quantum Hall bilayer in the presence of interlayer tunneling. To minimize finite-size effects, we assume bosonic particles at filling $\nu=1$ and study the transition between the two-component 220 Halperin (i.e., akin to two decoupled Laughlin states at $\nu=1/2$) and a single-component Moore--Read state. We numerically demonstrate the closing of the gap and extract the operator content of the CFT, including both integer and half-integer spin sectors, as summarized in Fig.~\ref{fig:summary}(c). 
These results establish fingerprints of an emergent Majorana fermion in a setting applicable to quantum Hall experiments. Beyond quantum Hall, they also open a new direction for the fuzzy sphere studies of 3D CFTs, which so far have been restricted to bosonic fields.

{\bf \em Gauged Majorana field theory.---}
The continuous Halperin-Pfaffian transition is driven by charge-neutral pairs of quasiparticles and quasiholes shared between the two layers~\cite{Read00, Wen00}. The critical theory is defined by the Lagrangian
\begin{equation}
		\mathcal{L} = \bar{\psi} (i \gamma_\mu\partial^\mu - m) \psi \, ,
\end{equation}
in which the Majorana fermion $\psi$ appears as a 2-component spinor that is indistinguishable from its antiparticle. The massless $m=0$ case enjoys an enlarged conformal symmetry. This symmetry imposes strong constraints on the operator content, organizing operators in conformal towers corresponding to different primaries of the theory \cite{DiFrancesco97, Zhu23}, as summarized in \cref{fig:summary}(c). The non-interacting Majorana field brings additional constraints. On the one hand, the equation of motion, $\gamma_{\mu}\partial^\mu \psi = 0$, fixes the descendants of $\psi$ to only $\partial_{\mu_1}\dots\partial_{\mu_l}\psi$, with scaling dimension $\Delta = l+1$ and SO(3) angular momentum $L = l + 1/2$. On the other hand, the reality condition also reduces the number of operators. For example, this theory does not have a conserved vector current, as $J^\mu = \bar{\psi}\gamma^\mu\psi$ vanishes identically. Through Fierz identities, this also implies that any simple multiple-fermion terms beyond $\psi$ and $\bar{\psi}\psi$ vanish. 

The emergence of conformal symmetry can be observed in microscopic models defined on spherical manifolds $S^{d-1}\times\mathbb{R}$ through the state-operator correspondence~\cite{Cardy84, Cardy85}. While historically this has presented difficulties for lattice models in $d>2$, the standard setup of FQH spherical geometry with a magnetic monopole placed at the center of the sphere~\cite{Haldane83} elegantly removes these problems. 
This new ``fuzzy sphere'' method of regularization~\cite{Zhu23} has already proven successful at extracting spectral data of many bosonic 3D CFTs~\cite{Zhu23, Hu23, Hu25, Han23, Hofmann24, Zhou25c,  Fardelli25, Han24, Fan25, Cruz25, Miro25, Dedushenko24, Zhou25a, Zhou24a, Yang25, Zhou25b, Taylor25, He25}. These studies realized CFT critical points by tuning an integer quantum Hall state of electrons through a spin transition. However, as shown in Ref.~\cite{Voinea25}, the fuzzy sphere regularization remains powerful at fractional fillings, which also allows to use bosons as underlying degrees of freedom. We will follow the latter approach to directly extract the universal description of the Halperin-to-Pfaffian transition from the fuzzy sphere numerics.  

We note that the Halperin-Pfaffian critical point differs in a few subtle respects from the conventional 3D Majorana CFT. The gauging of the global $\mathbb{Z}_2$ layer-exchange symmetry~\cite{Barkeshli11Bilayer,Teo15} constrains the excitations of half-integer spin, such as $\psi$, to live in a different Hilbert space compared to the integer-spin ones, such as $\bar{\psi}\psi$ and $T^{\mu\nu}$. This is equivalent to the insertion of a gauge charge at the center of the sphere; specifically, increasing both the magnetic flux  and the number of particles by $1$, the insertion of a $2\pi$-flux of $U(1)$ electric charge is bounded with a $\mathbb{Z}_2$ gauge charge. For this reason, the elementary field $\psi$ is a non-local fermion that carries gauge charge, and our critical theory is the \emph{gauged} 3D Majorana fermion. In our numerics below, we will identify the Hilbert space sectors and focus on the bare fermion field $\psi$, the fermion mass term $\bar{\psi}\psi$, the stress energy tensor $T^{\mu\nu} = \bar{\psi} \gamma^\mu \partial^\nu \psi$, and also the next half-integer spin primary $\bar{\psi} \psi \partial^\mu \psi$, see Supplementary Material (SM)~\cite{SM} for further details.

{\bf \em Model and phase diagram.---}We consider bosons at Landau level filling $\nu=1$ on a fuzzy sphere. The monopole strength $Q$ is related to the number of particles $N$ as $2Q = \nu^{-1}N - \mathcal{S}$, where $\mathcal{S}=2$ is the Wen-Zee shift~\cite{Wen92}. As illustrated in Fig.~\ref{fig:summary}(a), the Hamiltonian
$H=H_\text{intra} + H_\text{inter} + H_\text{t}$ contains intra- and interlayer interactions, as well as the tunneling term. Using the bosonic operator $b_a^\dagger(\mathbf{\Omega})$---which creates a particle in layer $a=\uparrow,\downarrow$ at a spherical angle $\mathbf{\Omega}$---and the corresponding density operator,  $n_a (\mathbf{\Omega}) \equiv b_a^\dagger(\mathbf{\Omega}) b_a(\mathbf{\Omega})$, the three terms in the Hamiltonian are given by
\begin{align}
\label{eq:hamiltonian}
H_\text{intra}  &= \sum_{a = \uparrow,\downarrow} \int V^\text{intra}(\mathbf{\Omega}_{12}) \, :n_a (\mathbf{\Omega}_1) n_a (\mathbf{\Omega}_2): \mathrm{d}^2 \mathbf{\Omega}_1 \mathrm{d}^2 \mathbf{\Omega}_2 \,, \nonumber  \\
    H_\text{inter} &= 2\int V^\text{inter}(\mathbf{\Omega}_{12}) \, n_\uparrow (\mathbf{\Omega}_1) n_\downarrow(\mathbf{\Omega}_2) \, \mathrm{d}^2 \mathbf{\Omega}_1 \mathrm{d}^2 \mathbf{\Omega}_2  \,, \\
    H_\text{t} &= - h\int  \left( b_\uparrow^\dagger(\mathbf{\Omega})b_\downarrow(\mathbf{\Omega}) + \mathrm{h{.}c{.}} \right) \mathrm{d}^2 \mathbf{\Omega} \,. \nonumber   
\end{align}    
Here, $h$ is the tunneling amplitude, and we take the interactions to be short-ranged, parametrized exclusively by $V_0^\text{intra}$ (set to $1$) and $V_0^\text{inter}$ Haldane pseudopotentials \cite{Haldane83}. The model above is invariant under SO(3) rotations, while finite $h$ leaves an additional $\mathbb{Z}_2$ symmetry that exchanges the two layers. 

At weak tunneling $h \ll V^\text{intra}, V^\text{inter}$, the ground state is the 220 Halperin state~\cite{Halperin83}:
\begin{equation}\label{eq:220 wf}
    \Psi_{220}(\{ u_i^a, v_i^a\}) = \prod_{i<j}^{N/2}(u^\uparrow_i v^\uparrow_j - v^\uparrow_i u^\uparrow_j)^2 \prod_{i<j}^{N/2}(u^\downarrow_i v^\downarrow_j - v^\downarrow_i u^\downarrow_j)^2 \, ,
\end{equation}
written in terms of standard spinor coordinates $u_j = \cos(\theta_j/2)\exp(i\phi_j/2)$, $v_j = \sin(\theta_j/2)\exp(-i\phi_j/2)$ on the fuzzy sphere. This state is a direct generalization of the Laughlin state~\cite{Laughlin83} to two species of particles and, for our model, it is the \emph{exact} ground state in the limit $h=0$ and $V_0^\text{inter}=0$~\cite{Haldane83}, while there is strong numerical evidence for its stability over a wider range of interactions~\cite{Liu16}.  

Tunneling can be viewed as a symmetrizer over the layer indices. Symmetrization of Eq.~(\ref{eq:220 wf}), via a Cauchy identity~\cite{Ho95}, results in the
Moore--Read Pfaffian state~\cite{Moore91}:
\begin{equation}\label{eq:pf wf}
    \Psi_\mathrm{Pf}(\{u_i, v_i\}) = \mathrm{Pf} \left( \frac{1}{u_i v_j - v_i u_j} \right) \prod_{i<j}^{N}(u_i v_j - v_i u_j) \, .
\end{equation}
One expects this wave function to describe the ground state in the limit $h \gg V^\text{intra}, V^\text{inter}$ since layer symmetrization leads to an effective single-component system described by the interaction $V_0 = (V_0^\text{inter} + V_0^\text{intra})/2$~\cite{Papic10}, whose ground state is in the Moore--Read phase~\cite{Cooper01, Regnault03}.

\begin{figure}[tb]
    \centering
    \includegraphics[width=0.48\textwidth]{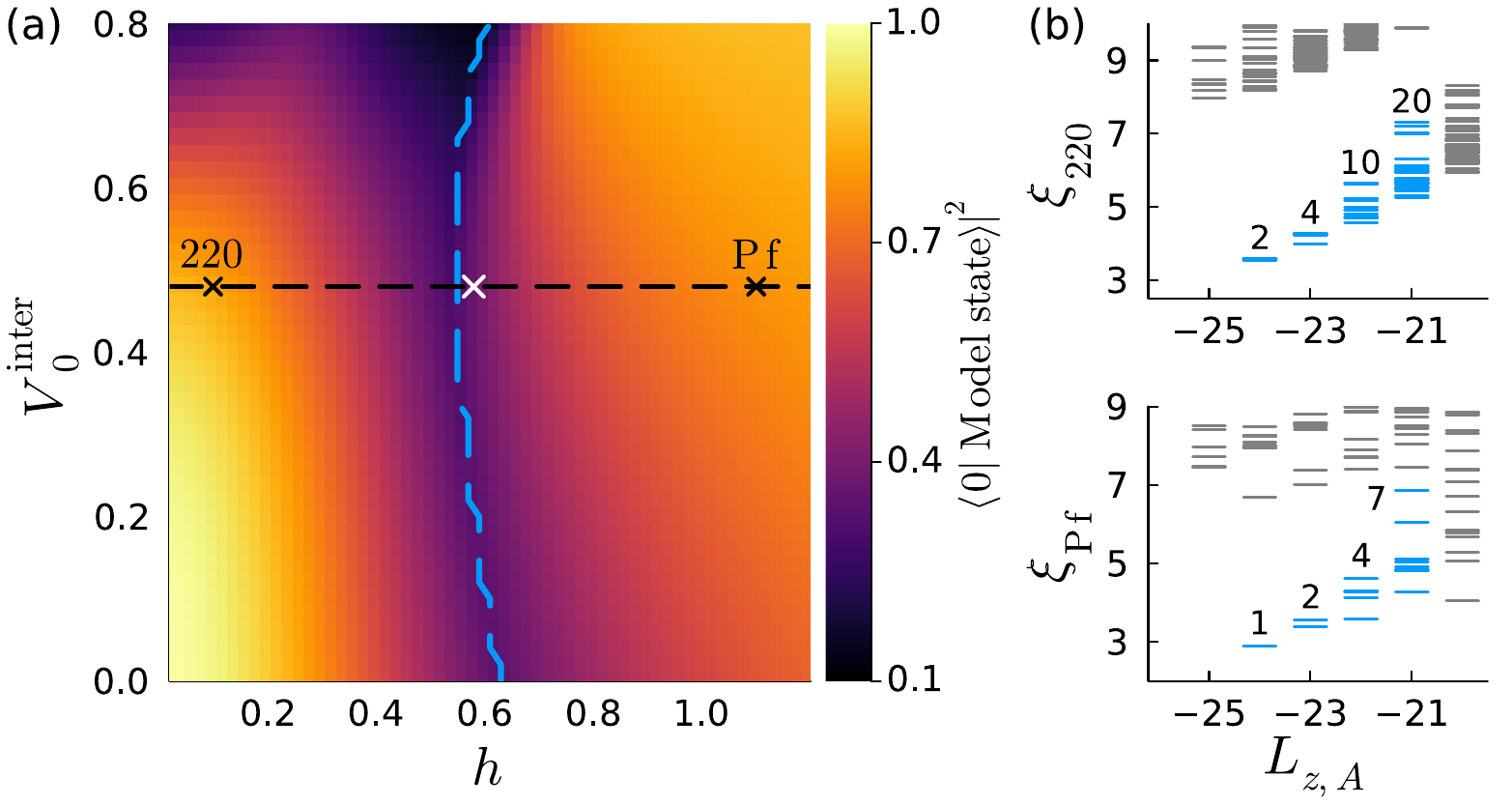}
    \caption{(a) Phase diagram of the model in Eq.~(\ref{eq:hamiltonian}), calculated using overlaps with the states in Eqs.~(\ref{eq:220 wf})-(\ref{eq:pf wf}). The color intensity represents the maximum overlap, $\max \left(|\langle 0 | \Psi_{220} \rangle|^2, |\langle 0 | \Psi_\mathrm{Pf} \rangle|^2 \right)$, while the blue dashed line is the approximate phase boundary where the two overlaps are equal. The white cross at $(V_0^\text{inter},h) = (0.48, 0.58)$ marks the gap closing point in Fig.~\ref{fig:cost_functions}.  (b) The real-space entanglement spectra for one point inside each phase, denoted by black crosses in (a). The multiplicities of the low-lying entanglement levels $\xi$, resolved by the $z$-component of angular momentum in the subsystem $A$, are indicated by numbers and match those of the respective model states (see text).
    Data in (a) are obtained by exact diagonalization for $N=12$ bosons, while (b) is for $N=14$, the largest accessible system size with Hlibert space dimension 87,150,620.
    }
    \label{fig:overlap_phase_diagram}
\end{figure}

The topological transition between the states in Eqs.~(\ref{eq:220 wf})-(\ref{eq:pf wf}), driven by $h$, lacks a local order parameter. We use the overlap with the model wave functions to map out the phase diagram in \cref{fig:overlap_phase_diagram}(a). The phase is identified by $\max \left(|\langle 0 | \Psi_{220} \rangle|^2, |\langle 0 | \Psi_\mathrm{Pf} \rangle|^2 \right)$, where $|0\rangle$ is the exact ground state obtained by exact diagonalization, and we draw an approximate phase boundary where the two overlaps are equal. 

As additional evidence, we identify each of the topological phases from their entanglement spectra~\cite{Li08}. Under a real-space bipartition~\cite{Dubail12,Sterdyniak12}, the counting of the energy levels of the entanglement Hamiltonian should reflect the edge theory of the bulk phase. Accordingly, the 220 Halperin phase exhibits two chiral bosons on the edge, while the Pfaffian phase has a chiral boson and Majorana fermion~\cite{XGWen1993,Read1996}; the replacement of the neutral bosonic edge mode by a single Majorana field at the transition is consistent with the 3D gauged Majorana description of the critical point. We checked the sector with half the total number of bosons in a hemisphere, where we obtain the expected state countings, e.g. $(2,4,10,20,\dots)$ for the Halperin state and $(1,2,4,7,\dots)$ for the Pfaffian state when each subsystem contains an odd number of particles, see Fig.~\ref{fig:overlap_phase_diagram}(b). 
We note that the exact ground state is generally expected to reproduce these countings only in a finite number of sectors, below the so-called ``entanglement gap''~\cite{Li08}, as seen in Fig.~\ref{fig:overlap_phase_diagram}(b). 

{\bf \em Critical point.---} At finite $h$, the charge-neutral pairs of quasiholes and quasiparticles lead to two species of decoupled fermions,
which are even and odd, respectively, under $\mathbb{Z}_2$ layer exchange. At the critical point, one species becomes massless, with its parity selected by the sign of $h$. For $h>0$ in Eq.~(\ref{eq:hamiltonian}), the gapless fermion lies in the $\mathbb{Z}_2$-odd sector. Consequently, to identify the operator spectrum of the 3D Majorana CFT we focus on two different sectors: the odd-particle, $\mathbb{Z}_2$-odd sector, corresponding to fermionic operators, and the even-particle, $\mathbb{Z}_2$-even sector, containing bosonic operators.

\begin{figure}[tb]
    \centering
    \includegraphics[width=0.48\textwidth]{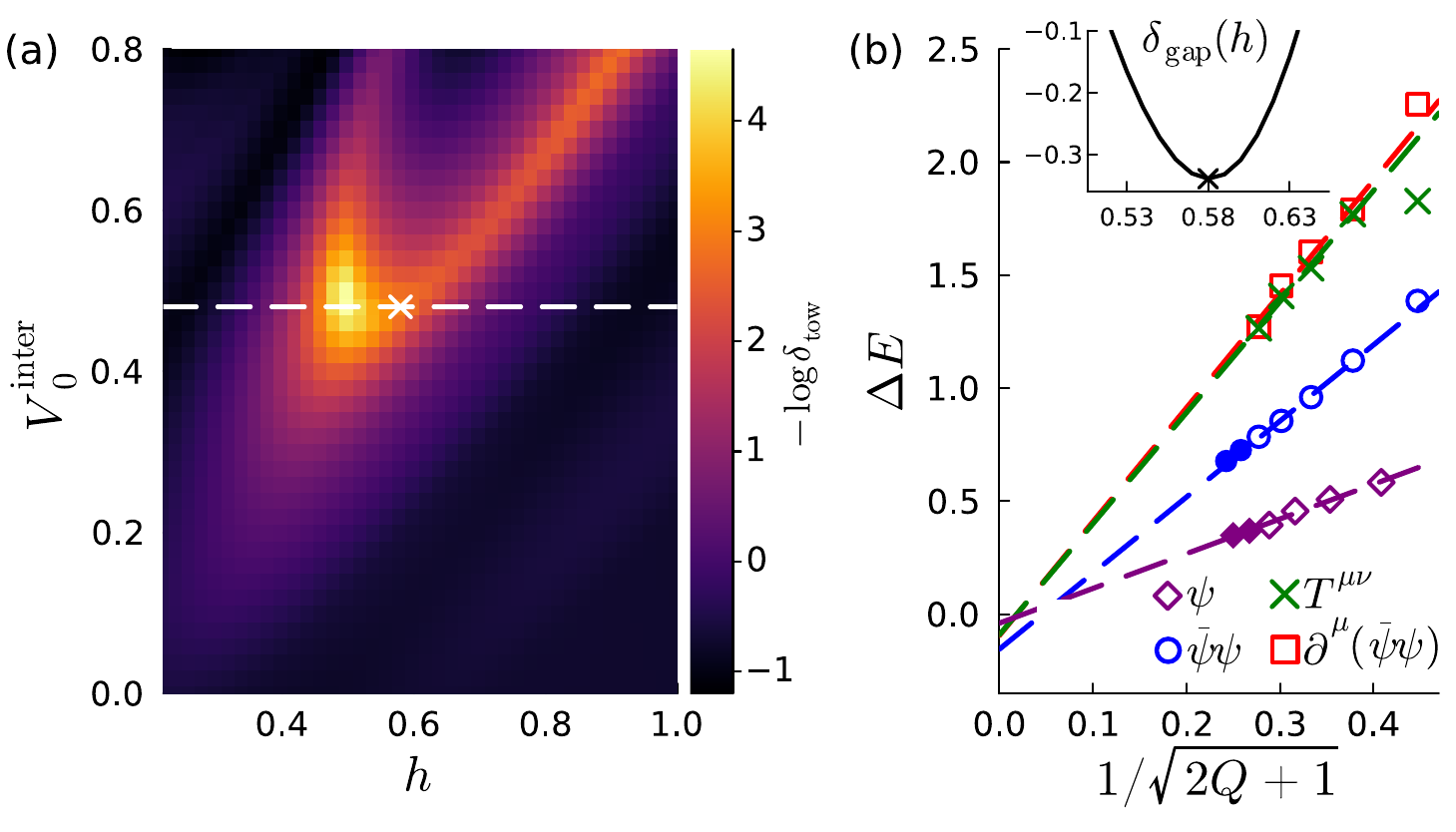}
    \caption{(a) The cost function $\delta_\mathrm{tow}$ quantifying the conformal structure in the low-lying spectrum across the phase diagram at a fixed system size $N=12$. A sharp peak occurs along the dashed line at $V_0^\text{inter}=0.48$. The optimal point (white cross) is further determined by the gap vanishing in (b).  (b) Extrapolated gaps of $\{ \bar{\psi}\psi, \partial(\bar{\psi}\psi), T \}$ at the optimal point $(V_0^\text{inter},h) = (0.48, 0.58)$, identified from the minimum of the gaplessness cost function $\delta_\text{gap}$ shown in the inset. Although not explicitly enforced, the gap of the neutral fermion $\psi$ also converges to zero at this point, a signature of a quantum critical point in both even- and odd-particle sectors. Empty markers correspond to exact diagonalization data, while solid markers ($N=15-18$) are obtained from DMRG and are not used in the extrapolations.}
    \label{fig:cost_functions}
\end{figure}

Before analyzing the spectrum, we need to identify the optimal point with minimal finite-size effects along the phase boundary in \cref{fig:overlap_phase_diagram}(a). We use two complementary approaches for this. First, to narrow down the space of parameters, we perform an optimization over the spectrum's conformal structure. We consider states in the (fixed) $N$-even sector with $\Delta\leq4$ and $L\leq2$, with the conformal tower cost function $\delta_\text{tow} = |\boldsymbol{\Delta}|^2 - |\boldsymbol{\Delta}\cdot \mathbf{E}|^2/|\mathbf{E}|^2$, where $\boldsymbol{\Delta}$ is the vector containing the operators' CFT scaling dimensions and $\mathbf{E}$ is the vector of the corresponding microscopic energies.  We plot $\delta_\text{tow}$ across the phase diagram in \cref{fig:cost_functions}(a), revealing a strong peak in the vicinity of $V_0^\text{inter} \approx 0.48$. We fix this value of $V_0^\text{inter}$ as optimal, and study the finite-size scaling of the excitation gap to determine the optimal $h$. 

Based on CFT prediction, $\Delta E {\sim} 1/R {\sim} 1/\sqrt{2Q+1}$, we construct the cost function $\delta_\text{gap} = \sum_i \Delta E_i(N {\to}\infty)$ that minimizes the sum of linear $1/R$ extrapolations for a set of gaps corresponding to $\{ \bar{\psi}\psi, \partial^\mu( \bar{\psi}\psi), T^{\mu\nu}\}$. The minimum of $\delta_\text{gap}$ is achieved at the optimal $h=0.58$, for which the gaps of previously mentioned states vanish, \cref{fig:cost_functions}(b). The (slightly) negative values of these extrapolated gaps in \cref{fig:cost_functions}(b) are attributed to the remnant finite-size effects which, as expected, are more pronounced for higher-energy states. As a consistency check, we also computed the lowest gap in the $N$-odd sector, corresponding to the neutral fermion $\psi$. Since this sector contains no conformal vacuum, the gap is calculated with respect to the averaged vacuum energy of the adjacent even system sizes ($N\pm 1$), and the convergence to zero is remarkably accurate.

\begin{figure}[tb]
    \centering
    \includegraphics[width=0.49\linewidth]{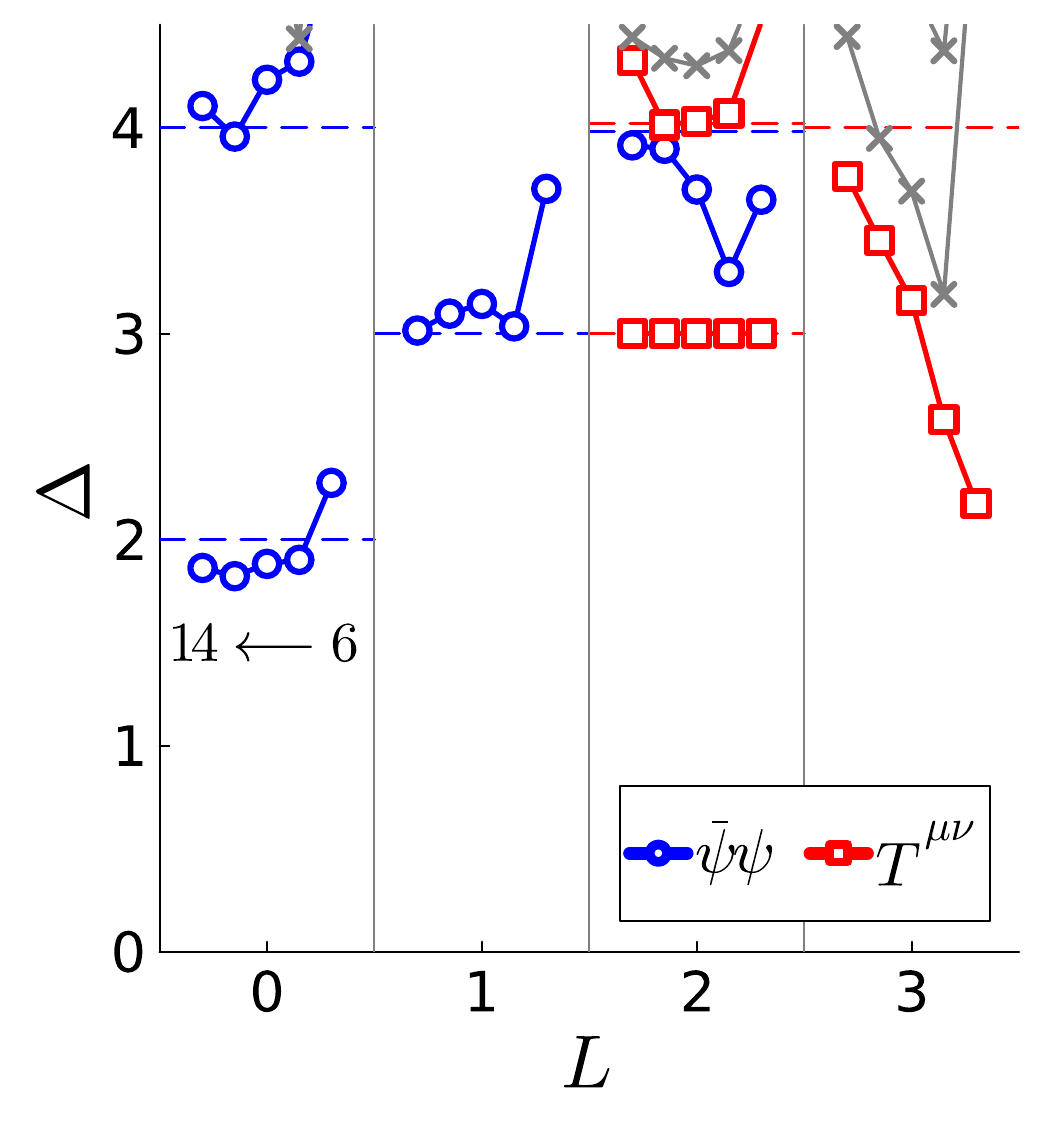}
    \includegraphics[width=0.49\linewidth]{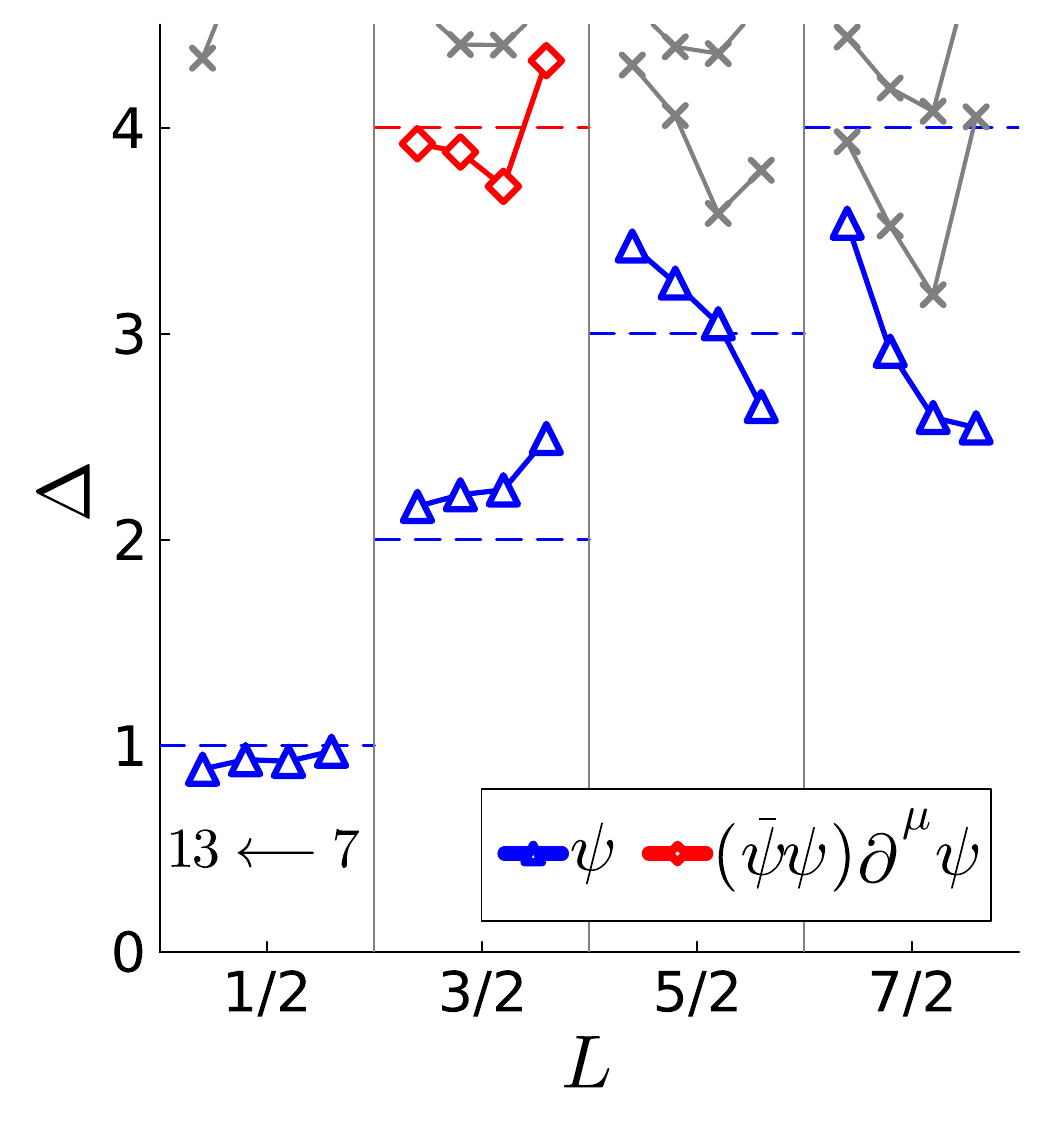}    
    \caption{Comparison between CFT data and the Hamiltonian spectrum at the optimal point $(V_0^\text{inter},h) = (0.48, 0.58)$. (a) Even-particle sector for $N=6,8,10,12,14$, containing integer angular momentum states. The lowest energy state in $L=2$ sector is taken to be the stress-energy tensor, whose energy was fixed to $\Delta = 3$. 
    (b) Odd-particle sector for $N=7,9,11,13$, containing half-integer angular momentum states. Energy levels are normalized with respect to the mean energies of the vacuum and stress-energy tensors in the adjacent, $N\pm 1$, even-particle sectors. 
    Both panels show the complete energy spectra for levels with $\Delta \leq 4.5$, $L \leq 7/2$. The expected CFT operators are labeled, with their scaling dimensions shown by dashed lines. Other signatures of the free Majorana CFT in these spectra also include the \emph{absence} of certain states, such as the conserved vector current, $J^\mu = \bar{\psi}\gamma^\mu\psi$, at $(L=1, \Delta = 2)$, and the state at $(L=1/2, \Delta=2)$, which is absent due to the equation of motion $\gamma_\mu \partial^\mu \psi = 0$. }
    \label{fig:spectra}
\end{figure}

{\bf \em Conformal spectrum.---}State-operator correspondence at the optimal critical point is shown in \cref{fig:spectra}. The ($N$-even, $\mathbb{Z}_2$-even) sector can be seen in \cref{fig:spectra}(a), where all energies have been rescaled such that the stress-energy tensor---the lowest-lying state at $L=2$---has $\Delta = 3$~\cite{Zhu23}. This sector is identified with the integer angular momentum sector of the 3D Majorana CFT, and a good agreement is obtained for the primaries $\bar{\psi}\psi$, $T^{\mu\nu}$, and their descendants. The absence of a conserved vector current, $J^\mu = \bar{\psi}\gamma^\mu\psi$ at $(L=1, \Delta=2)$ is also notable, further supporting the Majorana CFT nature of our critical point. 

Similarly, \cref{fig:spectra}(b) shows the ($N$-odd, $\mathbb{Z}_2$-odd) sector, which we identify as the half-integer angular momentum sector of the CFT. This sector contains no conformal vacuum, hence we rescale each system size using the average energies of the vacuum and stress-energy tensor of the adjacent \emph{even} sizes ($N\pm 1$). State-operator correspondence is clearly manifested for the primaries $\psi$ and $\bar{\psi}\psi\partial^\mu\psi$ and a few of their descendants. The non-interacting nature of the critical point is also highlighted by the absence of the $\gamma_\mu \partial^\mu \psi$ descendant at $(L=1/2, \Delta=2)$. 

A few comments are in order. For higher levels, $\Delta \geq 4$, the state-operator correspondence becomes less clear, with strong finite-size effects and additional states mixing in with the conformal towers. In other theories such as the 3D Ising CFT, conformal perturbation theory was useful in quantifying such finite-size effects~\cite{Lao23, Lauchli25}. However, the sparse operator-product expansion structure of the free Majorana CFT implies that the fermion mass $\bar{\psi}\psi$---the only relevant scalar operator---does not produce corrections to the lowest excited states in either even- or odd-particle sectors. Hence, the conformal perturbation approach cannot be directly applied to improve the matching of microscopic spectra with CFT beyond Fig.~\ref{fig:spectra}. Furthermore, as in previous work~\cite{Voinea25}, our model realizes a CFT which is space-time parity symmetric, but with no corresponding symmetry at the microscopic level. A potential operator that implements the parity symmetry may be related to the particle-hole (PH) symmetry of spinful bosons at $\nu=1$~\cite{Wang16, Geraedts17}, which is only emergent in the thermodynamic limit. Moreover, the microscopic parity operator appears non-trivial to write down since our calculations are performed at a non-PH-symmetric flux value, dictated by the Wen-Zee shift of the Pfaffian and Halperin states. 

{\bf \em Experimental implications.---}Droplets of bosonic FQH states have recently been realized in ultracold atoms~\cite{Leonard2023} and cavity-QED experiments~\cite{CanWang2024}. While the ultra short-range $V_0$ interactions in our model are native to such platforms, the two-component and Moore--Read states have yet to be realized. Moreover, such platforms are currently limited to a few bosons and far from the scaling regime required to observe the critical behavior. 

In solid state materials, the fermionic version of the transition studied above occurs at filling $\nu=1/2$ between the 331 Halperin and (fermionic) Moore--Read state. There is extensive experimental evidence for the 331 Halperin state and transition in both GaAs bilayers and wide quantum wells~\cite{Suen92, Eisenstein92, Suen94b, Shabani13}. Recent studies found evidence for the one-component state being the Moore-Read state based on the observation of Pfaffian daughter states surrounding $\nu=1/2$~\cite{Singh24,Singh25}, the nature of the one-component state past the transition has remained unclear. However, these studies were limited to transport, which probes the charge gap of the system and the latter may remain open throughout the transition~\cite{Zhu23}. By contrast, light-scattering probes, as in the recent experiment~\cite{Liang2024}, directly probe neutral excitations and would detect the transition. In a bilayer heterostructure with independent electric contacts to each layer~\cite{Spielman00}, it is possible to directly measure the interlayer tunneling current. The temperature dependence of this current is predicted to be $\langle \varphi^\dagger_{\uparrow} \varphi_{\downarrow}(x) \rangle \sim T^2$, since the tunneling term effectively relates to the mass of Majorana fermions, $\bar{\psi}\psi (x)$. Verification of this scaling would not only serve as evidence for the Majorana nature of the critical point, but it would also lend strong support to the identification of the Moore--Read state in the gapped phase. 

{\bf \em Conclusions.---}We have microscopically established the emergence of a gauged 3D Majorana fermion at the Halperin–Pfaffian transition. Using the fuzzy sphere regularization, we have identified the hallmarks of the corresponding CFT: gap closing and operator content. These results not only offer a fresh perspective on the long-standing problem of non-Abelian states in quantum Hall bilayers, but also provide the first unbiased microscopic dissection of a continuous FQH transition, which could be directly applied to other universality classes~\cite{Barkeshli11Bilayer,Zhang23XYstar,Barkeshli_2014}. It would also be important to study interfaces between Halperin and Pfaffian states~\cite{Yang17, Ma22, Crepel19}, which are directly tied to the critical theory discussed here. Beyond the quantum Hall setting, our work realizes the emergent fermionic fields within the fuzzy sphere framework, setting the stage for exploring interacting theories such as Yukawa CFTs~\cite{Fei16,gao2025interactingcherninsulatortransition}.

\textit{Note Added:} During the completion of this work, we became aware of Ref.~\cite{Zhou25_majorana} that realized the free Majorana fermion CFT in a different microscopic model. 

\begin{acknowledgments}

{\bf \em Acknowledgements.---}We thank Zheng Zhou and Yin-Chen He for useful discussions, and for sharing their unpublished results with us. C.V. and Z.P. acknowledge support by the Leverhulme Trust Research Leadership Award RL-2019-015 and EPSRC Grant EP/Z533634/1, UKRI1337. This research was supported in part by grant NSF PHY-2309135 to the Kavli Institute for Theoretical Physics (KITP). Computational portions of this research have made use of DiagHam~\cite{diagham} and FuzzifiED~\cite{FuzzifiED} software libraries, and they were carried out on Aire, part of the High-Performance Computing facilities at the University of Leeds.
W.Z. was supported by NSFC under No. 12474144. The Flatiron Institute is a division of the Simons Foundation.
\end{acknowledgments}

{\bf \em Data availability.---}The data that support the findings of this article are openly available~\cite{data}.

\bibliography{bibliography}

\begin{thebibliography}{118}%
\makeatletter
\providecommand \@ifxundefined [1]{%
 \@ifx{#1\undefined}
}%
\providecommand \@ifnum [1]{%
 \ifnum #1\expandafter \@firstoftwo
 \else \expandafter \@secondoftwo
 \fi
}%
\providecommand \@ifx [1]{%
 \ifx #1\expandafter \@firstoftwo
 \else \expandafter \@secondoftwo
 \fi
}%
\providecommand \natexlab [1]{#1}%
\providecommand \enquote  [1]{``#1''}%
\providecommand \bibnamefont  [1]{#1}%
\providecommand \bibfnamefont [1]{#1}%
\providecommand \citenamefont [1]{#1}%
\providecommand \href@noop [0]{\@secondoftwo}%
\providecommand \href [0]{\begingroup \@sanitize@url \@href}%
\providecommand \@href[1]{\@@startlink{#1}\@@href}%
\providecommand \@@href[1]{\endgroup#1\@@endlink}%
\providecommand \@sanitize@url [0]{\catcode `\\12\catcode `\$12\catcode
  `\&12\catcode `\#12\catcode `\^12\catcode `\_12\catcode `\%12\relax}%
\providecommand \@@startlink[1]{}%
\providecommand \@@endlink[0]{}%
\providecommand \url  [0]{\begingroup\@sanitize@url \@url }%
\providecommand \@url [1]{\endgroup\@href {#1}{\urlprefix }}%
\providecommand \urlprefix  [0]{URL }%
\providecommand \Eprint [0]{\href }%
\providecommand \doibase [0]{https://doi.org/}%
\providecommand \selectlanguage [0]{\@gobble}%
\providecommand \bibinfo  [0]{\@secondoftwo}%
\providecommand \bibfield  [0]{\@secondoftwo}%
\providecommand \translation [1]{[#1]}%
\providecommand \BibitemOpen [0]{}%
\providecommand \bibitemStop [0]{}%
\providecommand \bibitemNoStop [0]{.\EOS\space}%
\providecommand \EOS [0]{\spacefactor3000\relax}%
\providecommand \BibitemShut  [1]{\csname bibitem#1\endcsname}%
\let\auto@bib@innerbib\@empty
\bibitem [{\citenamefont {Tsui}\ \emph {et~al.}(1982)\citenamefont {Tsui},
  \citenamefont {Stormer},\ and\ \citenamefont {Gossard}}]{Tsui82}%
  \BibitemOpen
  \bibfield  {author} {\bibinfo {author} {\bibfnamefont {D.~C.}\ \bibnamefont
  {Tsui}}, \bibinfo {author} {\bibfnamefont {H.~L.}\ \bibnamefont {Stormer}},\
  and\ \bibinfo {author} {\bibfnamefont {A.~C.}\ \bibnamefont {Gossard}},\
  }\bibfield  {title} {\bibinfo {title} {Two-dimensional magnetotransport in
  the extreme quantum limit},\ }\href
  {https://doi.org/10.1103/PhysRevLett.48.1559} {\bibfield  {journal} {\bibinfo
   {journal} {Phys. Rev. Lett.}\ }\textbf {\bibinfo {volume} {48}},\ \bibinfo
  {pages} {1559} (\bibinfo {year} {1982})}\BibitemShut {NoStop}%
\bibitem [{\citenamefont {Laughlin}(1983)}]{Laughlin83}%
  \BibitemOpen
  \bibfield  {author} {\bibinfo {author} {\bibfnamefont {R.~B.}\ \bibnamefont
  {Laughlin}},\ }\bibfield  {title} {\bibinfo {title} {Anomalous quantum {Hall}
  effect: An incompressible quantum fluid with fractionally charged
  excitations},\ }\href {https://doi.org/10.1103/PhysRevLett.50.1395}
  {\bibfield  {journal} {\bibinfo  {journal} {Phys. Rev. Lett.}\ }\textbf
  {\bibinfo {volume} {50}},\ \bibinfo {pages} {1395} (\bibinfo {year}
  {1983})}\BibitemShut {NoStop}%
\bibitem [{\citenamefont {Leinaas}\ and\ \citenamefont
  {Myrheim}(1977)}]{Leinaas77}%
  \BibitemOpen
  \bibfield  {author} {\bibinfo {author} {\bibfnamefont {J.}~\bibnamefont
  {Leinaas}}\ and\ \bibinfo {author} {\bibfnamefont {J.}~\bibnamefont
  {Myrheim}},\ }\bibfield  {title} {\bibinfo {title} {On the theory of
  identical particles},\ }\href {https://doi.org/10.1007/BF02727953} {\bibfield
   {journal} {\bibinfo  {journal} {Il Nuovo Cimento B Series 11}\ }\textbf
  {\bibinfo {volume} {37}},\ \bibinfo {pages} {1} (\bibinfo {year}
  {1977})}\BibitemShut {NoStop}%
\bibitem [{\citenamefont {Wilczek}(1982)}]{Wilczek82}%
  \BibitemOpen
  \bibfield  {author} {\bibinfo {author} {\bibfnamefont {F.}~\bibnamefont
  {Wilczek}},\ }\bibfield  {title} {\bibinfo {title} {Quantum mechanics of
  fractional-spin particles},\ }\href
  {https://doi.org/10.1103/PhysRevLett.49.957} {\bibfield  {journal} {\bibinfo
  {journal} {Phys. Rev. Lett.}\ }\textbf {\bibinfo {volume} {49}},\ \bibinfo
  {pages} {957} (\bibinfo {year} {1982})}\BibitemShut {NoStop}%
\bibitem [{\citenamefont {Nayak}\ \emph {et~al.}(2008)\citenamefont {Nayak},
  \citenamefont {Simon}, \citenamefont {Stern}, \citenamefont {Freedman},\ and\
  \citenamefont {Das~Sarma}}]{Nayak08}%
  \BibitemOpen
  \bibfield  {author} {\bibinfo {author} {\bibfnamefont {C.}~\bibnamefont
  {Nayak}}, \bibinfo {author} {\bibfnamefont {S.~H.}\ \bibnamefont {Simon}},
  \bibinfo {author} {\bibfnamefont {A.}~\bibnamefont {Stern}}, \bibinfo
  {author} {\bibfnamefont {M.}~\bibnamefont {Freedman}},\ and\ \bibinfo
  {author} {\bibfnamefont {S.}~\bibnamefont {Das~Sarma}},\ }\bibfield  {title}
  {\bibinfo {title} {Non-{Abelian} anyons and topological quantum
  computation},\ }\href {https://doi.org/10.1103/RevModPhys.80.1083} {\bibfield
   {journal} {\bibinfo  {journal} {Rev. Mod. Phys.}\ }\textbf {\bibinfo
  {volume} {80}},\ \bibinfo {pages} {1083} (\bibinfo {year}
  {2008})}\BibitemShut {NoStop}%
\bibitem [{\citenamefont {Pachos}(2012)}]{pachos2012introduction}%
  \BibitemOpen
  \bibfield  {author} {\bibinfo {author} {\bibfnamefont {J.~K.}\ \bibnamefont
  {Pachos}},\ }\href@noop {} {\emph {\bibinfo {title} {Introduction to
  topological quantum computation}}}\ (\bibinfo  {publisher} {Cambridge
  University Press},\ \bibinfo {year} {2012})\BibitemShut {NoStop}%
\bibitem [{\citenamefont {Willett}\ \emph {et~al.}(1987)\citenamefont
  {Willett}, \citenamefont {Eisenstein}, \citenamefont {St\"ormer},
  \citenamefont {Tsui}, \citenamefont {Gossard},\ and\ \citenamefont
  {English}}]{Willett87}%
  \BibitemOpen
  \bibfield  {author} {\bibinfo {author} {\bibfnamefont {R.}~\bibnamefont
  {Willett}}, \bibinfo {author} {\bibfnamefont {J.~P.}\ \bibnamefont
  {Eisenstein}}, \bibinfo {author} {\bibfnamefont {H.~L.}\ \bibnamefont
  {St\"ormer}}, \bibinfo {author} {\bibfnamefont {D.~C.}\ \bibnamefont {Tsui}},
  \bibinfo {author} {\bibfnamefont {A.~C.}\ \bibnamefont {Gossard}},\ and\
  \bibinfo {author} {\bibfnamefont {J.~H.}\ \bibnamefont {English}},\
  }\bibfield  {title} {\bibinfo {title} {Observation of an even-denominator
  quantum number in the fractional quantum {{Hall}} effect},\ }\href
  {https://doi.org/10.1103/PhysRevLett.59.1776} {\bibfield  {journal} {\bibinfo
   {journal} {Phys. Rev. Lett.}\ }\textbf {\bibinfo {volume} {59}},\ \bibinfo
  {pages} {1776} (\bibinfo {year} {1987})}\BibitemShut {NoStop}%
\bibitem [{\citenamefont {Moore}\ and\ \citenamefont {Read}(1991)}]{Moore91}%
  \BibitemOpen
  \bibfield  {author} {\bibinfo {author} {\bibfnamefont {G.}~\bibnamefont
  {Moore}}\ and\ \bibinfo {author} {\bibfnamefont {N.}~\bibnamefont {Read}},\
  }\bibfield  {title} {\bibinfo {title} {Nonabelions in the fractional quantum
  {Hall} effect},\ }\href {https://doi.org/10.1016/0550-3213(91)90407-O}
  {\bibfield  {journal} {\bibinfo  {journal} {Nucl. Phys. B}\ }\textbf
  {\bibinfo {volume} {360}},\ \bibinfo {pages} {362 } (\bibinfo {year}
  {1991})}\BibitemShut {NoStop}%
\bibitem [{\citenamefont {Nayak}\ and\ \citenamefont
  {Wilczek}(1996)}]{Nayak96}%
  \BibitemOpen
  \bibfield  {author} {\bibinfo {author} {\bibfnamefont {C.}~\bibnamefont
  {Nayak}}\ and\ \bibinfo {author} {\bibfnamefont {F.}~\bibnamefont
  {Wilczek}},\ }\bibfield  {title} {\bibinfo {title} {2n-quasihole states
  realize 2(n-1)-dimensional spinor braiding statistics in paired quantum
  {Hall} states},\ }\href {https://doi.org/10.1016/0550-3213(96)00430-0}
  {\bibfield  {journal} {\bibinfo  {journal} {Nucl. Phys. B}\ }\textbf
  {\bibinfo {volume} {479}},\ \bibinfo {pages} {529} (\bibinfo {year}
  {1996})}\BibitemShut {NoStop}%
\bibitem [{\citenamefont {Read}\ and\ \citenamefont {Green}(2000)}]{Read00}%
  \BibitemOpen
  \bibfield  {author} {\bibinfo {author} {\bibfnamefont {N.}~\bibnamefont
  {Read}}\ and\ \bibinfo {author} {\bibfnamefont {D.}~\bibnamefont {Green}},\
  }\bibfield  {title} {\bibinfo {title} {Paired states of fermions in two
  dimensions with breaking of parity and time-reversal symmetries and the
  fractional quantum {Hall} effect},\ }\href
  {https://doi.org/10.1103/PhysRevB.61.10267} {\bibfield  {journal} {\bibinfo
  {journal} {Phys. Rev. B}\ }\textbf {\bibinfo {volume} {61}},\ \bibinfo
  {pages} {10267} (\bibinfo {year} {2000})}\BibitemShut {NoStop}%
\bibitem [{\citenamefont {Morf}(1998)}]{Morf98}%
  \BibitemOpen
  \bibfield  {author} {\bibinfo {author} {\bibfnamefont {R.~H.}\ \bibnamefont
  {Morf}},\ }\bibfield  {title} {\bibinfo {title} {Transition from quantum
  {Hall} to compressible states in the second {Landau} level: New light on the
  $\nu=5/2$ enigma},\ }\href {https://doi.org/10.1103/PhysRevLett.80.1505}
  {\bibfield  {journal} {\bibinfo  {journal} {Phys. Rev. Lett.}\ }\textbf
  {\bibinfo {volume} {80}},\ \bibinfo {pages} {1505} (\bibinfo {year}
  {1998})}\BibitemShut {NoStop}%
\bibitem [{\citenamefont {Rezayi}\ and\ \citenamefont
  {Haldane}(2000)}]{Rezayi00}%
  \BibitemOpen
  \bibfield  {author} {\bibinfo {author} {\bibfnamefont {E.~H.}\ \bibnamefont
  {Rezayi}}\ and\ \bibinfo {author} {\bibfnamefont {F.~D.~M.}\ \bibnamefont
  {Haldane}},\ }\bibfield  {title} {\bibinfo {title} {Incompressible paired
  {Hall} state, stripe order, and the composite fermion liquid phase in
  half-filled {Landau} levels},\ }\href
  {https://doi.org/10.1103/PhysRevLett.84.4685} {\bibfield  {journal} {\bibinfo
   {journal} {Phys. Rev. Lett.}\ }\textbf {\bibinfo {volume} {84}},\ \bibinfo
  {pages} {4685} (\bibinfo {year} {2000})}\BibitemShut {NoStop}%
\bibitem [{\citenamefont {Banerjee}\ \emph {et~al.}(2018)\citenamefont
  {Banerjee}, \citenamefont {Heiblum}, \citenamefont {Umansky}, \citenamefont
  {Feldman}, \citenamefont {Oreg},\ and\ \citenamefont {Stern}}]{Banerjee2018}%
  \BibitemOpen
  \bibfield  {author} {\bibinfo {author} {\bibfnamefont {M.}~\bibnamefont
  {Banerjee}}, \bibinfo {author} {\bibfnamefont {M.}~\bibnamefont {Heiblum}},
  \bibinfo {author} {\bibfnamefont {V.}~\bibnamefont {Umansky}}, \bibinfo
  {author} {\bibfnamefont {D.~E.}\ \bibnamefont {Feldman}}, \bibinfo {author}
  {\bibfnamefont {Y.}~\bibnamefont {Oreg}},\ and\ \bibinfo {author}
  {\bibfnamefont {A.}~\bibnamefont {Stern}},\ }\bibfield  {title} {\bibinfo
  {title} {Observation of half-integer thermal {Hall} conductance},\ }\href
  {https://doi.org/10.1038/s41586-018-0184-1} {\bibfield  {journal} {\bibinfo
  {journal} {Nature}\ }\textbf {\bibinfo {volume} {559}},\ \bibinfo {pages}
  {205} (\bibinfo {year} {2018})}\BibitemShut {NoStop}%
\bibitem [{\citenamefont {{Zibrov}}\ \emph {et~al.}(2017)\citenamefont
  {{Zibrov}}, \citenamefont {{Kometter}}, \citenamefont {{Zhou}}, \citenamefont
  {{Spanton}}, \citenamefont {{Taniguchi}}, \citenamefont {{Watanabe}},
  \citenamefont {{Zaletel}},\ and\ \citenamefont {{Young}}}]{Zibrov16}%
  \BibitemOpen
  \bibfield  {author} {\bibinfo {author} {\bibfnamefont {A.~A.}\ \bibnamefont
  {{Zibrov}}}, \bibinfo {author} {\bibfnamefont {C.~R.}\ \bibnamefont
  {{Kometter}}}, \bibinfo {author} {\bibfnamefont {H.}~\bibnamefont {{Zhou}}},
  \bibinfo {author} {\bibfnamefont {E.~M.}\ \bibnamefont {{Spanton}}}, \bibinfo
  {author} {\bibfnamefont {T.}~\bibnamefont {{Taniguchi}}}, \bibinfo {author}
  {\bibfnamefont {K.}~\bibnamefont {{Watanabe}}}, \bibinfo {author}
  {\bibfnamefont {M.~P.}\ \bibnamefont {{Zaletel}}},\ and\ \bibinfo {author}
  {\bibfnamefont {A.~F.}\ \bibnamefont {{Young}}},\ }\bibfield  {title}
  {\bibinfo {title} {Tunable interacting composite fermion phases in a
  half-filled bilayer-graphene {Landau} level},\ }\href
  {https://doi.org/10.1038/nature23893} {\bibfield  {journal} {\bibinfo
  {journal} {Nature}\ }\textbf {\bibinfo {volume} {549}},\ \bibinfo {pages}
  {360} (\bibinfo {year} {2017})}\BibitemShut {NoStop}%
\bibitem [{\citenamefont {Li}\ \emph {et~al.}(2017)\citenamefont {Li},
  \citenamefont {Tan}, \citenamefont {Chen}, \citenamefont {Zeng},
  \citenamefont {Taniguchi}, \citenamefont {Watanabe}, \citenamefont {Hone},\
  and\ \citenamefont {Dean}}]{Li17}%
  \BibitemOpen
  \bibfield  {author} {\bibinfo {author} {\bibfnamefont {J.~I.~A.}\
  \bibnamefont {Li}}, \bibinfo {author} {\bibfnamefont {C.}~\bibnamefont
  {Tan}}, \bibinfo {author} {\bibfnamefont {S.}~\bibnamefont {Chen}}, \bibinfo
  {author} {\bibfnamefont {Y.}~\bibnamefont {Zeng}}, \bibinfo {author}
  {\bibfnamefont {T.}~\bibnamefont {Taniguchi}}, \bibinfo {author}
  {\bibfnamefont {K.}~\bibnamefont {Watanabe}}, \bibinfo {author}
  {\bibfnamefont {J.}~\bibnamefont {Hone}},\ and\ \bibinfo {author}
  {\bibfnamefont {C.~R.}\ \bibnamefont {Dean}},\ }\bibfield  {title} {\bibinfo
  {title} {Even denominator fractional quantum {Hall} states in bilayer
  graphene},\ }\bibfield  {journal} {\bibinfo  {journal} {Science}\ }\href
  {https://doi.org/10.1126/science.aao2521} {10.1126/science.aao2521} (\bibinfo
  {year} {2017})\BibitemShut {NoStop}%
\bibitem [{\citenamefont {Zibrov}\ \emph {et~al.}(2018)\citenamefont {Zibrov},
  \citenamefont {Spanton}, \citenamefont {Zhou}, \citenamefont {Kometter},
  \citenamefont {Taniguchi}, \citenamefont {Watanabe},\ and\ \citenamefont
  {Young}}]{Zibrov17}%
  \BibitemOpen
  \bibfield  {author} {\bibinfo {author} {\bibfnamefont {A.~A.}\ \bibnamefont
  {Zibrov}}, \bibinfo {author} {\bibfnamefont {E.~M.}\ \bibnamefont {Spanton}},
  \bibinfo {author} {\bibfnamefont {H.}~\bibnamefont {Zhou}}, \bibinfo {author}
  {\bibfnamefont {C.}~\bibnamefont {Kometter}}, \bibinfo {author}
  {\bibfnamefont {T.}~\bibnamefont {Taniguchi}}, \bibinfo {author}
  {\bibfnamefont {K.}~\bibnamefont {Watanabe}},\ and\ \bibinfo {author}
  {\bibfnamefont {A.~F.}\ \bibnamefont {Young}},\ }\bibfield  {title} {\bibinfo
  {title} {Even-denominator fractional quantum {Hall} states at an isospin
  transition in monolayer graphene},\ }\href
  {https://doi.org/10.1038/s41567-018-0190-0} {\bibfield  {journal} {\bibinfo
  {journal} {Nature Physics}\ }\textbf {\bibinfo {volume} {14}},\ \bibinfo
  {pages} {930} (\bibinfo {year} {2018})}\BibitemShut {NoStop}%
\bibitem [{\citenamefont {Falson}\ \emph {et~al.}(2018)\citenamefont {Falson},
  \citenamefont {Tabrea}, \citenamefont {Zhang}, \citenamefont {Sodemann},
  \citenamefont {Kozuka}, \citenamefont {Tsukazaki}, \citenamefont {Kawasaki},
  \citenamefont {von Klitzing},\ and\ \citenamefont {Smet}}]{Falson18}%
  \BibitemOpen
  \bibfield  {author} {\bibinfo {author} {\bibfnamefont {J.}~\bibnamefont
  {Falson}}, \bibinfo {author} {\bibfnamefont {D.}~\bibnamefont {Tabrea}},
  \bibinfo {author} {\bibfnamefont {D.}~\bibnamefont {Zhang}}, \bibinfo
  {author} {\bibfnamefont {I.}~\bibnamefont {Sodemann}}, \bibinfo {author}
  {\bibfnamefont {Y.}~\bibnamefont {Kozuka}}, \bibinfo {author} {\bibfnamefont
  {A.}~\bibnamefont {Tsukazaki}}, \bibinfo {author} {\bibfnamefont
  {M.}~\bibnamefont {Kawasaki}}, \bibinfo {author} {\bibfnamefont
  {K.}~\bibnamefont {von Klitzing}},\ and\ \bibinfo {author} {\bibfnamefont
  {J.~H.}\ \bibnamefont {Smet}},\ }\bibfield  {title} {\bibinfo {title} {A
  cascade of phase transitions in an orbitally mixed half-filled landau
  level},\ }\href {https://doi.org/10.1126/sciadv.aat8742} {\bibfield
  {journal} {\bibinfo  {journal} {Science Advances}\ }\textbf {\bibinfo
  {volume} {4}},\ \bibinfo {pages} {eaat8742} (\bibinfo {year}
  {2018})}\BibitemShut {NoStop}%
\bibitem [{\citenamefont {Dutta}\ \emph {et~al.}(2021)\citenamefont {Dutta},
  \citenamefont {Yang}, \citenamefont {Melcer}, \citenamefont {Kundu},
  \citenamefont {Heiblum}, \citenamefont {Umansky}, \citenamefont {Oreg},
  \citenamefont {Stern},\ and\ \citenamefont {Mross}}]{Dutta21}%
  \BibitemOpen
  \bibfield  {author} {\bibinfo {author} {\bibfnamefont {B.}~\bibnamefont
  {Dutta}}, \bibinfo {author} {\bibfnamefont {W.}~\bibnamefont {Yang}},
  \bibinfo {author} {\bibfnamefont {R.}~\bibnamefont {Melcer}}, \bibinfo
  {author} {\bibfnamefont {H.~K.}\ \bibnamefont {Kundu}}, \bibinfo {author}
  {\bibfnamefont {M.}~\bibnamefont {Heiblum}}, \bibinfo {author} {\bibfnamefont
  {V.}~\bibnamefont {Umansky}}, \bibinfo {author} {\bibfnamefont
  {Y.}~\bibnamefont {Oreg}}, \bibinfo {author} {\bibfnamefont {A.}~\bibnamefont
  {Stern}},\ and\ \bibinfo {author} {\bibfnamefont {D.}~\bibnamefont {Mross}},\
  }\bibfield  {title} {\bibinfo {title} {Distinguishing between non-abelian
  topological orders in a quantum {Hall} system},\ }\href
  {https://doi.org/10.1126/science.abg6116} {\bibfield  {journal} {\bibinfo
  {journal} {Science}\ }\textbf {\bibinfo {volume} {0}},\ \bibinfo {pages}
  {eabg6116} (\bibinfo {year} {2021})}\BibitemShut {NoStop}%
\bibitem [{\citenamefont {Huang}\ \emph {et~al.}(2022)\citenamefont {Huang},
  \citenamefont {Fu}, \citenamefont {Hickey}, \citenamefont {Alem},
  \citenamefont {Lin}, \citenamefont {Watanabe}, \citenamefont {Taniguchi},\
  and\ \citenamefont {Zhu}}]{Huang23}%
  \BibitemOpen
  \bibfield  {author} {\bibinfo {author} {\bibfnamefont {K.}~\bibnamefont
  {Huang}}, \bibinfo {author} {\bibfnamefont {H.}~\bibnamefont {Fu}}, \bibinfo
  {author} {\bibfnamefont {D.~R.}\ \bibnamefont {Hickey}}, \bibinfo {author}
  {\bibfnamefont {N.}~\bibnamefont {Alem}}, \bibinfo {author} {\bibfnamefont
  {X.}~\bibnamefont {Lin}}, \bibinfo {author} {\bibfnamefont {K.}~\bibnamefont
  {Watanabe}}, \bibinfo {author} {\bibfnamefont {T.}~\bibnamefont
  {Taniguchi}},\ and\ \bibinfo {author} {\bibfnamefont {J.}~\bibnamefont
  {Zhu}},\ }\bibfield  {title} {\bibinfo {title} {Valley isospin controlled
  fractional quantum {Hall} states in bilayer graphene},\ }\href
  {https://doi.org/10.1103/PhysRevX.12.031019} {\bibfield  {journal} {\bibinfo
  {journal} {Phys. Rev. X}\ }\textbf {\bibinfo {volume} {12}},\ \bibinfo
  {pages} {031019} (\bibinfo {year} {2022})}\BibitemShut {NoStop}%
\bibitem [{\citenamefont {Liu}\ \emph {et~al.}(2022)\citenamefont {Liu},
  \citenamefont {Farahi}, \citenamefont {Chiu}, \citenamefont {Papic},
  \citenamefont {Watanabe}, \citenamefont {Taniguchi}, \citenamefont
  {Zaletel},\ and\ \citenamefont {Yazdani}}]{Xiaomeng2022}%
  \BibitemOpen
  \bibfield  {author} {\bibinfo {author} {\bibfnamefont {X.}~\bibnamefont
  {Liu}}, \bibinfo {author} {\bibfnamefont {G.}~\bibnamefont {Farahi}},
  \bibinfo {author} {\bibfnamefont {C.-L.}\ \bibnamefont {Chiu}}, \bibinfo
  {author} {\bibfnamefont {Z.}~\bibnamefont {Papic}}, \bibinfo {author}
  {\bibfnamefont {K.}~\bibnamefont {Watanabe}}, \bibinfo {author}
  {\bibfnamefont {T.}~\bibnamefont {Taniguchi}}, \bibinfo {author}
  {\bibfnamefont {M.~P.}\ \bibnamefont {Zaletel}},\ and\ \bibinfo {author}
  {\bibfnamefont {A.}~\bibnamefont {Yazdani}},\ }\bibfield  {title} {\bibinfo
  {title} {Visualizing broken symmetry and topological defects in a quantum
  {Hall} ferromagnet},\ }\href {https://doi.org/10.1126/science.abm3770}
  {\bibfield  {journal} {\bibinfo  {journal} {Science}\ }\textbf {\bibinfo
  {volume} {375}},\ \bibinfo {pages} {321} (\bibinfo {year}
  {2022})}\BibitemShut {NoStop}%
\bibitem [{\citenamefont {Farahi}\ \emph {et~al.}(2023)\citenamefont {Farahi},
  \citenamefont {Chiu}, \citenamefont {Liu}, \citenamefont {Papic},
  \citenamefont {Watanabe}, \citenamefont {Taniguchi}, \citenamefont
  {Zaletel},\ and\ \citenamefont {Yazdani}}]{Farahi2023}%
  \BibitemOpen
  \bibfield  {author} {\bibinfo {author} {\bibfnamefont {G.}~\bibnamefont
  {Farahi}}, \bibinfo {author} {\bibfnamefont {C.-L.}\ \bibnamefont {Chiu}},
  \bibinfo {author} {\bibfnamefont {X.}~\bibnamefont {Liu}}, \bibinfo {author}
  {\bibfnamefont {Z.}~\bibnamefont {Papic}}, \bibinfo {author} {\bibfnamefont
  {K.}~\bibnamefont {Watanabe}}, \bibinfo {author} {\bibfnamefont
  {T.}~\bibnamefont {Taniguchi}}, \bibinfo {author} {\bibfnamefont {M.~P.}\
  \bibnamefont {Zaletel}},\ and\ \bibinfo {author} {\bibfnamefont
  {A.}~\bibnamefont {Yazdani}},\ }\bibfield  {title} {\bibinfo {title} {Broken
  symmetries and excitation spectra of interacting electrons in partially
  filled {Landau} levels},\ }\bibfield  {journal} {\bibinfo  {journal} {Nature
  Physics}\ }\href {https://doi.org/10.1038/s41567-023-02126-z}
  {10.1038/s41567-023-02126-z} (\bibinfo {year} {2023})\BibitemShut {NoStop}%
\bibitem [{\citenamefont {Willett}\ \emph {et~al.}(2023)\citenamefont
  {Willett}, \citenamefont {Shtengel}, \citenamefont {Nayak}, \citenamefont
  {Pfeiffer}, \citenamefont {Chung}, \citenamefont {Peabody}, \citenamefont
  {Baldwin},\ and\ \citenamefont {West}}]{Willett2023}%
  \BibitemOpen
  \bibfield  {author} {\bibinfo {author} {\bibfnamefont {R.~L.}\ \bibnamefont
  {Willett}}, \bibinfo {author} {\bibfnamefont {K.}~\bibnamefont {Shtengel}},
  \bibinfo {author} {\bibfnamefont {C.}~\bibnamefont {Nayak}}, \bibinfo
  {author} {\bibfnamefont {L.~N.}\ \bibnamefont {Pfeiffer}}, \bibinfo {author}
  {\bibfnamefont {Y.~J.}\ \bibnamefont {Chung}}, \bibinfo {author}
  {\bibfnamefont {M.~L.}\ \bibnamefont {Peabody}}, \bibinfo {author}
  {\bibfnamefont {K.~W.}\ \bibnamefont {Baldwin}},\ and\ \bibinfo {author}
  {\bibfnamefont {K.~W.}\ \bibnamefont {West}},\ }\bibfield  {title} {\bibinfo
  {title} {Interference measurements of non-abelian $e/4$ \& abelian $e/2$
  quasiparticle braiding},\ }\href {https://doi.org/10.1103/PhysRevX.13.011028}
  {\bibfield  {journal} {\bibinfo  {journal} {Phys. Rev. X}\ }\textbf {\bibinfo
  {volume} {13}},\ \bibinfo {pages} {011028} (\bibinfo {year}
  {2023})}\BibitemShut {NoStop}%
\bibitem [{\citenamefont {Hu}\ \emph {et~al.}(2025{\natexlab{a}})\citenamefont
  {Hu}, \citenamefont {Tsui}, \citenamefont {He}, \citenamefont {Kamber},
  \citenamefont {Wang}, \citenamefont {Mohammadi}, \citenamefont {Watanabe},
  \citenamefont {Taniguchi}, \citenamefont {Papi{\'{c}}}, \citenamefont
  {Zaletel},\ and\ \citenamefont {Yazdani}}]{Hu2025}%
  \BibitemOpen
  \bibfield  {author} {\bibinfo {author} {\bibfnamefont {Y.}~\bibnamefont
  {Hu}}, \bibinfo {author} {\bibfnamefont {Y.-C.}\ \bibnamefont {Tsui}},
  \bibinfo {author} {\bibfnamefont {M.}~\bibnamefont {He}}, \bibinfo {author}
  {\bibfnamefont {U.}~\bibnamefont {Kamber}}, \bibinfo {author} {\bibfnamefont
  {T.}~\bibnamefont {Wang}}, \bibinfo {author} {\bibfnamefont {A.~S.}\
  \bibnamefont {Mohammadi}}, \bibinfo {author} {\bibfnamefont {K.}~\bibnamefont
  {Watanabe}}, \bibinfo {author} {\bibfnamefont {T.}~\bibnamefont {Taniguchi}},
  \bibinfo {author} {\bibfnamefont {Z.}~\bibnamefont {Papi{\'{c}}}}, \bibinfo
  {author} {\bibfnamefont {M.~P.}\ \bibnamefont {Zaletel}},\ and\ \bibinfo
  {author} {\bibfnamefont {A.}~\bibnamefont {Yazdani}},\ }\bibfield  {title}
  {\bibinfo {title} {High-resolution tunnelling spectroscopy of fractional
  quantum {Hall} states},\ }\href {https://doi.org/10.1038/s41567-025-02830-y}
  {\bibfield  {journal} {\bibinfo  {journal} {Nature Physics}\ }\textbf
  {\bibinfo {volume} {21}},\ \bibinfo {pages} {716} (\bibinfo {year}
  {2025}{\natexlab{a}})}\BibitemShut {NoStop}%
\bibitem [{\citenamefont {Girvin}\ and\ \citenamefont
  {MacDonald}(2007)}]{Girvin07}%
  \BibitemOpen
  \bibfield  {author} {\bibinfo {author} {\bibfnamefont {S.~M.}\ \bibnamefont
  {Girvin}}\ and\ \bibinfo {author} {\bibfnamefont {A.~H.}\ \bibnamefont
  {MacDonald}},\ }\bibinfo {title} {Multicomponent quantum {Hall} systems: The
  sum of their parts and more},\ in\ \href
  {http://dx.doi.org/10.1002/9783527617258.ch5} {\emph {\bibinfo {booktitle}
  {Perspectives in Quantum {Hall} Effects}}}\ (\bibinfo  {publisher} {Wiley-VCH
  Verlag GmbH},\ \bibinfo {year} {2007})\ pp.\ \bibinfo {pages}
  {161--224}\BibitemShut {NoStop}%
\bibitem [{\citenamefont {Eisenstein}(2014)}]{Eisenstein14}%
  \BibitemOpen
  \bibfield  {author} {\bibinfo {author} {\bibfnamefont {J.}~\bibnamefont
  {Eisenstein}},\ }\bibfield  {title} {\bibinfo {title} {Exciton condensation
  in bilayer quantum {Hall} systems},\ }\href
  {https://doi.org/https://doi.org/10.1146/annurev-conmatphys-031113-133832}
  {\bibfield  {journal} {\bibinfo  {journal} {Annual Review of Condensed Matter
  Physics}\ }\textbf {\bibinfo {volume} {5}},\ \bibinfo {pages} {159} (\bibinfo
  {year} {2014})}\BibitemShut {NoStop}%
\bibitem [{\citenamefont {Suen}\ \emph {et~al.}(1992)\citenamefont {Suen},
  \citenamefont {Engel}, \citenamefont {Santos}, \citenamefont {Shayegan},\
  and\ \citenamefont {Tsui}}]{Suen92}%
  \BibitemOpen
  \bibfield  {author} {\bibinfo {author} {\bibfnamefont {Y.~W.}\ \bibnamefont
  {Suen}}, \bibinfo {author} {\bibfnamefont {L.~W.}\ \bibnamefont {Engel}},
  \bibinfo {author} {\bibfnamefont {M.~B.}\ \bibnamefont {Santos}}, \bibinfo
  {author} {\bibfnamefont {M.}~\bibnamefont {Shayegan}},\ and\ \bibinfo
  {author} {\bibfnamefont {D.~C.}\ \bibnamefont {Tsui}},\ }\bibfield  {title}
  {\bibinfo {title} {Observation of a $\nu=1/2$ fractional quantum {Hall} state
  in a double-layer electron system},\ }\href
  {https://doi.org/10.1103/PhysRevLett.68.1379} {\bibfield  {journal} {\bibinfo
   {journal} {Phys. Rev. Lett.}\ }\textbf {\bibinfo {volume} {68}},\ \bibinfo
  {pages} {1379} (\bibinfo {year} {1992})}\BibitemShut {NoStop}%
\bibitem [{\citenamefont {Eisenstein}\ \emph {et~al.}(1992)\citenamefont
  {Eisenstein}, \citenamefont {Boebinger}, \citenamefont {Pfeiffer},
  \citenamefont {West},\ and\ \citenamefont {He}}]{Eisenstein92}%
  \BibitemOpen
  \bibfield  {author} {\bibinfo {author} {\bibfnamefont {J.~P.}\ \bibnamefont
  {Eisenstein}}, \bibinfo {author} {\bibfnamefont {G.~S.}\ \bibnamefont
  {Boebinger}}, \bibinfo {author} {\bibfnamefont {L.~N.}\ \bibnamefont
  {Pfeiffer}}, \bibinfo {author} {\bibfnamefont {K.~W.}\ \bibnamefont {West}},\
  and\ \bibinfo {author} {\bibfnamefont {S.}~\bibnamefont {He}},\ }\bibfield
  {title} {\bibinfo {title} {New fractional quantum {Hall} state in
  double-layer two-dimensional electron systems},\ }\href
  {https://doi.org/10.1103/PhysRevLett.68.1383} {\bibfield  {journal} {\bibinfo
   {journal} {Phys. Rev. Lett.}\ }\textbf {\bibinfo {volume} {68}},\ \bibinfo
  {pages} {1383} (\bibinfo {year} {1992})}\BibitemShut {NoStop}%
\bibitem [{\citenamefont {Suen}\ \emph {et~al.}(1994)\citenamefont {Suen},
  \citenamefont {Manoharan}, \citenamefont {Ying}, \citenamefont {Santos},\
  and\ \citenamefont {Shayegan}}]{Suen94b}%
  \BibitemOpen
  \bibfield  {author} {\bibinfo {author} {\bibfnamefont {Y.~W.}\ \bibnamefont
  {Suen}}, \bibinfo {author} {\bibfnamefont {H.~C.}\ \bibnamefont {Manoharan}},
  \bibinfo {author} {\bibfnamefont {X.}~\bibnamefont {Ying}}, \bibinfo {author}
  {\bibfnamefont {M.~B.}\ \bibnamefont {Santos}},\ and\ \bibinfo {author}
  {\bibfnamefont {M.}~\bibnamefont {Shayegan}},\ }\bibfield  {title} {\bibinfo
  {title} {Origin of the $\nu=1/2$ fractional quantum {Hall} state in wide
  single quantum wells},\ }\href {https://doi.org/10.1103/PhysRevLett.72.3405}
  {\bibfield  {journal} {\bibinfo  {journal} {Phys. Rev. Lett.}\ }\textbf
  {\bibinfo {volume} {72}},\ \bibinfo {pages} {3405} (\bibinfo {year}
  {1994})}\BibitemShut {NoStop}%
\bibitem [{\citenamefont {Shabani}\ \emph {et~al.}(2013)\citenamefont
  {Shabani}, \citenamefont {Liu}, \citenamefont {Shayegan}, \citenamefont
  {Pfeiffer}, \citenamefont {West},\ and\ \citenamefont {Baldwin}}]{Shabani13}%
  \BibitemOpen
  \bibfield  {author} {\bibinfo {author} {\bibfnamefont {J.}~\bibnamefont
  {Shabani}}, \bibinfo {author} {\bibfnamefont {Y.}~\bibnamefont {Liu}},
  \bibinfo {author} {\bibfnamefont {M.}~\bibnamefont {Shayegan}}, \bibinfo
  {author} {\bibfnamefont {L.~N.}\ \bibnamefont {Pfeiffer}}, \bibinfo {author}
  {\bibfnamefont {K.~W.}\ \bibnamefont {West}},\ and\ \bibinfo {author}
  {\bibfnamefont {K.~W.}\ \bibnamefont {Baldwin}},\ }\bibfield  {title}
  {\bibinfo {title} {Phase diagrams for the stability of the
  $\ensuremath{\nu}=\frac{1}{2}$ fractional quantum {Hall} effect in electron
  systems confined to symmetric, wide {Ga}{As} quantum wells},\ }\href
  {https://doi.org/10.1103/PhysRevB.88.245413} {\bibfield  {journal} {\bibinfo
  {journal} {Phys. Rev. B}\ }\textbf {\bibinfo {volume} {88}},\ \bibinfo
  {pages} {245413} (\bibinfo {year} {2013})}\BibitemShut {NoStop}%
\bibitem [{\citenamefont {Singh}\ \emph {et~al.}(2024)\citenamefont {Singh},
  \citenamefont {Wang}, \citenamefont {Tai}, \citenamefont {Calhoun},
  \citenamefont {Villegas~Rosales}, \citenamefont {Madathil}, \citenamefont
  {Gupta}, \citenamefont {Baldwin}, \citenamefont {Pfeiffer},\ and\
  \citenamefont {Shayegan}}]{Singh24}%
  \BibitemOpen
  \bibfield  {author} {\bibinfo {author} {\bibfnamefont {S.~K.}\ \bibnamefont
  {Singh}}, \bibinfo {author} {\bibfnamefont {C.}~\bibnamefont {Wang}},
  \bibinfo {author} {\bibfnamefont {C.~T.}\ \bibnamefont {Tai}}, \bibinfo
  {author} {\bibfnamefont {C.~S.}\ \bibnamefont {Calhoun}}, \bibinfo {author}
  {\bibfnamefont {K.~A.}\ \bibnamefont {Villegas~Rosales}}, \bibinfo {author}
  {\bibfnamefont {P.~T.}\ \bibnamefont {Madathil}}, \bibinfo {author}
  {\bibfnamefont {A.}~\bibnamefont {Gupta}}, \bibinfo {author} {\bibfnamefont
  {K.~W.}\ \bibnamefont {Baldwin}}, \bibinfo {author} {\bibfnamefont {L.~N.}\
  \bibnamefont {Pfeiffer}},\ and\ \bibinfo {author} {\bibfnamefont
  {M.}~\bibnamefont {Shayegan}},\ }\bibfield  {title} {\bibinfo {title}
  {{Topological phase transition between Jain states and daughter states of the
  $\nu{=}1/2$ fractional quantum Hall state}},\ }\href
  {https://doi.org/10.1038/s41567-024-02517-w} {\bibfield  {journal} {\bibinfo
  {journal} {Nature Physics}\ }\textbf {\bibinfo {volume} {20}},\ \bibinfo
  {pages} {1247} (\bibinfo {year} {2024})}\BibitemShut {NoStop}%
\bibitem [{\citenamefont {Singh}\ \emph {et~al.}(2025)\citenamefont {Singh},
  \citenamefont {Wang}, \citenamefont {Gupta}, \citenamefont {Baldwin},
  \citenamefont {Pfeiffer},\ and\ \citenamefont {Shayegan}}]{Singh25}%
  \BibitemOpen
  \bibfield  {author} {\bibinfo {author} {\bibfnamefont {S.~K.}\ \bibnamefont
  {Singh}}, \bibinfo {author} {\bibfnamefont {C.}~\bibnamefont {Wang}},
  \bibinfo {author} {\bibfnamefont {A.}~\bibnamefont {Gupta}}, \bibinfo
  {author} {\bibfnamefont {K.~W.}\ \bibnamefont {Baldwin}}, \bibinfo {author}
  {\bibfnamefont {L.~N.}\ \bibnamefont {Pfeiffer}},\ and\ \bibinfo {author}
  {\bibfnamefont {M.}~\bibnamefont {Shayegan}},\ }\bibfield  {title} {\bibinfo
  {title} {Fractional quantum hall state at
  $\ensuremath{\nu}\text{}=\text{}1/2$ with energy gap up to 6 k and possible
  transition from the one- to two-component state},\ }\href
  {https://doi.org/10.1103/ywpx-qm7d} {\bibfield  {journal} {\bibinfo
  {journal} {Phys. Rev. Lett.}\ }\textbf {\bibinfo {volume} {135}},\ \bibinfo
  {pages} {246603} (\bibinfo {year} {2025})}\BibitemShut {NoStop}%
\bibitem [{\citenamefont {Halperin}(1983)}]{Halperin83}%
  \BibitemOpen
  \bibfield  {author} {\bibinfo {author} {\bibfnamefont {B.~I.}\ \bibnamefont
  {Halperin}},\ }\bibfield  {title} {\bibinfo {title} {Theory of the quantized
  {Hall} conductance},\ }\href {https://doi.org/https://doi.org/
  10.5169/seals-115362} {\bibfield  {journal} {\bibinfo  {journal} {Helvetica
  Physica Acta}\ }\textbf {\bibinfo {volume} {56}},\ \bibinfo {pages} {75}
  (\bibinfo {year} {1983})}\BibitemShut {NoStop}%
\bibitem [{\citenamefont {He}\ \emph {et~al.}(1993)\citenamefont {He},
  \citenamefont {Das~Sarma},\ and\ \citenamefont {Xie}}]{He93}%
  \BibitemOpen
  \bibfield  {author} {\bibinfo {author} {\bibfnamefont {S.}~\bibnamefont
  {He}}, \bibinfo {author} {\bibfnamefont {S.}~\bibnamefont {Das~Sarma}},\ and\
  \bibinfo {author} {\bibfnamefont {X.~C.}\ \bibnamefont {Xie}},\ }\bibfield
  {title} {\bibinfo {title} {Quantized {Hall} effect and quantum phase
  transitions in coupled two-layer electron systems},\ }\href
  {https://doi.org/10.1103/PhysRevB.47.4394} {\bibfield  {journal} {\bibinfo
  {journal} {Phys. Rev. B}\ }\textbf {\bibinfo {volume} {47}},\ \bibinfo
  {pages} {4394} (\bibinfo {year} {1993})}\BibitemShut {NoStop}%
\bibitem [{\citenamefont {Nomura}\ and\ \citenamefont
  {Yoshioka}(2004)}]{Nomura04}%
  \BibitemOpen
  \bibfield  {author} {\bibinfo {author} {\bibfnamefont {K.}~\bibnamefont
  {Nomura}}\ and\ \bibinfo {author} {\bibfnamefont {D.}~\bibnamefont
  {Yoshioka}},\ }\bibfield  {title} {\bibinfo {title} {Gap evolution in
  $\nu{=}1/2$ bilayer quantum {Hall} systems},\ }\href
  {https://doi.org/10.1143/JPSJ.73.2612} {\bibfield  {journal} {\bibinfo
  {journal} {Journal of the Physical Society of Japan}\ }\textbf {\bibinfo
  {volume} {73}},\ \bibinfo {pages} {2612} (\bibinfo {year} {2004})},\ \Eprint
  {https://arxiv.org/abs/https://doi.org/10.1143/JPSJ.73.2612}
  {https://doi.org/10.1143/JPSJ.73.2612} \BibitemShut {NoStop}%
\bibitem [{\citenamefont {Papi\ifmmode~\acute{c}\else \'{c}\fi{}}\ \emph
  {et~al.}(2010)\citenamefont {Papi\ifmmode~\acute{c}\else \'{c}\fi{}},
  \citenamefont {Goerbig}, \citenamefont {Regnault},\ and\ \citenamefont
  {Milovanovi\ifmmode~\acute{c}\else \'{c}\fi{}}}]{Papic10}%
  \BibitemOpen
  \bibfield  {author} {\bibinfo {author} {\bibfnamefont {Z.}~\bibnamefont
  {Papi\ifmmode~\acute{c}\else \'{c}\fi{}}}, \bibinfo {author} {\bibfnamefont
  {M.~O.}\ \bibnamefont {Goerbig}}, \bibinfo {author} {\bibfnamefont
  {N.}~\bibnamefont {Regnault}},\ and\ \bibinfo {author} {\bibfnamefont
  {M.~V.}\ \bibnamefont {Milovanovi\ifmmode~\acute{c}\else \'{c}\fi{}}},\
  }\bibfield  {title} {\bibinfo {title} {Tunneling-driven breakdown of the 331
  state and the emergent {Pfaffian} and composite {Fermi} liquid phases},\
  }\href {https://doi.org/10.1103/PhysRevB.82.075302} {\bibfield  {journal}
  {\bibinfo  {journal} {Phys. Rev. B}\ }\textbf {\bibinfo {volume} {82}},\
  \bibinfo {pages} {075302} (\bibinfo {year} {2010})}\BibitemShut {NoStop}%
\bibitem [{\citenamefont {Peterson}\ and\ \citenamefont
  {Das~Sarma}(2010)}]{Peterson10a}%
  \BibitemOpen
  \bibfield  {author} {\bibinfo {author} {\bibfnamefont {M.~R.}\ \bibnamefont
  {Peterson}}\ and\ \bibinfo {author} {\bibfnamefont {S.}~\bibnamefont
  {Das~Sarma}},\ }\bibfield  {title} {\bibinfo {title} {Quantum hall phase
  diagram of half-filled bilayers in the lowest and the second orbital landau
  levels: Abelian versus non-abelian incompressible fractional quantum hall
  states},\ }\href {https://doi.org/10.1103/PhysRevB.81.165304} {\bibfield
  {journal} {\bibinfo  {journal} {Phys. Rev. B}\ }\textbf {\bibinfo {volume}
  {81}},\ \bibinfo {pages} {165304} (\bibinfo {year} {2010})}\BibitemShut
  {NoStop}%
\bibitem [{\citenamefont {Peterson}\ \emph {et~al.}(2010)\citenamefont
  {Peterson}, \citenamefont {Papi\ifmmode~\acute{c}\else \'{c}\fi{}},\ and\
  \citenamefont {Das~Sarma}}]{Peterson10b}%
  \BibitemOpen
  \bibfield  {author} {\bibinfo {author} {\bibfnamefont {M.~R.}\ \bibnamefont
  {Peterson}}, \bibinfo {author} {\bibfnamefont {Z.}~\bibnamefont
  {Papi\ifmmode~\acute{c}\else \'{c}\fi{}}},\ and\ \bibinfo {author}
  {\bibfnamefont {S.}~\bibnamefont {Das~Sarma}},\ }\bibfield  {title} {\bibinfo
  {title} {Fractional quantum {Hall} effects in bilayers in the presence of
  interlayer tunneling and charge imbalance},\ }\href
  {https://doi.org/10.1103/PhysRevB.82.235312} {\bibfield  {journal} {\bibinfo
  {journal} {Phys. Rev. B}\ }\textbf {\bibinfo {volume} {82}},\ \bibinfo
  {pages} {235312} (\bibinfo {year} {2010})}\BibitemShut {NoStop}%
\bibitem [{\citenamefont {Liu}\ \emph {et~al.}(2016)\citenamefont {Liu},
  \citenamefont {Vaezi}, \citenamefont {Repellin},\ and\ \citenamefont
  {Regnault}}]{Liu16}%
  \BibitemOpen
  \bibfield  {author} {\bibinfo {author} {\bibfnamefont {Z.}~\bibnamefont
  {Liu}}, \bibinfo {author} {\bibfnamefont {A.}~\bibnamefont {Vaezi}}, \bibinfo
  {author} {\bibfnamefont {C.}~\bibnamefont {Repellin}},\ and\ \bibinfo
  {author} {\bibfnamefont {N.}~\bibnamefont {Regnault}},\ }\bibfield  {title}
  {\bibinfo {title} {Phase diagram of
  $\ensuremath{\nu}=\frac{1}{2}+\frac{1}{2}$ bilayer bosons with interlayer
  couplings},\ }\href {https://doi.org/10.1103/PhysRevB.93.085115} {\bibfield
  {journal} {\bibinfo  {journal} {Phys. Rev. B}\ }\textbf {\bibinfo {volume}
  {93}},\ \bibinfo {pages} {085115} (\bibinfo {year} {2016})}\BibitemShut
  {NoStop}%
\bibitem [{\citenamefont {Zhu}\ \emph {et~al.}(2016)\citenamefont {Zhu},
  \citenamefont {Liu}, \citenamefont {Haldane},\ and\ \citenamefont
  {Sheng}}]{Zhu16}%
  \BibitemOpen
  \bibfield  {author} {\bibinfo {author} {\bibfnamefont {W.}~\bibnamefont
  {Zhu}}, \bibinfo {author} {\bibfnamefont {Z.}~\bibnamefont {Liu}}, \bibinfo
  {author} {\bibfnamefont {F.~D.~M.}\ \bibnamefont {Haldane}},\ and\ \bibinfo
  {author} {\bibfnamefont {D.~N.}\ \bibnamefont {Sheng}},\ }\bibfield  {title}
  {\bibinfo {title} {Fractional quantum {Hall} bilayers at half filling:
  Tunneling-driven non-{Abelian} phase},\ }\href
  {https://doi.org/10.1103/PhysRevB.94.245147} {\bibfield  {journal} {\bibinfo
  {journal} {Phys. Rev. B}\ }\textbf {\bibinfo {volume} {94}},\ \bibinfo
  {pages} {245147} (\bibinfo {year} {2016})}\BibitemShut {NoStop}%
\bibitem [{\citenamefont {Cabra}\ \emph {et~al.}(2001)\citenamefont {Cabra},
  \citenamefont {Lopez},\ and\ \citenamefont {Rossini}}]{Cabra2001}%
  \BibitemOpen
  \bibfield  {author} {\bibinfo {author} {\bibfnamefont {D.}~\bibnamefont
  {Cabra}}, \bibinfo {author} {\bibfnamefont {A.}~\bibnamefont {Lopez}},\ and\
  \bibinfo {author} {\bibfnamefont {G.}~\bibnamefont {Rossini}},\ }\bibfield
  {title} {\bibinfo {title} {Transition from abelian to non-{Abelian} {FQHE}
  states},\ }\href {https://doi.org/10.1007/s100510170346} {\bibfield
  {journal} {\bibinfo  {journal} {Eur. Phys. J. B}\ }\textbf {\bibinfo {volume}
  {19}},\ \bibinfo {pages} {21} (\bibinfo {year} {2001})}\BibitemShut {NoStop}%
\bibitem [{\citenamefont {Cappelli}\ \emph {et~al.}(2001)\citenamefont
  {Cappelli}, \citenamefont {Georgiev},\ and\ \citenamefont
  {Todorov}}]{CAPPELLI2001}%
  \BibitemOpen
  \bibfield  {author} {\bibinfo {author} {\bibfnamefont {A.}~\bibnamefont
  {Cappelli}}, \bibinfo {author} {\bibfnamefont {L.~S.}\ \bibnamefont
  {Georgiev}},\ and\ \bibinfo {author} {\bibfnamefont {I.~T.}\ \bibnamefont
  {Todorov}},\ }\bibfield  {title} {\bibinfo {title} {Parafermion hall states
  from coset projections of abelian conformal theories},\ }\href
  {https://doi.org/https://doi.org/10.1016/S0550-3213(00)00774-4} {\bibfield
  {journal} {\bibinfo  {journal} {Nuclear Physics B}\ }\textbf {\bibinfo
  {volume} {599}},\ \bibinfo {pages} {499} (\bibinfo {year}
  {2001})}\BibitemShut {NoStop}%
\bibitem [{\citenamefont {Repellin}\ \emph {et~al.}(2015)\citenamefont
  {Repellin}, \citenamefont {Neupert}, \citenamefont {Bernevig},\ and\
  \citenamefont {Regnault}}]{Repellin2015}%
  \BibitemOpen
  \bibfield  {author} {\bibinfo {author} {\bibfnamefont {C.}~\bibnamefont
  {Repellin}}, \bibinfo {author} {\bibfnamefont {T.}~\bibnamefont {Neupert}},
  \bibinfo {author} {\bibfnamefont {B.~A.}\ \bibnamefont {Bernevig}},\ and\
  \bibinfo {author} {\bibfnamefont {N.}~\bibnamefont {Regnault}},\ }\bibfield
  {title} {\bibinfo {title} {Projective construction of the ${\mathbb{z}}_{k}$
  read-rezayi fractional quantum hall states and their excitations on the torus
  geometry},\ }\href {https://doi.org/10.1103/PhysRevB.92.115128} {\bibfield
  {journal} {\bibinfo  {journal} {Phys. Rev. B}\ }\textbf {\bibinfo {volume}
  {92}},\ \bibinfo {pages} {115128} (\bibinfo {year} {2015})}\BibitemShut
  {NoStop}%
\bibitem [{\citenamefont {Cr\'epel}\ \emph
  {et~al.}(2019{\natexlab{a}})\citenamefont {Cr\'epel}, \citenamefont
  {Estienne},\ and\ \citenamefont {Regnault}}]{Crepel2019}%
  \BibitemOpen
  \bibfield  {author} {\bibinfo {author} {\bibfnamefont {V.}~\bibnamefont
  {Cr\'epel}}, \bibinfo {author} {\bibfnamefont {B.}~\bibnamefont {Estienne}},\
  and\ \bibinfo {author} {\bibfnamefont {N.}~\bibnamefont {Regnault}},\
  }\bibfield  {title} {\bibinfo {title} {Variational ansatz for an abelian to
  non-abelian topological phase transition in $\ensuremath{\nu}=1/2+1/2$
  bilayers},\ }\href {https://doi.org/10.1103/PhysRevLett.123.126804}
  {\bibfield  {journal} {\bibinfo  {journal} {Phys. Rev. Lett.}\ }\textbf
  {\bibinfo {volume} {123}},\ \bibinfo {pages} {126804} (\bibinfo {year}
  {2019}{\natexlab{a}})}\BibitemShut {NoStop}%
\bibitem [{\citenamefont {Jain}(1989)}]{Jain89}%
  \BibitemOpen
  \bibfield  {author} {\bibinfo {author} {\bibfnamefont {J.~K.}\ \bibnamefont
  {Jain}},\ }\bibfield  {title} {\bibinfo {title} {Composite-fermion approach
  for the fractional quantum {Hall} effect},\ }\href
  {https://doi.org/10.1103/PhysRevLett.63.199} {\bibfield  {journal} {\bibinfo
  {journal} {Phys. Rev. Lett.}\ }\textbf {\bibinfo {volume} {63}},\ \bibinfo
  {pages} {199} (\bibinfo {year} {1989})}\BibitemShut {NoStop}%
\bibitem [{\citenamefont {Wen}(2000)}]{Wen00}%
  \BibitemOpen
  \bibfield  {author} {\bibinfo {author} {\bibfnamefont {X.-G.}\ \bibnamefont
  {Wen}},\ }\bibfield  {title} {\bibinfo {title} {Continuous topological phase
  transitions between clean quantum {Hall} states},\ }\href
  {https://doi.org/10.1103/PhysRevLett.84.3950} {\bibfield  {journal} {\bibinfo
   {journal} {Phys. Rev. Lett.}\ }\textbf {\bibinfo {volume} {84}},\ \bibinfo
  {pages} {3950} (\bibinfo {year} {2000})}\BibitemShut {NoStop}%
\bibitem [{\citenamefont {Barkeshli}\ and\ \citenamefont
  {Wen}(2010)}]{Barkeshli10Bilayer}%
  \BibitemOpen
  \bibfield  {author} {\bibinfo {author} {\bibfnamefont {M.}~\bibnamefont
  {Barkeshli}}\ and\ \bibinfo {author} {\bibfnamefont {X.-G.}\ \bibnamefont
  {Wen}},\ }\bibfield  {title} {\bibinfo {title} {Anyon condensation and
  continuous topological phase transitions in non-abelian fractional quantum
  {Hall} states},\ }\href {https://doi.org/10.1103/PhysRevLett.105.216804}
  {\bibfield  {journal} {\bibinfo  {journal} {Phys. Rev. Lett.}\ }\textbf
  {\bibinfo {volume} {105}},\ \bibinfo {pages} {216804} (\bibinfo {year}
  {2010})}\BibitemShut {NoStop}%
\bibitem [{\citenamefont {Barkeshli}\ and\ \citenamefont
  {Wen}(2011)}]{Barkeshli11Bilayer}%
  \BibitemOpen
  \bibfield  {author} {\bibinfo {author} {\bibfnamefont {M.}~\bibnamefont
  {Barkeshli}}\ and\ \bibinfo {author} {\bibfnamefont {X.-G.}\ \bibnamefont
  {Wen}},\ }\bibfield  {title} {\bibinfo {title} {Bilayer quantum hall phase
  transitions and the orbifold non-abelian fractional quantum {Hall} states},\
  }\href {https://doi.org/10.1103/PhysRevB.84.115121} {\bibfield  {journal}
  {\bibinfo  {journal} {Phys. Rev. B}\ }\textbf {\bibinfo {volume} {84}},\
  \bibinfo {pages} {115121} (\bibinfo {year} {2011})}\BibitemShut {NoStop}%
\bibitem [{\citenamefont {Elliott}\ and\ \citenamefont
  {Franz}(2015)}]{Elliott15}%
  \BibitemOpen
  \bibfield  {author} {\bibinfo {author} {\bibfnamefont {S.~R.}\ \bibnamefont
  {Elliott}}\ and\ \bibinfo {author} {\bibfnamefont {M.}~\bibnamefont
  {Franz}},\ }\bibfield  {title} {\bibinfo {title} {Colloquium: Majorana
  fermions in nuclear, particle, and solid-state physics},\ }\href
  {https://doi.org/10.1103/RevModPhys.87.137} {\bibfield  {journal} {\bibinfo
  {journal} {Rev. Mod. Phys.}\ }\textbf {\bibinfo {volume} {87}},\ \bibinfo
  {pages} {137} (\bibinfo {year} {2015})}\BibitemShut {NoStop}%
\bibitem [{\citenamefont {Yazdani}\ \emph {et~al.}(2023)\citenamefont
  {Yazdani}, \citenamefont {von Oppen}, \citenamefont {Halperin},\ and\
  \citenamefont {Yacoby}}]{YazdaniMajoranaReview}%
  \BibitemOpen
  \bibfield  {author} {\bibinfo {author} {\bibfnamefont {A.}~\bibnamefont
  {Yazdani}}, \bibinfo {author} {\bibfnamefont {F.}~\bibnamefont {von Oppen}},
  \bibinfo {author} {\bibfnamefont {B.~I.}\ \bibnamefont {Halperin}},\ and\
  \bibinfo {author} {\bibfnamefont {A.}~\bibnamefont {Yacoby}},\ }\bibfield
  {title} {\bibinfo {title} {{Hunting for Majoranas}},\ }\href
  {https://doi.org/10.1126/science.ade0850} {\bibfield  {journal} {\bibinfo
  {journal} {Science}\ }\textbf {\bibinfo {volume} {380}},\ \bibinfo {pages}
  {eade0850} (\bibinfo {year} {2023})}\BibitemShut {NoStop}%
\bibitem [{\citenamefont {Zhu}\ \emph {et~al.}(2023)\citenamefont {Zhu},
  \citenamefont {Han}, \citenamefont {Huffman}, \citenamefont {Hofmann},\ and\
  \citenamefont {He}}]{Zhu23}%
  \BibitemOpen
  \bibfield  {author} {\bibinfo {author} {\bibfnamefont {W.}~\bibnamefont
  {Zhu}}, \bibinfo {author} {\bibfnamefont {C.}~\bibnamefont {Han}}, \bibinfo
  {author} {\bibfnamefont {E.}~\bibnamefont {Huffman}}, \bibinfo {author}
  {\bibfnamefont {J.~S.}\ \bibnamefont {Hofmann}},\ and\ \bibinfo {author}
  {\bibfnamefont {Y.-C.}\ \bibnamefont {He}},\ }\bibfield  {title} {\bibinfo
  {title} {{Uncovering Conformal Symmetry in the 3D Ising Transition:
  State-Operator Correspondence from a Quantum Fuzzy Sphere Regularization}},\
  }\href {https://doi.org/10.1103/PhysRevX.13.021009} {\bibfield  {journal}
  {\bibinfo  {journal} {Phys. Rev. X}\ }\textbf {\bibinfo {volume} {13}},\
  \bibinfo {pages} {021009} (\bibinfo {year} {2023})}\BibitemShut {NoStop}%
\bibitem [{\citenamefont {Francesco}\ \emph {et~al.}(1997)\citenamefont
  {Francesco}, \citenamefont {Mathieu},\ and\ \citenamefont
  {Senechal}}]{DiFrancesco97}%
  \BibitemOpen
  \bibfield  {author} {\bibinfo {author} {\bibfnamefont {P.~D.}\ \bibnamefont
  {Francesco}}, \bibinfo {author} {\bibfnamefont {P.}~\bibnamefont {Mathieu}},\
  and\ \bibinfo {author} {\bibfnamefont {D.}~\bibnamefont {Senechal}},\ }\href
  {http://books.google.com/books?id=keUrdME5rhIC} {\emph {\bibinfo {title}
  {Conformal Field Theory}}},\ Graduate Texts in Contemporary Physics\
  (\bibinfo  {publisher} {Springer},\ \bibinfo {year} {1997})\BibitemShut
  {NoStop}%
\bibitem [{\citenamefont {Cardy}(1984)}]{Cardy84}%
  \BibitemOpen
  \bibfield  {author} {\bibinfo {author} {\bibfnamefont {J.~L.}\ \bibnamefont
  {Cardy}},\ }\bibfield  {title} {\bibinfo {title} {Conformal invariance and
  universality in finite-size scaling},\ }\href
  {https://doi.org/10.1088/0305-4470/17/7/003} {\bibfield  {journal} {\bibinfo
  {journal} {Journal of Physics A: Mathematical and General}\ }\textbf
  {\bibinfo {volume} {17}},\ \bibinfo {pages} {L385} (\bibinfo {year}
  {1984})}\BibitemShut {NoStop}%
\bibitem [{\citenamefont {Cardy}(1985)}]{Cardy85}%
  \BibitemOpen
  \bibfield  {author} {\bibinfo {author} {\bibfnamefont {J.~L.}\ \bibnamefont
  {Cardy}},\ }\bibfield  {title} {\bibinfo {title} {Universal amplitudes in
  finite-size scaling: generalisation to arbitrary dimensionality},\ }\href
  {https://doi.org/10.1088/0305-4470/18/13/005} {\bibfield  {journal} {\bibinfo
   {journal} {J. Phys. A: Math. Gen.}\ }\textbf {\bibinfo {volume} {18}},\
  \bibinfo {pages} {L757} (\bibinfo {year} {1985})}\BibitemShut {NoStop}%
\bibitem [{\citenamefont {Haldane}(1983)}]{Haldane83}%
  \BibitemOpen
  \bibfield  {author} {\bibinfo {author} {\bibfnamefont {F.~D.~M.}\
  \bibnamefont {Haldane}},\ }\bibfield  {title} {\bibinfo {title} {Fractional
  quantization of the {Hall} effect: A hierarchy of incompressible quantum
  fluid states},\ }\href {https://doi.org/10.1103/PhysRevLett.51.605}
  {\bibfield  {journal} {\bibinfo  {journal} {Phys. Rev. Lett.}\ }\textbf
  {\bibinfo {volume} {51}},\ \bibinfo {pages} {605} (\bibinfo {year}
  {1983})}\BibitemShut {NoStop}%
\bibitem [{\citenamefont {Hu}\ \emph {et~al.}(2023)\citenamefont {Hu},
  \citenamefont {He},\ and\ \citenamefont {Zhu}}]{Hu23}%
  \BibitemOpen
  \bibfield  {author} {\bibinfo {author} {\bibfnamefont {L.}~\bibnamefont
  {Hu}}, \bibinfo {author} {\bibfnamefont {Y.-C.}\ \bibnamefont {He}},\ and\
  \bibinfo {author} {\bibfnamefont {W.}~\bibnamefont {Zhu}},\ }\bibfield
  {title} {\bibinfo {title} {{Operator Product Expansion Coefficients of the 3D
  Ising Criticality via Quantum Fuzzy Spheres}},\ }\href
  {https://doi.org/10.1103/PhysRevLett.131.031601} {\bibfield  {journal}
  {\bibinfo  {journal} {Phys. Rev. Lett.}\ }\textbf {\bibinfo {volume} {131}},\
  \bibinfo {pages} {031601} (\bibinfo {year} {2023})}\BibitemShut {NoStop}%
\bibitem [{\citenamefont {Hu}\ \emph {et~al.}(2025{\natexlab{b}})\citenamefont
  {Hu}, \citenamefont {Zhu},\ and\ \citenamefont {He}}]{Hu25}%
  \BibitemOpen
  \bibfield  {author} {\bibinfo {author} {\bibfnamefont {L.}~\bibnamefont
  {Hu}}, \bibinfo {author} {\bibfnamefont {W.}~\bibnamefont {Zhu}},\ and\
  \bibinfo {author} {\bibfnamefont {Y.-C.}\ \bibnamefont {He}},\ }\bibfield
  {title} {\bibinfo {title} {Entropic $f$ function of three-dimensional ising
  conformal field theory via fuzzy sphere regularization},\ }\href
  {https://doi.org/10.1103/PhysRevB.111.155151} {\bibfield  {journal} {\bibinfo
   {journal} {Phys. Rev. B}\ }\textbf {\bibinfo {volume} {111}},\ \bibinfo
  {pages} {155151} (\bibinfo {year} {2025}{\natexlab{b}})}\BibitemShut
  {NoStop}%
\bibitem [{\citenamefont {Han}\ \emph {et~al.}(2023)\citenamefont {Han},
  \citenamefont {Hu}, \citenamefont {Zhu},\ and\ \citenamefont {He}}]{Han23}%
  \BibitemOpen
  \bibfield  {author} {\bibinfo {author} {\bibfnamefont {C.}~\bibnamefont
  {Han}}, \bibinfo {author} {\bibfnamefont {L.}~\bibnamefont {Hu}}, \bibinfo
  {author} {\bibfnamefont {W.}~\bibnamefont {Zhu}},\ and\ \bibinfo {author}
  {\bibfnamefont {Y.-C.}\ \bibnamefont {He}},\ }\bibfield  {title} {\bibinfo
  {title} {{Conformal four-point correlators of the three-dimensional Ising
  transition via the quantum fuzzy sphere}},\ }\href
  {https://doi.org/10.1103/PhysRevB.108.235123} {\bibfield  {journal} {\bibinfo
   {journal} {Phys. Rev. B}\ }\textbf {\bibinfo {volume} {108}},\ \bibinfo
  {pages} {235123} (\bibinfo {year} {2023})}\BibitemShut {NoStop}%
\bibitem [{\citenamefont {Hofmann}\ \emph {et~al.}(2024)\citenamefont
  {Hofmann}, \citenamefont {Goth}, \citenamefont {Zhu}, \citenamefont {He},\
  and\ \citenamefont {Huffman}}]{Hofmann24}%
  \BibitemOpen
  \bibfield  {author} {\bibinfo {author} {\bibfnamefont {J.~S.}\ \bibnamefont
  {Hofmann}}, \bibinfo {author} {\bibfnamefont {F.}~\bibnamefont {Goth}},
  \bibinfo {author} {\bibfnamefont {W.}~\bibnamefont {Zhu}}, \bibinfo {author}
  {\bibfnamefont {Y.-C.}\ \bibnamefont {He}},\ and\ \bibinfo {author}
  {\bibfnamefont {E.}~\bibnamefont {Huffman}},\ }\bibfield  {title} {\bibinfo
  {title} {{Quantum Monte Carlo simulation of the 3D Ising transition on the
  fuzzy sphere}},\ }\href {https://doi.org/10.21468/SciPostPhysCore.7.2.028}
  {\bibfield  {journal} {\bibinfo  {journal} {SciPost Phys. Core}\ }\textbf
  {\bibinfo {volume} {7}},\ \bibinfo {pages} {028} (\bibinfo {year}
  {2024})}\BibitemShut {NoStop}%
\bibitem [{\citenamefont {Zhou}\ and\ \citenamefont {Zou}(2025)}]{Zhou25c}%
  \BibitemOpen
  \bibfield  {author} {\bibinfo {author} {\bibfnamefont {Z.}~\bibnamefont
  {Zhou}}\ and\ \bibinfo {author} {\bibfnamefont {Y.}~\bibnamefont {Zou}},\
  }\bibfield  {title} {\bibinfo {title} {{Studying the 3d Ising surface CFTs on
  the fuzzy sphere}},\ }\href {https://doi.org/10.21468/SciPostPhys.18.1.031}
  {\bibfield  {journal} {\bibinfo  {journal} {SciPost Phys.}\ }\textbf
  {\bibinfo {volume} {18}},\ \bibinfo {pages} {031} (\bibinfo {year}
  {2025})}\BibitemShut {NoStop}%
\bibitem [{\citenamefont {Fardelli}\ \emph {et~al.}(2025)\citenamefont
  {Fardelli}, \citenamefont {Fitzpatrick},\ and\ \citenamefont
  {Katz}}]{Fardelli25}%
  \BibitemOpen
  \bibfield  {author} {\bibinfo {author} {\bibfnamefont {G.}~\bibnamefont
  {Fardelli}}, \bibinfo {author} {\bibfnamefont {A.~L.}\ \bibnamefont
  {Fitzpatrick}},\ and\ \bibinfo {author} {\bibfnamefont {E.}~\bibnamefont
  {Katz}},\ }\bibfield  {title} {\bibinfo {title} {{Constructing the infrared
  conformal generators on the fuzzy sphere}},\ }\href
  {https://doi.org/10.21468/SciPostPhys.18.3.086} {\bibfield  {journal}
  {\bibinfo  {journal} {SciPost Phys.}\ }\textbf {\bibinfo {volume} {18}},\
  \bibinfo {pages} {086} (\bibinfo {year} {2025})}\BibitemShut {NoStop}%
\bibitem [{\citenamefont {Han}\ \emph {et~al.}(2024)\citenamefont {Han},
  \citenamefont {Hu},\ and\ \citenamefont {Zhu}}]{Han24}%
  \BibitemOpen
  \bibfield  {author} {\bibinfo {author} {\bibfnamefont {C.}~\bibnamefont
  {Han}}, \bibinfo {author} {\bibfnamefont {L.}~\bibnamefont {Hu}},\ and\
  \bibinfo {author} {\bibfnamefont {W.}~\bibnamefont {Zhu}},\ }\bibfield
  {title} {\bibinfo {title} {{Conformal operator content of the Wilson-Fisher
  transition on fuzzy sphere bilayers}},\ }\href
  {https://doi.org/10.1103/PhysRevB.110.115113} {\bibfield  {journal} {\bibinfo
   {journal} {Phys. Rev. B}\ }\textbf {\bibinfo {volume} {110}},\ \bibinfo
  {pages} {115113} (\bibinfo {year} {2024})}\BibitemShut {NoStop}%
\bibitem [{\citenamefont {Fan}\ \emph {et~al.}(2025)\citenamefont {Fan},
  \citenamefont {Dong},\ and\ \citenamefont {Vishwanath}}]{Fan25}%
  \BibitemOpen
  \bibfield  {author} {\bibinfo {author} {\bibfnamefont {R.}~\bibnamefont
  {Fan}}, \bibinfo {author} {\bibfnamefont {J.}~\bibnamefont {Dong}},\ and\
  \bibinfo {author} {\bibfnamefont {A.}~\bibnamefont {Vishwanath}},\ }\href
  {https://arxiv.org/abs/2505.06342} {\bibinfo {title} {{Simulating the
  non-unitary Yang-Lee conformal field theory on the fuzzy sphere}}} (\bibinfo
  {year} {2025}),\ \Eprint {https://arxiv.org/abs/2505.06342} {arXiv:2505.06342
  [cond-mat.str-el]} \BibitemShut {NoStop}%
\bibitem [{\citenamefont {Arguello~Cruz}\ \emph {et~al.}(2026)\citenamefont
  {Arguello~Cruz}, \citenamefont {Klebanov}, \citenamefont {Tarnopolsky},\ and\
  \citenamefont {Xin}}]{Cruz25}%
  \BibitemOpen
  \bibfield  {author} {\bibinfo {author} {\bibfnamefont {E.}~\bibnamefont
  {Arguello~Cruz}}, \bibinfo {author} {\bibfnamefont {I.~R.}\ \bibnamefont
  {Klebanov}}, \bibinfo {author} {\bibfnamefont {G.}~\bibnamefont
  {Tarnopolsky}},\ and\ \bibinfo {author} {\bibfnamefont {Y.}~\bibnamefont
  {Xin}},\ }\bibfield  {title} {\bibinfo {title} {Yang-lee quantum criticality
  in various dimensions},\ }\href {https://doi.org/10.1103/w4qg-2xwn}
  {\bibfield  {journal} {\bibinfo  {journal} {Phys. Rev. X}\ }\textbf {\bibinfo
  {volume} {16}},\ \bibinfo {pages} {011022} (\bibinfo {year}
  {2026})}\BibitemShut {NoStop}%
\bibitem [{\citenamefont {Miró}\ and\ \citenamefont
  {Delouche}(2025)}]{Miro25}%
  \BibitemOpen
  \bibfield  {author} {\bibinfo {author} {\bibfnamefont {J.~E.}\ \bibnamefont
  {Miró}}\ and\ \bibinfo {author} {\bibfnamefont {O.}~\bibnamefont
  {Delouche}},\ }\bibfield  {title} {\bibinfo {title} {{Flowing from the Ising
  model on the fuzzy sphere to the 3D Lee-Yang CFT}},\ }\href
  {https://doi.org/10.1007/JHEP10(2025)037} {\bibfield  {journal} {\bibinfo
  {journal} {Journal of High Energy Physics}\ }\textbf {\bibinfo {volume}
  {2025}},\ \bibinfo {pages} {37} (\bibinfo {year} {2025})}\BibitemShut
  {NoStop}%
\bibitem [{\citenamefont {Dedushenko}(2024)}]{Dedushenko24}%
  \BibitemOpen
  \bibfield  {author} {\bibinfo {author} {\bibfnamefont {M.}~\bibnamefont
  {Dedushenko}},\ }\href {https://arxiv.org/abs/2407.15948} {\bibinfo {title}
  {{Ising {BCFT} from Fuzzy Hemisphere}}} (\bibinfo {year} {2024}),\ \Eprint
  {https://arxiv.org/abs/2407.15948} {arXiv:2407.15948 [hep-th]} \BibitemShut
  {NoStop}%
\bibitem [{\citenamefont {Zhou}\ and\ \citenamefont {He}(2025)}]{Zhou25a}%
  \BibitemOpen
  \bibfield  {author} {\bibinfo {author} {\bibfnamefont {Z.}~\bibnamefont
  {Zhou}}\ and\ \bibinfo {author} {\bibfnamefont {Y.-C.}\ \bibnamefont {He}},\
  }\bibfield  {title} {\bibinfo {title} {3d conformal field theories with
  $\mathrm{Sp}(n)$ global symmetry on a fuzzy sphere},\ }\href
  {https://doi.org/10.1103/xstj-xvcy} {\bibfield  {journal} {\bibinfo
  {journal} {Phys. Rev. Lett.}\ }\textbf {\bibinfo {volume} {135}},\ \bibinfo
  {pages} {026504} (\bibinfo {year} {2025})}\BibitemShut {NoStop}%
\bibitem [{\citenamefont {Zhou}\ \emph {et~al.}(2024)\citenamefont {Zhou},
  \citenamefont {Hu}, \citenamefont {Zhu},\ and\ \citenamefont {He}}]{Zhou24a}%
  \BibitemOpen
  \bibfield  {author} {\bibinfo {author} {\bibfnamefont {Z.}~\bibnamefont
  {Zhou}}, \bibinfo {author} {\bibfnamefont {L.}~\bibnamefont {Hu}}, \bibinfo
  {author} {\bibfnamefont {W.}~\bibnamefont {Zhu}},\ and\ \bibinfo {author}
  {\bibfnamefont {Y.-C.}\ \bibnamefont {He}},\ }\bibfield  {title} {\bibinfo
  {title} {{SO(5) Deconfined Phase Transition under the Fuzzy-Sphere
  Microscope: Approximate Conformal Symmetry, Pseudo-Criticality, and Operator
  Spectrum}},\ }\href {https://doi.org/10.1103/PhysRevX.14.021044} {\bibfield
  {journal} {\bibinfo  {journal} {Phys. Rev. X}\ }\textbf {\bibinfo {volume}
  {14}},\ \bibinfo {pages} {021044} (\bibinfo {year} {2024})}\BibitemShut
  {NoStop}%
\bibitem [{\citenamefont {Yang}\ \emph {et~al.}(2026)\citenamefont {Yang},
  \citenamefont {Hu}, \citenamefont {Han}, \citenamefont {Zhu},\ and\
  \citenamefont {Chen}}]{Yang25}%
  \BibitemOpen
  \bibfield  {author} {\bibinfo {author} {\bibfnamefont {S.}~\bibnamefont
  {Yang}}, \bibinfo {author} {\bibfnamefont {L.-d.}\ \bibnamefont {Hu}},
  \bibinfo {author} {\bibfnamefont {C.}~\bibnamefont {Han}}, \bibinfo {author}
  {\bibfnamefont {W.}~\bibnamefont {Zhu}},\ and\ \bibinfo {author}
  {\bibfnamefont {Y.}~\bibnamefont {Chen}},\ }\bibfield  {title} {\bibinfo
  {title} {{Conformal Operator Flows of the Deconfined Quantum Criticality from
  $\mathrm{SO}(5)$ to $\mathrm{O}(4)$}},\ }\href
  {https://doi.org/10.1103/l6vw-6z79} {\bibfield  {journal} {\bibinfo
  {journal} {Phys. Rev. Lett.}\ }\textbf {\bibinfo {volume} {136}},\ \bibinfo
  {pages} {076505} (\bibinfo {year} {2026})}\BibitemShut {NoStop}%
\bibitem [{\citenamefont {Zhou}\ \emph {et~al.}(2025)\citenamefont {Zhou},
  \citenamefont {Wang},\ and\ \citenamefont {He}}]{Zhou25b}%
  \BibitemOpen
  \bibfield  {author} {\bibinfo {author} {\bibfnamefont {Z.}~\bibnamefont
  {Zhou}}, \bibinfo {author} {\bibfnamefont {C.}~\bibnamefont {Wang}},\ and\
  \bibinfo {author} {\bibfnamefont {Y.-C.}\ \bibnamefont {He}},\ }\href
  {https://arxiv.org/abs/2507.19580} {\bibinfo {title} {Chern-simons-matter
  conformal field theory on fuzzy sphere: Confinement transition of
  {Kalmeyer}-{Laughlin} chiral spin liquid}} (\bibinfo {year} {2025}),\ \Eprint
  {https://arxiv.org/abs/2507.19580} {arXiv:2507.19580 [cond-mat.str-el]}
  \BibitemShut {NoStop}%
\bibitem [{\citenamefont {Taylor}\ \emph {et~al.}(2026)\citenamefont {Taylor},
  \citenamefont {Voinea}, \citenamefont {Papi\ifmmode~\acute{c}\else
  \'{c}\fi{}},\ and\ \citenamefont {Fan}}]{Taylor25}%
  \BibitemOpen
  \bibfield  {author} {\bibinfo {author} {\bibfnamefont {J.}~\bibnamefont
  {Taylor}}, \bibinfo {author} {\bibfnamefont {C.}~\bibnamefont {Voinea}},
  \bibinfo {author} {\bibfnamefont {Z.}~\bibnamefont
  {Papi\ifmmode~\acute{c}\else \'{c}\fi{}}},\ and\ \bibinfo {author}
  {\bibfnamefont {R.}~\bibnamefont {Fan}},\ }\bibfield  {title} {\bibinfo
  {title} {{Conformal Scalar Field Theory from Ising Tricriticality on the
  Fuzzy Sphere}},\ }\href {https://doi.org/10.1103/cj3l-cf58} {\bibfield
  {journal} {\bibinfo  {journal} {Phys. Rev. Lett.}\ }\textbf {\bibinfo
  {volume} {136}},\ \bibinfo {pages} {056503} (\bibinfo {year}
  {2026})}\BibitemShut {NoStop}%
\bibitem [{\citenamefont {He}(2025)}]{He25}%
  \BibitemOpen
  \bibfield  {author} {\bibinfo {author} {\bibfnamefont {Y.-C.}\ \bibnamefont
  {He}},\ }\href {https://arxiv.org/abs/2506.14904} {\bibinfo {title} {Free
  real scalar cft on fuzzy sphere: spectrum, algebra and wavefunction ansatz}}
  (\bibinfo {year} {2025}),\ \Eprint {https://arxiv.org/abs/2506.14904}
  {arXiv:2506.14904 [hep-th]} \BibitemShut {NoStop}%
\bibitem [{\citenamefont {Voinea}\ \emph {et~al.}(2025)\citenamefont {Voinea},
  \citenamefont {Fan}, \citenamefont {Regnault},\ and\ \citenamefont
  {Papi\ifmmode~\acute{c}\else \'{c}\fi{}}}]{Voinea25}%
  \BibitemOpen
  \bibfield  {author} {\bibinfo {author} {\bibfnamefont {C.}~\bibnamefont
  {Voinea}}, \bibinfo {author} {\bibfnamefont {R.}~\bibnamefont {Fan}},
  \bibinfo {author} {\bibfnamefont {N.}~\bibnamefont {Regnault}},\ and\
  \bibinfo {author} {\bibfnamefont {Z.}~\bibnamefont
  {Papi\ifmmode~\acute{c}\else \'{c}\fi{}}},\ }\bibfield  {title} {\bibinfo
  {title} {Regularizing {3D} conformal field theories via anyons on the fuzzy
  sphere},\ }\href {https://doi.org/10.1103/bf4k-phl9} {\bibfield  {journal}
  {\bibinfo  {journal} {Phys. Rev. X}\ }\textbf {\bibinfo {volume} {15}},\
  \bibinfo {pages} {031007} (\bibinfo {year} {2025})}\BibitemShut {NoStop}%
\bibitem [{\citenamefont {Teo}\ \emph {et~al.}(2015)\citenamefont {Teo},
  \citenamefont {Hughes},\ and\ \citenamefont {Fradkin}}]{Teo15}%
  \BibitemOpen
  \bibfield  {author} {\bibinfo {author} {\bibfnamefont {J.~C.}\ \bibnamefont
  {Teo}}, \bibinfo {author} {\bibfnamefont {T.~L.}\ \bibnamefont {Hughes}},\
  and\ \bibinfo {author} {\bibfnamefont {E.}~\bibnamefont {Fradkin}},\
  }\bibfield  {title} {\bibinfo {title} {Theory of twist liquids: Gauging an
  anyonic symmetry},\ }\href
  {https://doi.org/https://doi.org/10.1016/j.aop.2015.05.012} {\bibfield
  {journal} {\bibinfo  {journal} {Annals of Physics}\ }\textbf {\bibinfo
  {volume} {360}},\ \bibinfo {pages} {349} (\bibinfo {year}
  {2015})}\BibitemShut {NoStop}%
\bibitem [{SM()}]{SM}%
  \BibitemOpen
  \href@noop {} {}\bibinfo {note} {See the Supplemental Online Material for
  details of the derivations and further numerical results.}\BibitemShut
  {Stop}%
\bibitem [{\citenamefont {Wen}\ and\ \citenamefont {Zee}(1992)}]{Wen92}%
  \BibitemOpen
  \bibfield  {author} {\bibinfo {author} {\bibfnamefont {X.~G.}\ \bibnamefont
  {Wen}}\ and\ \bibinfo {author} {\bibfnamefont {A.}~\bibnamefont {Zee}},\
  }\bibfield  {title} {\bibinfo {title} {Shift and spin vector: New topological
  quantum numbers for the {Hall} fluids},\ }\href
  {https://doi.org/10.1103/PhysRevLett.69.953} {\bibfield  {journal} {\bibinfo
  {journal} {Phys. Rev. Lett.}\ }\textbf {\bibinfo {volume} {69}},\ \bibinfo
  {pages} {953} (\bibinfo {year} {1992})}\BibitemShut {NoStop}%
\bibitem [{\citenamefont {Ho}(1995)}]{Ho95}%
  \BibitemOpen
  \bibfield  {author} {\bibinfo {author} {\bibfnamefont {T.-L.}\ \bibnamefont
  {Ho}},\ }\bibfield  {title} {\bibinfo {title} {Broken symmetry of
  two-component $\mathit{\ensuremath{\nu}}=1/2$ quantum hall states},\ }\href
  {https://doi.org/10.1103/PhysRevLett.75.1186} {\bibfield  {journal} {\bibinfo
   {journal} {Phys. Rev. Lett.}\ }\textbf {\bibinfo {volume} {75}},\ \bibinfo
  {pages} {1186} (\bibinfo {year} {1995})}\BibitemShut {NoStop}%
\bibitem [{\citenamefont {Cooper}\ \emph {et~al.}(2001)\citenamefont {Cooper},
  \citenamefont {Wilkin},\ and\ \citenamefont {Gunn}}]{Cooper01}%
  \BibitemOpen
  \bibfield  {author} {\bibinfo {author} {\bibfnamefont {N.~R.}\ \bibnamefont
  {Cooper}}, \bibinfo {author} {\bibfnamefont {N.~K.}\ \bibnamefont {Wilkin}},\
  and\ \bibinfo {author} {\bibfnamefont {J.~M.~F.}\ \bibnamefont {Gunn}},\
  }\bibfield  {title} {\bibinfo {title} {Quantum phases of vortices in rotating
  bose-einstein condensates},\ }\href
  {https://doi.org/10.1103/PhysRevLett.87.120405} {\bibfield  {journal}
  {\bibinfo  {journal} {Phys. Rev. Lett.}\ }\textbf {\bibinfo {volume} {87}},\
  \bibinfo {pages} {120405} (\bibinfo {year} {2001})}\BibitemShut {NoStop}%
\bibitem [{\citenamefont {Regnault}\ and\ \citenamefont
  {Jolicoeur}(2003)}]{Regnault03}%
  \BibitemOpen
  \bibfield  {author} {\bibinfo {author} {\bibfnamefont {N.}~\bibnamefont
  {Regnault}}\ and\ \bibinfo {author} {\bibfnamefont {T.}~\bibnamefont
  {Jolicoeur}},\ }\bibfield  {title} {\bibinfo {title} {Quantum {Hall}
  fractions in rotating bose-einstein condensates},\ }\href
  {https://doi.org/10.1103/PhysRevLett.91.030402} {\bibfield  {journal}
  {\bibinfo  {journal} {Phys. Rev. Lett.}\ }\textbf {\bibinfo {volume} {91}},\
  \bibinfo {pages} {030402} (\bibinfo {year} {2003})}\BibitemShut {NoStop}%
\bibitem [{\citenamefont {Li}\ and\ \citenamefont {Haldane}(2008)}]{Li08}%
  \BibitemOpen
  \bibfield  {author} {\bibinfo {author} {\bibfnamefont {H.}~\bibnamefont
  {Li}}\ and\ \bibinfo {author} {\bibfnamefont {F.~D.~M.}\ \bibnamefont
  {Haldane}},\ }\bibfield  {title} {\bibinfo {title} {Entanglement spectrum as
  a generalization of entanglement entropy: Identification of topological order
  in non-{Abelian} fractional quantum {Hall} effect states},\ }\href
  {https://doi.org/10.1103/PhysRevLett.101.010504} {\bibfield  {journal}
  {\bibinfo  {journal} {Phys. Rev. Lett.}\ }\textbf {\bibinfo {volume} {101}},\
  \bibinfo {pages} {010504} (\bibinfo {year} {2008})}\BibitemShut {NoStop}%
\bibitem [{\citenamefont {Dubail}\ \emph {et~al.}(2012)\citenamefont {Dubail},
  \citenamefont {Read},\ and\ \citenamefont {Rezayi}}]{Dubail12}%
  \BibitemOpen
  \bibfield  {author} {\bibinfo {author} {\bibfnamefont {J.}~\bibnamefont
  {Dubail}}, \bibinfo {author} {\bibfnamefont {N.}~\bibnamefont {Read}},\ and\
  \bibinfo {author} {\bibfnamefont {E.~H.}\ \bibnamefont {Rezayi}},\ }\bibfield
   {title} {\bibinfo {title} {Real-space entanglement spectrum of quantum
  {Hall} systems},\ }\href {https://doi.org/10.1103/PhysRevB.85.115321}
  {\bibfield  {journal} {\bibinfo  {journal} {Phys. Rev. B}\ }\textbf {\bibinfo
  {volume} {85}},\ \bibinfo {pages} {115321} (\bibinfo {year}
  {2012})}\BibitemShut {NoStop}%
\bibitem [{\citenamefont {Sterdyniak}\ \emph {et~al.}(2012)\citenamefont
  {Sterdyniak}, \citenamefont {Chandran}, \citenamefont {Regnault},
  \citenamefont {Bernevig},\ and\ \citenamefont {Bonderson}}]{Sterdyniak12}%
  \BibitemOpen
  \bibfield  {author} {\bibinfo {author} {\bibfnamefont {A.}~\bibnamefont
  {Sterdyniak}}, \bibinfo {author} {\bibfnamefont {A.}~\bibnamefont
  {Chandran}}, \bibinfo {author} {\bibfnamefont {N.}~\bibnamefont {Regnault}},
  \bibinfo {author} {\bibfnamefont {B.~A.}\ \bibnamefont {Bernevig}},\ and\
  \bibinfo {author} {\bibfnamefont {P.}~\bibnamefont {Bonderson}},\ }\bibfield
  {title} {\bibinfo {title} {Real-space entanglement spectrum of quantum {Hall}
  states},\ }\href {https://doi.org/10.1103/PhysRevB.85.125308} {\bibfield
  {journal} {\bibinfo  {journal} {Phys. Rev. B}\ }\textbf {\bibinfo {volume}
  {85}},\ \bibinfo {pages} {125308} (\bibinfo {year} {2012})}\BibitemShut
  {NoStop}%
\bibitem [{\citenamefont {Wen}(1993)}]{XGWen1993}%
  \BibitemOpen
  \bibfield  {author} {\bibinfo {author} {\bibfnamefont {X.-G.}\ \bibnamefont
  {Wen}},\ }\bibfield  {title} {\bibinfo {title} {Topological order and edge
  structure of \ensuremath{\nu}=1/2 quantum {Hall} state},\ }\href
  {https://doi.org/10.1103/PhysRevLett.70.355} {\bibfield  {journal} {\bibinfo
  {journal} {Phys. Rev. Lett.}\ }\textbf {\bibinfo {volume} {70}},\ \bibinfo
  {pages} {355} (\bibinfo {year} {1993})}\BibitemShut {NoStop}%
\bibitem [{\citenamefont {Milovanovi\ifmmode~\acute{c}\else \'{c}\fi{}}\ and\
  \citenamefont {Read}(1996)}]{Read1996}%
  \BibitemOpen
  \bibfield  {author} {\bibinfo {author} {\bibfnamefont {M.}~\bibnamefont
  {Milovanovi\ifmmode~\acute{c}\else \'{c}\fi{}}}\ and\ \bibinfo {author}
  {\bibfnamefont {N.}~\bibnamefont {Read}},\ }\bibfield  {title} {\bibinfo
  {title} {Edge excitations of paired fractional quantum {Hall} states},\
  }\href {https://doi.org/10.1103/PhysRevB.53.13559} {\bibfield  {journal}
  {\bibinfo  {journal} {Phys. Rev. B}\ }\textbf {\bibinfo {volume} {53}},\
  \bibinfo {pages} {13559} (\bibinfo {year} {1996})}\BibitemShut {NoStop}%
\bibitem [{\citenamefont {Lao}\ and\ \citenamefont {Rychkov}(2023)}]{Lao23}%
  \BibitemOpen
  \bibfield  {author} {\bibinfo {author} {\bibfnamefont {B.-X.}\ \bibnamefont
  {Lao}}\ and\ \bibinfo {author} {\bibfnamefont {S.}~\bibnamefont {Rychkov}},\
  }\bibfield  {title} {\bibinfo {title} {{3D Ising CFT and exact
  diagonalization on icosahedron: The power of conformal perturbation
  theory}},\ }\href {https://doi.org/10.21468/SciPostPhys.15.6.243} {\bibfield
  {journal} {\bibinfo  {journal} {SciPost Phys.}\ }\textbf {\bibinfo {volume}
  {15}},\ \bibinfo {pages} {243} (\bibinfo {year} {2023})}\BibitemShut
  {NoStop}%
\bibitem [{\citenamefont {Läuchli}\ \emph {et~al.}(2025)\citenamefont
  {Läuchli}, \citenamefont {Herviou}, \citenamefont {Wilhelm},\ and\
  \citenamefont {Rychkov}}]{Lauchli25}%
  \BibitemOpen
  \bibfield  {author} {\bibinfo {author} {\bibfnamefont {A.~M.}\ \bibnamefont
  {Läuchli}}, \bibinfo {author} {\bibfnamefont {L.}~\bibnamefont {Herviou}},
  \bibinfo {author} {\bibfnamefont {P.~H.}\ \bibnamefont {Wilhelm}},\ and\
  \bibinfo {author} {\bibfnamefont {S.}~\bibnamefont {Rychkov}},\ }\bibfield
  {title} {\bibinfo {title} {{Exact diagonalization, matrix product states and
  conformal perturbation theory study of a 3D Ising fuzzy sphere model}},\
  }\href {https://doi.org/10.21468/SciPostPhys.19.3.076} {\bibfield  {journal}
  {\bibinfo  {journal} {SciPost Phys.}\ }\textbf {\bibinfo {volume} {19}},\
  \bibinfo {pages} {076} (\bibinfo {year} {2025})}\BibitemShut {NoStop}%
\bibitem [{\citenamefont {Wang}\ and\ \citenamefont {Senthil}(2016)}]{Wang16}%
  \BibitemOpen
  \bibfield  {author} {\bibinfo {author} {\bibfnamefont {C.}~\bibnamefont
  {Wang}}\ and\ \bibinfo {author} {\bibfnamefont {T.}~\bibnamefont {Senthil}},\
  }\bibfield  {title} {\bibinfo {title} {Composite fermi liquids in the lowest
  landau level},\ }\href {https://doi.org/10.1103/PhysRevB.94.245107}
  {\bibfield  {journal} {\bibinfo  {journal} {Phys. Rev. B}\ }\textbf {\bibinfo
  {volume} {94}},\ \bibinfo {pages} {245107} (\bibinfo {year}
  {2016})}\BibitemShut {NoStop}%
\bibitem [{\citenamefont {Geraedts}\ \emph {et~al.}(2017)\citenamefont
  {Geraedts}, \citenamefont {Repellin}, \citenamefont {Wang}, \citenamefont
  {Mong}, \citenamefont {Senthil},\ and\ \citenamefont
  {Regnault}}]{Geraedts17}%
  \BibitemOpen
  \bibfield  {author} {\bibinfo {author} {\bibfnamefont {S.~D.}\ \bibnamefont
  {Geraedts}}, \bibinfo {author} {\bibfnamefont {C.}~\bibnamefont {Repellin}},
  \bibinfo {author} {\bibfnamefont {C.}~\bibnamefont {Wang}}, \bibinfo {author}
  {\bibfnamefont {R.~S.~K.}\ \bibnamefont {Mong}}, \bibinfo {author}
  {\bibfnamefont {T.}~\bibnamefont {Senthil}},\ and\ \bibinfo {author}
  {\bibfnamefont {N.}~\bibnamefont {Regnault}},\ }\bibfield  {title} {\bibinfo
  {title} {Emergent particle-hole symmetry in spinful bosonic quantum {Hall}
  systems},\ }\href {https://doi.org/10.1103/PhysRevB.96.075148} {\bibfield
  {journal} {\bibinfo  {journal} {Phys. Rev. B}\ }\textbf {\bibinfo {volume}
  {96}},\ \bibinfo {pages} {075148} (\bibinfo {year} {2017})}\BibitemShut
  {NoStop}%
\bibitem [{\citenamefont {L{\'e}onard}\ \emph {et~al.}(2023)\citenamefont
  {L{\'e}onard}, \citenamefont {Kim}, \citenamefont {Kwan}, \citenamefont
  {Segura}, \citenamefont {Grusdt}, \citenamefont {Repellin}, \citenamefont
  {Goldman},\ and\ \citenamefont {Greiner}}]{Leonard2023}%
  \BibitemOpen
  \bibfield  {author} {\bibinfo {author} {\bibfnamefont {J.}~\bibnamefont
  {L{\'e}onard}}, \bibinfo {author} {\bibfnamefont {S.}~\bibnamefont {Kim}},
  \bibinfo {author} {\bibfnamefont {J.}~\bibnamefont {Kwan}}, \bibinfo {author}
  {\bibfnamefont {P.}~\bibnamefont {Segura}}, \bibinfo {author} {\bibfnamefont
  {F.}~\bibnamefont {Grusdt}}, \bibinfo {author} {\bibfnamefont
  {C.}~\bibnamefont {Repellin}}, \bibinfo {author} {\bibfnamefont
  {N.}~\bibnamefont {Goldman}},\ and\ \bibinfo {author} {\bibfnamefont
  {M.}~\bibnamefont {Greiner}},\ }\bibfield  {title} {\bibinfo {title}
  {Realization of a fractional quantum {Hall} state with ultracold atoms},\
  }\href {https://doi.org/10.1038/s41586-023-06122-4} {\bibfield  {journal}
  {\bibinfo  {journal} {Nature}\ }\textbf {\bibinfo {volume} {619}},\ \bibinfo
  {pages} {495} (\bibinfo {year} {2023})}\BibitemShut {NoStop}%
\bibitem [{\citenamefont {Wang}\ \emph {et~al.}(2024)\citenamefont {Wang},
  \citenamefont {Liu}, \citenamefont {Chen}, \citenamefont {Chen},
  \citenamefont {Zhao}, \citenamefont {Ying}, \citenamefont {Shang},
  \citenamefont {Wang}, \citenamefont {Huo}, \citenamefont {Peng},
  \citenamefont {Zhu}, \citenamefont {Lu},\ and\ \citenamefont
  {Pan}}]{CanWang2024}%
  \BibitemOpen
  \bibfield  {author} {\bibinfo {author} {\bibfnamefont {C.}~\bibnamefont
  {Wang}}, \bibinfo {author} {\bibfnamefont {F.-M.}\ \bibnamefont {Liu}},
  \bibinfo {author} {\bibfnamefont {M.-C.}\ \bibnamefont {Chen}}, \bibinfo
  {author} {\bibfnamefont {H.}~\bibnamefont {Chen}}, \bibinfo {author}
  {\bibfnamefont {X.-H.}\ \bibnamefont {Zhao}}, \bibinfo {author}
  {\bibfnamefont {C.}~\bibnamefont {Ying}}, \bibinfo {author} {\bibfnamefont
  {Z.-X.}\ \bibnamefont {Shang}}, \bibinfo {author} {\bibfnamefont {J.-W.}\
  \bibnamefont {Wang}}, \bibinfo {author} {\bibfnamefont {Y.-H.}\ \bibnamefont
  {Huo}}, \bibinfo {author} {\bibfnamefont {C.-Z.}\ \bibnamefont {Peng}},
  \bibinfo {author} {\bibfnamefont {X.}~\bibnamefont {Zhu}}, \bibinfo {author}
  {\bibfnamefont {C.-Y.}\ \bibnamefont {Lu}},\ and\ \bibinfo {author}
  {\bibfnamefont {J.-W.}\ \bibnamefont {Pan}},\ }\bibfield  {title} {\bibinfo
  {title} {Realization of fractional quantum {Hall} state with interacting
  photons},\ }\href {https://doi.org/10.1126/science.ado3912} {\bibfield
  {journal} {\bibinfo  {journal} {Science}\ }\textbf {\bibinfo {volume}
  {384}},\ \bibinfo {pages} {579} (\bibinfo {year} {2024})}\BibitemShut
  {NoStop}%
\bibitem [{\citenamefont {Liang}\ \emph {et~al.}(2024)\citenamefont {Liang},
  \citenamefont {Liu}, \citenamefont {Yang}, \citenamefont {Huang},
  \citenamefont {Wurstbauer}, \citenamefont {Dean}, \citenamefont {West},
  \citenamefont {Pfeiffer}, \citenamefont {Du},\ and\ \citenamefont
  {Pinczuk}}]{Liang2024}%
  \BibitemOpen
  \bibfield  {author} {\bibinfo {author} {\bibfnamefont {J.}~\bibnamefont
  {Liang}}, \bibinfo {author} {\bibfnamefont {Z.}~\bibnamefont {Liu}}, \bibinfo
  {author} {\bibfnamefont {Z.}~\bibnamefont {Yang}}, \bibinfo {author}
  {\bibfnamefont {Y.}~\bibnamefont {Huang}}, \bibinfo {author} {\bibfnamefont
  {U.}~\bibnamefont {Wurstbauer}}, \bibinfo {author} {\bibfnamefont {C.~R.}\
  \bibnamefont {Dean}}, \bibinfo {author} {\bibfnamefont {K.~W.}\ \bibnamefont
  {West}}, \bibinfo {author} {\bibfnamefont {L.~N.}\ \bibnamefont {Pfeiffer}},
  \bibinfo {author} {\bibfnamefont {L.}~\bibnamefont {Du}},\ and\ \bibinfo
  {author} {\bibfnamefont {A.}~\bibnamefont {Pinczuk}},\ }\bibfield  {title}
  {\bibinfo {title} {Evidence for chiral graviton modes in fractional quantum
  {Hall} liquids},\ }\href {https://doi.org/10.1038/s41586-024-07201-w}
  {\bibfield  {journal} {\bibinfo  {journal} {Nature}\ }\textbf {\bibinfo
  {volume} {628}},\ \bibinfo {pages} {78} (\bibinfo {year} {2024})}\BibitemShut
  {NoStop}%
\bibitem [{\citenamefont {Spielman}\ \emph {et~al.}(2000)\citenamefont
  {Spielman}, \citenamefont {Eisenstein}, \citenamefont {Pfeiffer},\ and\
  \citenamefont {West}}]{Spielman00}%
  \BibitemOpen
  \bibfield  {author} {\bibinfo {author} {\bibfnamefont {I.~B.}\ \bibnamefont
  {Spielman}}, \bibinfo {author} {\bibfnamefont {J.~P.}\ \bibnamefont
  {Eisenstein}}, \bibinfo {author} {\bibfnamefont {L.~N.}\ \bibnamefont
  {Pfeiffer}},\ and\ \bibinfo {author} {\bibfnamefont {K.~W.}\ \bibnamefont
  {West}},\ }\bibfield  {title} {\bibinfo {title} {Resonantly enhanced
  tunneling in a double layer quantum {Hall} ferromagnet},\ }\href
  {https://doi.org/10.1103/PhysRevLett.84.5808} {\bibfield  {journal} {\bibinfo
   {journal} {Phys. Rev. Lett.}\ }\textbf {\bibinfo {volume} {84}},\ \bibinfo
  {pages} {5808} (\bibinfo {year} {2000})}\BibitemShut {NoStop}%
\bibitem [{\citenamefont {Zhang}\ \emph {et~al.}(2023)\citenamefont {Zhang},
  \citenamefont {Zhu},\ and\ \citenamefont {Vishwanath}}]{Zhang23XYstar}%
  \BibitemOpen
  \bibfield  {author} {\bibinfo {author} {\bibfnamefont {Y.-H.}\ \bibnamefont
  {Zhang}}, \bibinfo {author} {\bibfnamefont {Z.}~\bibnamefont {Zhu}},\ and\
  \bibinfo {author} {\bibfnamefont {A.}~\bibnamefont {Vishwanath}},\ }\bibfield
   {title} {\bibinfo {title} {{XY* Transition and Extraordinary Boundary
  Criticality from Fractional Exciton Condensation in Quantum Hall Bilayer}},\
  }\href {https://doi.org/10.1103/PhysRevX.13.031023} {\bibfield  {journal}
  {\bibinfo  {journal} {Phys. Rev. X}\ }\textbf {\bibinfo {volume} {13}},\
  \bibinfo {pages} {031023} (\bibinfo {year} {2023})}\BibitemShut {NoStop}%
\bibitem [{\citenamefont {Barkeshli}\ and\ \citenamefont
  {McGreevy}(2014)}]{Barkeshli_2014}%
  \BibitemOpen
  \bibfield  {author} {\bibinfo {author} {\bibfnamefont {M.}~\bibnamefont
  {Barkeshli}}\ and\ \bibinfo {author} {\bibfnamefont {J.}~\bibnamefont
  {McGreevy}},\ }\bibfield  {title} {\bibinfo {title} {Continuous transition
  between fractional quantum hall and superfluid states},\ }\bibfield
  {journal} {\bibinfo  {journal} {Physical Review B}\ }\textbf {\bibinfo
  {volume} {89}},\ \href {https://doi.org/10.1103/physrevb.89.235116}
  {10.1103/physrevb.89.235116} (\bibinfo {year} {2014})\BibitemShut {NoStop}%
\bibitem [{\citenamefont {Yang}(2017)}]{Yang17}%
  \BibitemOpen
  \bibfield  {author} {\bibinfo {author} {\bibfnamefont {K.}~\bibnamefont
  {Yang}},\ }\bibfield  {title} {\bibinfo {title} {Interface and phase
  transition between {Moore}-{Read} and {Halperin} 331 fractional quantum
  {Hall} states: Realization of chiral {Majorana} fermion},\ }\href
  {https://doi.org/10.1103/PhysRevB.96.241305} {\bibfield  {journal} {\bibinfo
  {journal} {Phys. Rev. B}\ }\textbf {\bibinfo {volume} {96}},\ \bibinfo
  {pages} {241305} (\bibinfo {year} {2017})}\BibitemShut {NoStop}%
\bibitem [{\citenamefont {Ma}\ and\ \citenamefont {Yang}(2022)}]{Ma22}%
  \BibitemOpen
  \bibfield  {author} {\bibinfo {author} {\bibfnamefont {K.~K.~W.}\
  \bibnamefont {Ma}}\ and\ \bibinfo {author} {\bibfnamefont {K.}~\bibnamefont
  {Yang}},\ }\bibfield  {title} {\bibinfo {title} {Simple analog of the
  black-hole information paradox in quantum {Hall} interfaces},\ }\href
  {https://doi.org/10.1103/PhysRevB.105.045306} {\bibfield  {journal} {\bibinfo
   {journal} {Phys. Rev. B}\ }\textbf {\bibinfo {volume} {105}},\ \bibinfo
  {pages} {045306} (\bibinfo {year} {2022})}\BibitemShut {NoStop}%
\bibitem [{\citenamefont {Cr\'epel}\ \emph
  {et~al.}(2019{\natexlab{b}})\citenamefont {Cr\'epel}, \citenamefont
  {Estienne},\ and\ \citenamefont {Regnault}}]{Crepel19}%
  \BibitemOpen
  \bibfield  {author} {\bibinfo {author} {\bibfnamefont {V.}~\bibnamefont
  {Cr\'epel}}, \bibinfo {author} {\bibfnamefont {B.}~\bibnamefont {Estienne}},\
  and\ \bibinfo {author} {\bibfnamefont {N.}~\bibnamefont {Regnault}},\
  }\bibfield  {title} {\bibinfo {title} {Variational ansatz for an {Abelian} to
  {Non-Abelian} topological phase transition in $\ensuremath{\nu}=1/2+1/2$
  bilayers},\ }\href {https://doi.org/10.1103/PhysRevLett.123.126804}
  {\bibfield  {journal} {\bibinfo  {journal} {Phys. Rev. Lett.}\ }\textbf
  {\bibinfo {volume} {123}},\ \bibinfo {pages} {126804} (\bibinfo {year}
  {2019}{\natexlab{b}})}\BibitemShut {NoStop}%
\bibitem [{\citenamefont {Fei}\ \emph {et~al.}(2016)\citenamefont {Fei},
  \citenamefont {Giombi}, \citenamefont {Klebanov},\ and\ \citenamefont
  {Tarnopolsky}}]{Fei16}%
  \BibitemOpen
  \bibfield  {author} {\bibinfo {author} {\bibfnamefont {L.}~\bibnamefont
  {Fei}}, \bibinfo {author} {\bibfnamefont {S.}~\bibnamefont {Giombi}},
  \bibinfo {author} {\bibfnamefont {I.~R.}\ \bibnamefont {Klebanov}},\ and\
  \bibinfo {author} {\bibfnamefont {G.}~\bibnamefont {Tarnopolsky}},\
  }\bibfield  {title} {\bibinfo {title} {Yukawa conformal field theories and
  emergent supersymmetry},\ }\href {https://doi.org/10.1093/ptep/ptw120}
  {\bibfield  {journal} {\bibinfo  {journal} {Progress of Theoretical and
  Experimental Physics}\ }\textbf {\bibinfo {volume} {2016}},\ \bibinfo {pages}
  {12C105} (\bibinfo {year} {2016})},\ \Eprint
  {https://arxiv.org/abs/https://academic.oup.com/ptep/article-pdf/2016/12/12C105/9620609/ptw120.pdf}
  {https://academic.oup.com/ptep/article-pdf/2016/12/12C105/9620609/ptw120.pdf}
  \BibitemShut {NoStop}%
\bibitem [{\citenamefont {Gao}\ \emph {et~al.}(2025)\citenamefont {Gao},
  \citenamefont {Wang},\ and\ \citenamefont
  {Lee}}]{gao2025interactingcherninsulatortransition}%
  \BibitemOpen
  \bibfield  {author} {\bibinfo {author} {\bibfnamefont {Z.-Q.}\ \bibnamefont
  {Gao}}, \bibinfo {author} {\bibfnamefont {T.}~\bibnamefont {Wang}},\ and\
  \bibinfo {author} {\bibfnamefont {D.-H.}\ \bibnamefont {Lee}},\ }\href
  {https://arxiv.org/abs/2504.15338} {\bibinfo {title} {Interacting chern
  insulator transition on the sphere: revealing the gross-neveu-yukawa
  criticality}} (\bibinfo {year} {2025}),\ \Eprint
  {https://arxiv.org/abs/2504.15338} {arXiv:2504.15338 [cond-mat.str-el]}
  \BibitemShut {NoStop}%
\bibitem [{\citenamefont {Zhou}\ \emph {et~al.}(2026)\citenamefont {Zhou},
  \citenamefont {Gaiotto},\ and\ \citenamefont {He}}]{Zhou25_majorana}%
  \BibitemOpen
  \bibfield  {author} {\bibinfo {author} {\bibfnamefont {Z.}~\bibnamefont
  {Zhou}}, \bibinfo {author} {\bibfnamefont {D.}~\bibnamefont {Gaiotto}},\ and\
  \bibinfo {author} {\bibfnamefont {Y.-C.}\ \bibnamefont {He}},\ }\href
  {https://arxiv.org/abs/2509.08038} {\bibinfo {title} {Free and interacting
  fermionic conformal field theories on the fuzzy sphere}} (\bibinfo {year}
  {2026}),\ \Eprint {https://arxiv.org/abs/2509.08038} {arXiv:2509.08038
  [hep-th]} \BibitemShut {NoStop}%
\bibitem [{dia()}]{diagham}%
  \BibitemOpen
  \href@noop {} {}\bibinfo {note} {Diag{H}am,
  \url{https://www.nick-ux.org/diagham}}\BibitemShut {NoStop}%
\bibitem [{\citenamefont {Zhou}(2025)}]{FuzzifiED}%
  \BibitemOpen
  \bibfield  {author} {\bibinfo {author} {\bibfnamefont {Z.}~\bibnamefont
  {Zhou}},\ }\href {https://arxiv.org/abs/2503.00100} {\bibinfo {title}
  {{FuzzifiED -- Julia package for numerics on the fuzzy sphere}}} (\bibinfo
  {year} {2025}),\ \Eprint {https://arxiv.org/abs/2503.00100} {arXiv:2503.00100
  [cond-mat.str-el]} \BibitemShut {NoStop}%
\bibitem [{dat()}]{data}%
  \BibitemOpen
  \href@noop {} {}\bibinfo {note} {Cristian Voinea, Wei Zhu, Nicolas Regnault,
  and Zlatko Papi\'c, \url{https://doi.org/10.5518/1817}}\BibitemShut {NoStop}%
\bibitem [{\citenamefont {Fradkin}\ \emph {et~al.}(1998)\citenamefont
  {Fradkin}, \citenamefont {Nayak}, \citenamefont {Tsvelik},\ and\
  \citenamefont {Wilczek}}]{Fradkin97}%
  \BibitemOpen
  \bibfield  {author} {\bibinfo {author} {\bibfnamefont {E.~H.}\ \bibnamefont
  {Fradkin}}, \bibinfo {author} {\bibfnamefont {C.}~\bibnamefont {Nayak}},
  \bibinfo {author} {\bibfnamefont {A.}~\bibnamefont {Tsvelik}},\ and\ \bibinfo
  {author} {\bibfnamefont {F.}~\bibnamefont {Wilczek}},\ }\bibfield  {title}
  {\bibinfo {title} {{A Chern-Simons effective field theory for the Pfaffian
  quantum Hall state}},\ }\href {https://doi.org/10.1016/S0550-3213(98)00111-4}
  {\bibfield  {journal} {\bibinfo  {journal} {Nucl. Phys. B}\ }\textbf
  {\bibinfo {volume} {516}},\ \bibinfo {pages} {704} (\bibinfo {year}
  {1998})},\ \Eprint {https://arxiv.org/abs/cond-mat/9711087}
  {arXiv:cond-mat/9711087} \BibitemShut {NoStop}%
\bibitem [{\citenamefont {Fradkin}\ \emph {et~al.}(1999)\citenamefont
  {Fradkin}, \citenamefont {Nayak},\ and\ \citenamefont
  {Schoutens}}]{Fradkin99}%
  \BibitemOpen
  \bibfield  {author} {\bibinfo {author} {\bibfnamefont {E.}~\bibnamefont
  {Fradkin}}, \bibinfo {author} {\bibfnamefont {C.}~\bibnamefont {Nayak}},\
  and\ \bibinfo {author} {\bibfnamefont {K.}~\bibnamefont {Schoutens}},\
  }\bibfield  {title} {\bibinfo {title} {Landau-ginzburg theories for
  non-abelian quantum hall states},\ }\href
  {https://doi.org/https://doi.org/10.1016/S0550-3213(99)00039-5} {\bibfield
  {journal} {\bibinfo  {journal} {Nuclear Physics B}\ }\textbf {\bibinfo
  {volume} {546}},\ \bibinfo {pages} {711} (\bibinfo {year}
  {1999})}\BibitemShut {NoStop}%
\bibitem [{\citenamefont {Barkeshli}\ \emph {et~al.}(2019)\citenamefont
  {Barkeshli}, \citenamefont {Bonderson}, \citenamefont {Cheng},\ and\
  \citenamefont {Wang}}]{Barkeshli19}%
  \BibitemOpen
  \bibfield  {author} {\bibinfo {author} {\bibfnamefont {M.}~\bibnamefont
  {Barkeshli}}, \bibinfo {author} {\bibfnamefont {P.}~\bibnamefont
  {Bonderson}}, \bibinfo {author} {\bibfnamefont {M.}~\bibnamefont {Cheng}},\
  and\ \bibinfo {author} {\bibfnamefont {Z.}~\bibnamefont {Wang}},\ }\bibfield
  {title} {\bibinfo {title} {Symmetry fractionalization, defects, and gauging
  of topological phases},\ }\href {https://doi.org/10.1103/PhysRevB.100.115147}
  {\bibfield  {journal} {\bibinfo  {journal} {Phys. Rev. B}\ }\textbf {\bibinfo
  {volume} {100}},\ \bibinfo {pages} {115147} (\bibinfo {year}
  {2019})}\BibitemShut {NoStop}%
\bibitem [{\citenamefont {Hsin}\ and\ \citenamefont {Seiberg}(2016)}]{Hsin16}%
  \BibitemOpen
  \bibfield  {author} {\bibinfo {author} {\bibfnamefont {P.-S.}\ \bibnamefont
  {Hsin}}\ and\ \bibinfo {author} {\bibfnamefont {N.}~\bibnamefont {Seiberg}},\
  }\bibfield  {title} {\bibinfo {title} {{Level/rank Duality and
  Chern-Simons-Matter Theories}},\ }\href
  {https://doi.org/10.1007/JHEP09(2016)095} {\bibfield  {journal} {\bibinfo
  {journal} {Journal of High Energy Physics}\ }\textbf {\bibinfo {volume}
  {09}},\ \bibinfo {pages} {095} (\bibinfo {year} {2016})}\BibitemShut
  {NoStop}%
\bibitem [{\citenamefont {Witten}(1989)}]{Witten89}%
  \BibitemOpen
  \bibfield  {author} {\bibinfo {author} {\bibfnamefont {E.}~\bibnamefont
  {Witten}},\ }\bibfield  {title} {\bibinfo {title} {{Quantum Field Theory and
  the Jones Polynomial}},\ }\href {https://doi.org/10.1007/BF01217730}
  {\bibfield  {journal} {\bibinfo  {journal} {Commun. Math. Phys.}\ }\textbf
  {\bibinfo {volume} {121}},\ \bibinfo {pages} {351} (\bibinfo {year}
  {1989})}\BibitemShut {NoStop}%
\bibitem [{\citenamefont {Teo}\ and\ \citenamefont {Kane}(2014)}]{Teo14}%
  \BibitemOpen
  \bibfield  {author} {\bibinfo {author} {\bibfnamefont {J.~C.~Y.}\
  \bibnamefont {Teo}}\ and\ \bibinfo {author} {\bibfnamefont {C.~L.}\
  \bibnamefont {Kane}},\ }\bibfield  {title} {\bibinfo {title} {From luttinger
  liquid to non-abelian quantum hall states},\ }\href
  {https://doi.org/10.1103/PhysRevB.89.085101} {\bibfield  {journal} {\bibinfo
  {journal} {Phys. Rev. B}\ }\textbf {\bibinfo {volume} {89}},\ \bibinfo
  {pages} {085101} (\bibinfo {year} {2014})}\BibitemShut {NoStop}%
\bibitem [{\citenamefont {M\"oller}\ \emph {et~al.}(2011)\citenamefont
  {M\"oller}, \citenamefont {W\'ojs},\ and\ \citenamefont {Cooper}}]{Moller11}%
  \BibitemOpen
  \bibfield  {author} {\bibinfo {author} {\bibfnamefont {G.}~\bibnamefont
  {M\"oller}}, \bibinfo {author} {\bibfnamefont {A.}~\bibnamefont {W\'ojs}},\
  and\ \bibinfo {author} {\bibfnamefont {N.~R.}\ \bibnamefont {Cooper}},\
  }\bibfield  {title} {\bibinfo {title} {Neutral fermion excitations in the
  {Moore}-{Read} state at filling factor $\ensuremath{\nu}=5/2$},\ }\href
  {https://doi.org/10.1103/PhysRevLett.107.036803} {\bibfield  {journal}
  {\bibinfo  {journal} {Phys. Rev. Lett.}\ }\textbf {\bibinfo {volume} {107}},\
  \bibinfo {pages} {036803} (\bibinfo {year} {2011})}\BibitemShut {NoStop}%
\bibitem [{\citenamefont {Bonderson}\ \emph {et~al.}(2011)\citenamefont
  {Bonderson}, \citenamefont {Feiguin},\ and\ \citenamefont
  {Nayak}}]{Bonderson11}%
  \BibitemOpen
  \bibfield  {author} {\bibinfo {author} {\bibfnamefont {P.}~\bibnamefont
  {Bonderson}}, \bibinfo {author} {\bibfnamefont {A.~E.}\ \bibnamefont
  {Feiguin}},\ and\ \bibinfo {author} {\bibfnamefont {C.}~\bibnamefont
  {Nayak}},\ }\bibfield  {title} {\bibinfo {title} {Numerical calculation of
  the neutral fermion gap at the $\ensuremath{\nu}=5/2$ fractional quantum hall
  state},\ }\href {https://doi.org/10.1103/PhysRevLett.106.186802} {\bibfield
  {journal} {\bibinfo  {journal} {Phys. Rev. Lett.}\ }\textbf {\bibinfo
  {volume} {106}},\ \bibinfo {pages} {186802} (\bibinfo {year}
  {2011})}\BibitemShut {NoStop}%
\bibitem [{\citenamefont {Renn}\ and\ \citenamefont {Roberts}(1993)}]{Renn93}%
  \BibitemOpen
  \bibfield  {author} {\bibinfo {author} {\bibfnamefont {S.~R.}\ \bibnamefont
  {Renn}}\ and\ \bibinfo {author} {\bibfnamefont {B.~W.}\ \bibnamefont
  {Roberts}},\ }\bibfield  {title} {\bibinfo {title} {Magnetorotons and the
  fractional quantum {Hall} effect in double-quantum-well systems},\ }\href
  {https://doi.org/10.1103/PhysRevB.48.10926} {\bibfield  {journal} {\bibinfo
  {journal} {Phys. Rev. B}\ }\textbf {\bibinfo {volume} {48}},\ \bibinfo
  {pages} {10926} (\bibinfo {year} {1993})}\BibitemShut {NoStop}%
\bibitem [{\citenamefont {MacDonald}\ and\ \citenamefont
  {Zhang}(1994)}]{MacDonald94}%
  \BibitemOpen
  \bibfield  {author} {\bibinfo {author} {\bibfnamefont {A.~H.}\ \bibnamefont
  {MacDonald}}\ and\ \bibinfo {author} {\bibfnamefont {S.-C.}\ \bibnamefont
  {Zhang}},\ }\bibfield  {title} {\bibinfo {title} {Collective excitations in
  double-layer quantum hall systems},\ }\href
  {https://doi.org/10.1103/PhysRevB.49.17208} {\bibfield  {journal} {\bibinfo
  {journal} {Phys. Rev. B}\ }\textbf {\bibinfo {volume} {49}},\ \bibinfo
  {pages} {17208} (\bibinfo {year} {1994})}\BibitemShut {NoStop}%
\bibitem [{\citenamefont {Shizuya}(2003)}]{Shizuya03}%
  \BibitemOpen
  \bibfield  {author} {\bibinfo {author} {\bibfnamefont {K.}~\bibnamefont
  {Shizuya}},\ }\bibfield  {title} {\bibinfo {title} {Single-mode approximation
  and effective chern–simons theories for quantum hall systems},\ }\href
  {https://doi.org/10.1142/S0217979203023446} {\bibfield  {journal} {\bibinfo
  {journal} {International Journal of Modern Physics B}\ }\textbf {\bibinfo
  {volume} {17}},\ \bibinfo {pages} {5875} (\bibinfo {year} {2003})},\ \Eprint
  {https://arxiv.org/abs/https://doi.org/10.1142/S0217979203023446}
  {https://doi.org/10.1142/S0217979203023446} \BibitemShut {NoStop}%
\bibitem [{\citenamefont {Girvin}\ \emph {et~al.}(1986)\citenamefont {Girvin},
  \citenamefont {MacDonald},\ and\ \citenamefont {Platzman}}]{Girvin86}%
  \BibitemOpen
  \bibfield  {author} {\bibinfo {author} {\bibfnamefont {S.~M.}\ \bibnamefont
  {Girvin}}, \bibinfo {author} {\bibfnamefont {A.~H.}\ \bibnamefont
  {MacDonald}},\ and\ \bibinfo {author} {\bibfnamefont {P.~M.}\ \bibnamefont
  {Platzman}},\ }\bibfield  {title} {\bibinfo {title} {Magneto-roton theory of
  collective excitations in the fractional quantum {Hall} effect},\ }\href
  {https://doi.org/10.1103/PhysRevB.33.2481} {\bibfield  {journal} {\bibinfo
  {journal} {Phys. Rev. B}\ }\textbf {\bibinfo {volume} {33}},\ \bibinfo
  {pages} {2481} (\bibinfo {year} {1986})}\BibitemShut {NoStop}%
\bibitem [{\citenamefont {Morf}\ \emph {et~al.}(2002)\citenamefont {Morf},
  \citenamefont {d'Ambrumenil},\ and\ \citenamefont {Das~Sarma}}]{Morf02}%
  \BibitemOpen
  \bibfield  {author} {\bibinfo {author} {\bibfnamefont {R.~H.}\ \bibnamefont
  {Morf}}, \bibinfo {author} {\bibfnamefont {N.}~\bibnamefont {d'Ambrumenil}},\
  and\ \bibinfo {author} {\bibfnamefont {S.}~\bibnamefont {Das~Sarma}},\
  }\bibfield  {title} {\bibinfo {title} {Excitation gaps in fractional quantum
  {Hall} states: An exact diagonalization study},\ }\href
  {https://doi.org/10.1103/PhysRevB.66.075408} {\bibfield  {journal} {\bibinfo
  {journal} {Phys. Rev. B}\ }\textbf {\bibinfo {volume} {66}},\ \bibinfo
  {pages} {075408} (\bibinfo {year} {2002})}\BibitemShut {NoStop}%
\bibitem [{\citenamefont {Jain}\ and\ \citenamefont {Kamilla}(1997)}]{Jain97}%
  \BibitemOpen
  \bibfield  {author} {\bibinfo {author} {\bibfnamefont {J.~K.}\ \bibnamefont
  {Jain}}\ and\ \bibinfo {author} {\bibfnamefont {R.~K.}\ \bibnamefont
  {Kamilla}},\ }\bibfield  {title} {\bibinfo {title} {Composite fermions in the
  {Hilbert} space of the lowest electronic {Landau} level},\ }\href
  {https://doi.org/10.1142/S0217979297001301} {\bibfield  {journal} {\bibinfo
  {journal} {Int. J. Mod. Phys. B}\ }\textbf {\bibinfo {volume} {11}},\
  \bibinfo {pages} {2621} (\bibinfo {year} {1997})}\BibitemShut {NoStop}%
\bibitem [{\citenamefont {Balram}\ and\ \citenamefont
  {W\'ojs}(2020)}]{Balram20b}%
  \BibitemOpen
  \bibfield  {author} {\bibinfo {author} {\bibfnamefont {A.~C.}\ \bibnamefont
  {Balram}}\ and\ \bibinfo {author} {\bibfnamefont {A.}~\bibnamefont
  {W\'ojs}},\ }\bibfield  {title} {\bibinfo {title} {Fractional quantum {Hall}
  effect at $\ensuremath{\nu}=2+4/9$},\ }\href
  {https://doi.org/10.1103/PhysRevResearch.2.032035} {\bibfield  {journal}
  {\bibinfo  {journal} {Phys. Rev. Research}\ }\textbf {\bibinfo {volume}
  {2}},\ \bibinfo {pages} {032035} (\bibinfo {year} {2020})}\BibitemShut
  {NoStop}%
\bibitem [{\citenamefont {Read}\ and\ \citenamefont {Rezayi}(1996)}]{Read96}%
  \BibitemOpen
  \bibfield  {author} {\bibinfo {author} {\bibfnamefont {N.}~\bibnamefont
  {Read}}\ and\ \bibinfo {author} {\bibfnamefont {E.}~\bibnamefont {Rezayi}},\
  }\bibfield  {title} {\bibinfo {title} {Quasiholes and fermionic zero modes of
  paired fractional quantum {Hall} states: The mechanism for non-abelian
  statistics},\ }\href {https://doi.org/10.1103/PhysRevB.54.16864} {\bibfield
  {journal} {\bibinfo  {journal} {Phys. Rev. B}\ }\textbf {\bibinfo {volume}
  {54}},\ \bibinfo {pages} {16864} (\bibinfo {year} {1996})}\BibitemShut
  {NoStop}%
\end{thebibliography}%

\newpage 
\cleardoublepage 

\setcounter{equation}{0}
\setcounter{figure}{0}
\setcounter{table}{0}
\setcounter{page}{1}
\setcounter{section}{0}
\makeatletter
\renewcommand{\theequation}{S\arabic{equation}}
\renewcommand{\thefigure}{S\arabic{figure}}
\renewcommand{\thesection}{S\Roman{section}}
\renewcommand{\thepage}{\arabic{page}}
\renewcommand{\thetable}{S\arabic{table}}

\onecolumngrid

\begin{center}
\textbf{\large Supplemental Online Material for ``Critical Majorana fermion at a topological quantum Hall bilayer transition" }\\[5pt]
\begin{center}
Cristian Voinea$^{1}$, Wei Zhu$^{2,3}$, Nicolas Regnault$^{4,5,6}$, and Zlatko Papi\'c$^{1}$\\
\emph{$^{1}$School of Physics and Astronomy, University of Leeds, Leeds LS2 9JT, United Kingdom}\\
\emph{$^{2}$Institute of Natural Sciences, Westlake Institute for Advanced Study, Hangzhou 310024, China}\\
\emph{$^{3}$Department of Physics, School of Science, Westlake University, Hangzhou 310030, China}\\
\emph{$^{4}$Center for Computational Quantum Physics, Flatiron Institute, 162 5th Avenue, New York, NY 10010, USA}\\
\emph{$^{5}$Department of Physics, Princeton University, Princeton, New Jersey 08544, USA}\\
\emph{$^{6}$Laboratoire de Physique de l'Ecole normale sup\'{e}rieure, ENS, Universit\'{e} PSL, CNRS, Sorbonne Universit\'{e}, Universit\'{e} Paris-Diderot, Sorbonne Paris Cit\'{e}, 75005 Paris, France}
\end{center}

\begin{quote}
{\small In this Supplementary Material, we provide (i) more details about the gauged 3D Majorana CFT, (ii) extended analysis of different types of neutral and charge gaps, and (iii) a further discussion of the phase diagram and operator content at the critical point.
}  \\[20pt]
\end{quote}
\end{center}

\section{S1. Halperin-to-Pfaffian transition}\label{sec:cft}

In this section, we start by reviewing the current understanding of the Halperin-to-Pfaffian transition from the perspective of topological quantum field theory. We then provide a more quantitative understanding of the emergent 3D CFT at the phase transition, namely the $\mathbb{Z}_2$ gauged free Majorana fermion. 

\subsection{Topological phase transition}

A general Halperin-$\text{(nnm)}$ phase can be described using a coupled Abelian Chern-Simons (CS) theory:
\begin{equation}\label{eq:cs-halperin}
    \mathcal{L}_\text{Halperin} = \frac{1}{4\pi}\int_{S^2} K_{IJ}a_I \mathrm{d}a_J \, , \quad \text{where} \quad K = \begin{pmatrix} n & m \\ m & n \end{pmatrix} \, .
\end{equation}
Here, $a_{I=1,2}$ are $\mathrm{U}(1)$ the statistical gauge fields resulting from flux attachment in each of the layers (with the condensed notation $a_I \mathrm{d}a_J \equiv \epsilon^{\mu\nu\lambda} a_{I,\mu} \partial_{\nu} a_{J,\lambda}$), and the $K$-matrix captures the topological content of the theory. For the $220$ state, this reduces down to two uncoupled $\mathrm{U}(1)$ CS theories at level $2$. There exists an additional global $\mathbb{Z}_2$ symmetry, corresponding to exchanging the two layers. 

At large tunneling, the emergent bosonic Pfaffian state is captured by the non-Abelian $\mathrm{SU}(2)_2$ CS theory \cite{Fradkin97, Fradkin99}
\begin{equation}\label{eq:cs-pfaffian}
    \mathcal{L}_\text{Pfaffian} = \frac{2}{4\pi}\int_{S^2} \left( a \mathrm{d}a + \frac{2}{3} a^3 \right) \, .
\end{equation}
Here, the gauge field $a$ has an element associated to each $\mathrm{SU}(2)$ algebra element, such that $a \mathrm{d}a \equiv \epsilon^{\mu\nu\lambda} a^a_{\mu} \partial_{\nu} a^a_{\lambda}$ and $a^3 \equiv \epsilon^{\mu\nu\lambda} f_{abc} a^a_{\mu} a^b_{\nu} a^c_{\lambda}$ ($f_{abc}$ are the structure constants of the Lie algebra). Although the initial $\mathbb{Z}_2$ symmetry is obscured in this description of the Pfaffian, it plays an important role. While initially only a global symmetry, the symmetrization of the gauge fields in \cref{eq:cs-halperin} promotes it to a local (gauge) symmetry; in other words, in the Pfaffian phase the $\mathbb{Z}_2$ field is deconfined, and the Halperin phase is the confined (Higgs) phase \cite{Teo15, Barkeshli19}.  

The explicit gauge symmetry breaking pattern from $\mathrm{U}(1)_2 \times \mathrm{U}(1)_2$ to $\mathrm{SU}(2)_2$ becomes apparent if we start  from the $\nu=1/2$ Laughlin states making up the Halperin 220 order. This can be equivalently written in terms of a $\mathrm{SU}(2)_1$ theory \cite{Fradkin99}; this symmetry becomes manifest in the special case of hard-core bosons, and is understood more generally as the level-rank duality \cite{Hsin16}. Consequently, the topological order in the absence of inter-layer tunneling can be described as $\mathrm{SU}(2)_1 \times \mathrm{SU}(2)_1$. The possible emergence of the Pfaffian order $\mathrm{SU}(2)_2$ can be inferred from the edge theory; here, the corresponding chiral Wess-Zumino-Witten (WZW) model \cite{Witten89} can be rewritten in terms of the cosets:
\begin{equation}
    \mathrm{SU}(2)_1 \times \mathrm{SU}(2)_1 \to \frac{\mathrm{SU}(2)_1 \times \mathrm{SU}(2)_1}{\mathrm{SU}(2)_2} \times \frac{\mathrm{SU}(2)_2}{\mathrm{U}(1)} \times \mathrm{U}(1)\,,
\end{equation}
i.e. a copy of the chiral Ising model, a Majorana fermion ($\mathbb{Z}_2$ parafermion) and an additional boson, respectively. At the phase transition considered here, the modes of the chiral Ising $\mathrm{SU}(2)_1 \times \mathrm{SU}(2)_1 / \mathrm{SU}(2)_2$ are gapped out; since in the coset construction the chiral central charge is additive, projecting out these modes takes us from $c=2$ to $c=3/2$, as expected \cite{Teo14}.

\subsection{Gauged Majorana fermion CFT}

The Lagrangian for the free Majorana fermion in 3D takes the form
\begin{equation} \label{eq:majorana lagrangian}
		\mathcal{L} = \bar{\psi} (i \gamma^\mu\partial_\mu - m) \psi\,,
	\end{equation}
 where $\psi=(\psi_1,\psi_2)^\mathrm{T}$ is a two-component Majorana fermion field with $\psi_{1,2}$ Grassmann fields, and $\bar{\psi}=\psi^\mathrm{T} (i\gamma^0)$. Using the Lorentzian signature $(-1,1,1)$, the chosen $\gamma$ matrix representation
\begin{equation}
    \gamma^{0} = i \sigma^2 \, , \quad \gamma^{1} = \sigma^1 \, , \quad \gamma^{2} = \sigma^3 \, ,
\end{equation}
obeys the known Clifford algebra $\{ \gamma^{\mu}, \gamma^{\nu}\} = 2 \eta^{\mu \nu}$.

Let us list the low-lying operator content of the theory. The fermion operator $\psi$ has angular momentum $L=1/2$ and scaling dimension $\Delta_{\psi} = 1$. Its two-point correlation function is 
\begin{equation}\label{eq:fermion-2pt}
    \langle \psi^{\alpha}(x_1) \psi_{\beta}(x_2) \rangle = \frac{ic_\psi (x_{12})^{\alpha}_{\ \beta}}{|x_{12}|^{2 \Delta_{\psi} + 1 }} \,,
\end{equation}
where we use the notation $x^{\alpha}_{\ \beta} = x^{\mu} (\gamma_{\mu})^{\alpha}_{\ \beta}$ and the constant $c_\psi$ is fixed (see below).

The equation of motion, $\gamma_\mu \partial^\mu \psi = 0$, imposes strong constraints on the conformal tower of states of the fermion $\psi$, such that the only allowed descendants are
\begin{equation}
 \partial_{\mu_1}\dots\partial_{\mu_l}\psi, \quad \text{with scaling dimension} \; \Delta = l+1 \quad \text{and SO(3) angular momentum} \; L = l + 1/2.   
\end{equation}
The fermion bilinear $\bar{\psi}\psi$, which is space-time parity odd, is the only relevant scalar with $\Delta_{\bar{\psi}\psi} = 2$ and $L=0$. Its descendant family can be similarly constructed as 
\begin{equation}
\partial_{\mu_1}\dots\partial_{\mu_l}\square^n\bar{\psi}\psi, \quad \text{with} \quad  \Delta = l + 2n + 2, \;    L = l\,.
\end{equation}
Note that, in order to properly normalize the two-point function of the singlet to $\langle \bar{\psi}\psi (x) \bar{\psi}\psi(0) \rangle = |x|^{-2\Delta_{\bar{\psi}\psi}} $, we set the normalization constant to $c_\psi = 1/2$.

Due to the charge-neutrality of the Majorana, the charge current $J^\mu = \bar{\psi} \gamma^\mu \psi$ vanishes identically; moreover, with no chiral symmetry in odd dimensions, this theory does not possess any conserved currents at $L = 1$. Therefore, the stress-energy tensor is the next spinning operator with $L=2$ and $\Delta_T = 3$. 

For general spinning operators, we can construct two types of descendant families. 
The first parity-preserving family is given by the operators 
\begin{equation}
\partial_{\nu_1}\dots\partial_{\nu_j}\partial_{\mu_1}\dots\partial_{\mu_i}\square^n O_{\mu_1 \dots \mu_l} \quad \text{with} \quad \Delta_O + 2n + i + j, \;\;\; L=l+j-i.
\end{equation}  
The second family, 
\begin{equation}
 \epsilon_{\mu_k \rho \tau} \partial_\rho \partial_{\nu_1}\dots\partial_{\nu_j}\partial_{\mu_1}\dots\partial_{\mu_i}\square^n O_{\mu_1 \dots \mu_l}, \quad \text{with} \quad  \Delta_O + 2n + i + j +1, \;\;\;   L=l+j-i\;,
\end{equation}
reverses the space-time parity of $O$. In the case of the stress energy tensor, the additional conservation law $\partial_\mu T_{\mu\nu} = 0$ restricts the above families of descendants to $i = 0$. 

Moving to higher multiple-fermion operators, many conformal families vanish altogether due to the Majorana nature of the fermion. Since there are only two independent components $\psi_1, \psi_2$, operators such as $(\bar{\psi}\psi)^2$ or $\psi (\bar{\psi}\psi)$ vanish identically. However, the field derivative $\partial \psi$ provides independent spinor fields, with which additional primaries can be constructed. The lowest-lying operator with such a form is $(\bar{\psi}\psi)\partial^\mu\psi$, with angular momentum $L=3/2$ and $\Delta_{(\bar{\psi}\psi)\partial\psi} = 4$. 

\cref{tab:rawdata} lists all operators belonging to the conformal families above with scaling dimension $\Delta \leq 4$ and angular momentum $L\leq 3$. The theoretical data is used to plot Fig. 1(c) in the main text. The exact data is compared to the fuzzy sphere data for the model in the main text at system size $N=14$, rescaled such that the stress-energy tensor has $\Delta_T = 3$~\cite{Zhu23}.

\begin{table}
		\begin{tabular}{c|c|c|c|c}
			\hline
			\hline
			spin ($L$)  & layer $\mathbb{Z}_2$ & Operator & $\Delta_{\text{theory}}$ & $\Delta_\text{FS}$ ($N=14$)   \\
			\hline		
			1/2 & -1 & $\psi$ & 1.000 & 0.886  \\	
			3/2 & -1 & $\partial_\mu\psi$ & 2.000 & 2.162  \\
            1/2 & -1 & $\gamma_\mu \partial^\mu \psi=0$ & NA & - \\
			5/2 & -1 & $\partial_\mu\partial_\nu \psi$ & 3.000 & 3.426 \\
			7/2 & -1 & $\partial_\mu\partial_\nu\partial_\rho \psi$ & 4.000 & 3.536 \\
			\hline
			3/2 & -1 & $\bar{\psi}\psi \partial_\mu\psi$ & 4.000 & 3.921 \\
			\hline 
			0 & 1 & $\bar{\psi}\psi$ & 2.000 &  1.862 \\	
			1 & 1 & $\partial_{\mu} (\bar{\psi}\psi)$ & 3.000 & 3.015  \\					
			2 & 1 & $\partial_\mu\partial_\nu (\bar{\psi}\psi)$ & 4.000 &  3.914  \\
			0 & 1 & $\square (\bar{\psi}\psi)$ & 4.000 & 4.105  \\
            \hline
            2 & 1 & $T_{\mu\nu}$ & 3.000 & 3.000 \\
            2 & 1 & $\epsilon_{\rho \alpha\beta}\partial_{\alpha}T_{\mu\beta}$ & 4.000 & 4.326 \\
            3 & 1 & $\partial_{\alpha}T_{\mu\nu}$ & 4.000 & 3.763 \\
			\hline
			\hline 
		\end{tabular}
        
		\caption{Scaling dimensions of the fields in the 3D gauged Majorana fermion CFT (we restrict to $\Delta \le 4$). Fields are labeled by their angular momentum $L$, scaling dimension $\Delta$, and charge under the layer $\mathbb{Z}_2$ symmetry. The data at $N=14$ shows good agreement with the exact scaling dimensions of the theory.}
        \label{tab:rawdata}
	\end{table}

\section{S2. Gap analysis at the critical point}

In this section, we provide additional numerical evidence for the critical point studied in the main text. In particular, we perform finite-size scaling analysis of the gaps in the CFT and non-CFT sectors of the Hilbert space.

\subsection{Non-CFT states in the neutral sector}

\begin{figure}[tbh]
    \centering
    \includegraphics[width=0.75\textwidth]{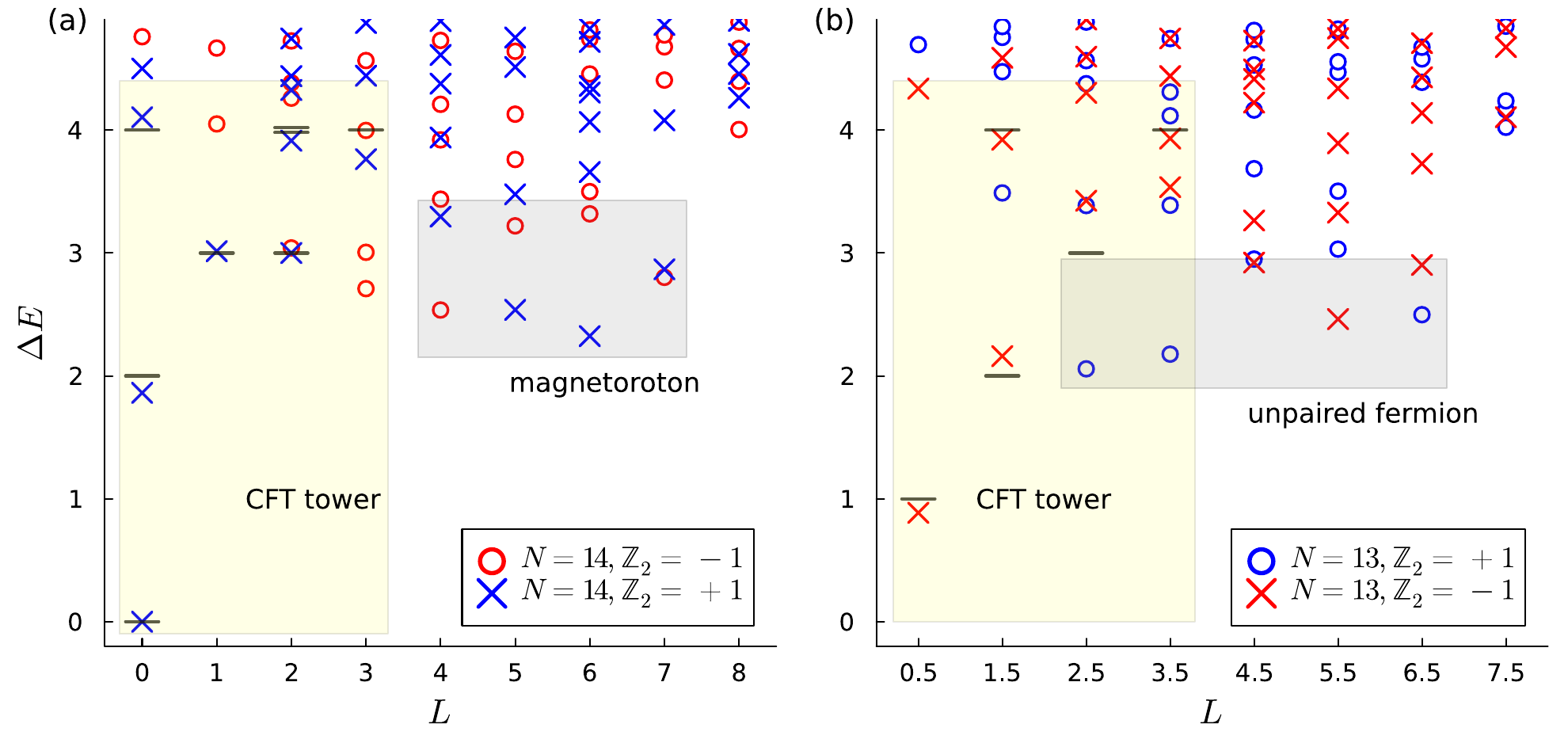}
    \caption{Finite-size spectrum of the model in the main text at its critical point, rescaled such that the stress-energy tensor has scaling dimension $\Delta_T=3$. (a) The even-particle sector at system size $N=14$.  At low energy and angular momentum, the expected CFT scaling dimensions (shown as black lines) are in good agreement with the microscopic energies of the $\mathbb{Z}_2$-even sector. The $\mathbb{Z}_2$-odd sector is decoupled from the critical theory. The collective excitation branches of the bilayer FQH state are also highlighted by the grey box. (b) Same as (a) but for the odd-particle sector at $N=13$. Similarly to the even-particle sector, the CFT tower of states is well captured by the $\mathbb{Z}_2$-odd sector, while the reverse parity sector does not participate to the transition. In this sector, the collective excitation corresponds to the unpaired fermion mode of the Moore--Read state~\cite{Moller11, Bonderson11}. }
    \label{fig:full_spectra}
\end{figure}

The complete spectrum of the model at the critical point $(V_0^\text{inter}=0.48,h=0.58)$ in the neutral sector (i.e. $2Q = N-2$) is shown in \cref{fig:full_spectra}. As detailed in the main text, we identify the ($N$-even, $\mathbb{Z}_2$-even) and ($N$-odd, $\mathbb{Z}_2$-odd) sectors as the CFT sectors, where we are able to accurately capture the operator content at low energies; the other layer-parity sectors are effectively decoupled from the phase transition. 

At higher angular momenta, the collective modes corresponding to gapped bilayer FQH states~\cite{Renn93, MacDonald94, Shizuya03} can also be observed in \cref{fig:full_spectra}. In the $N$-even sector, we identify the ``magnetoroton'' branch \cite{Girvin86}, corresponding to the density-wave distortion of the FQH droplet, while in the odd-particle sector we identify the Bogoliubov-type quasiparticle of the weakly paired states---the ``unpaired fermion'' branch \cite{Moller11, Bonderson11}.

\begin{figure}[tbh]
    \centering
    \includegraphics[width=0.8\textwidth]{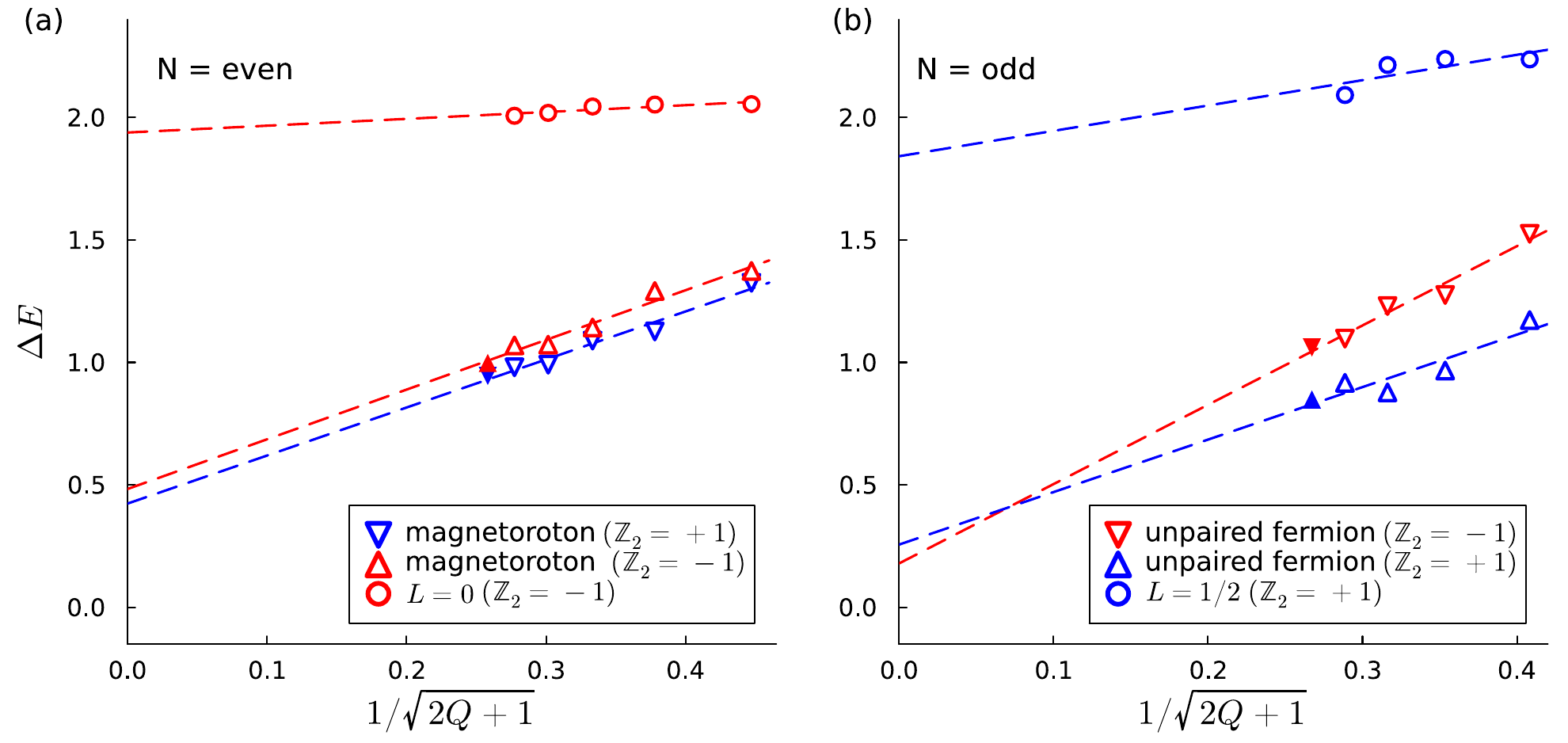}
    \caption{Finite-size scaling of energy gaps in the neutral sector. (a) The even-particle sector for system sizes $N=6-14$ (exact diagonalization) and $N=16$ (DMRG with bond dimension up to 16000 for the magnetoroton). The magnetoroton collective modes of the Halperin-$220$/Pfaffian states are identified as the lowest energy state outside the CFT tower region. The angular momentum values of these states are, as a function of system size, $L_{\mathbb{Z}_2 = +1} = (3,4,4,5,6, 6)$ and $L_{\mathbb{Z}_2 = -1} = (2, 3, 3, 3, 4, 4)$, with gaps extrapolating to a finite value. In the $\mathbb{Z}_2$-odd sector, we also confirm that the lowest singlet state has a finite energy gap. (b) The odd-particle sector for system sizes $N=7-13$ (exact diagonalization) and $N=15$ (DMRG with bond dimension up to 16000 for the unpaired fermion). Here, the unpaired fermion collective modes are also identified, taking angular momentum values $L_{\mathbb{Z}_2 = -1}=(7/2, 9/2, 9/2, 11/2, 13/2)$ and $L_{\mathbb{Z}_2 = +1}=(5/2, 5/2, 5/2, 5/2, 7/2)$ (note that this does not hinder the identification of CFT state as it is in the decoupled sector). Similarly, the $L=1/2$ state also extrapolates to a finite gap. 
    }
    \label{fig:neutral_sector_gapped}
\end{figure}

All excitations described above (which are not part of the CFT sector) are expected to display a finite gap in the thermodynamic limit, hence they will reside deep within the continuum of the spectrum whose low-energy part will be dominated by CFT states. 
In \cref{fig:neutral_sector_gapped}, we verify this picture by performing thermodynamic-limit extrapolations for several of the expected gapped levels in the spectrum. This confirms that all collective modes (i.e., the magnetoroton and unpaired fermion in all layer $\mathbb{Z}_2$ sectors) as well as the singlet states in the decoupled $\mathbb{Z}_2$ sectors converge to a finite energy gap. 
    
We note that the scenario in Fig.~\ref{fig:neutral_sector_gapped} is similar to Ref.~\cite{Voinea25}, where gapped collective excitations were constructed using a mean-field picture, allowing their identification in the spectrum even when they are no longer well-separated in energy from other eigenstates. This mean-field approach was facilitated by the fact that the charge sector remains in the same phase throughout the transition in Ref.~\cite{Voinea25}, while in the present case we have two distinct phases on either side of the transition. Nevertheless, one could attempt a similar identification here starting from the variational ansatz proposed in Ref.~\cite{Ho95}.

\subsection{Charge gap}

At the critical point, the $U(1)$ electric charge is expected to decouple (albeit with a remnant topological contribution); we probe this by studying the behaviour of the charge gap as a function of the tunneling field. This is particularly important, as the charge excitations of the Halperin-220 and Pfaffian states are inherently different, with the latter carrying non-Abelian exchange statistics. Previously, a sharp upward cusp in the charge gap at the transition was experimentally observed in wide quantum wells at $\nu=1/2$~\cite{Suen94b}, and it has also been reproduced in numerical simulations on realistic models~\cite{Nomura04, Peterson10a}. Therefore, non-analytic behaviour is expected at the critical point, that would signal a transition between these distinct charged quasiparticles.

\begin{figure}[tbh]
    \centering
    \includegraphics[width=0.7\textwidth]{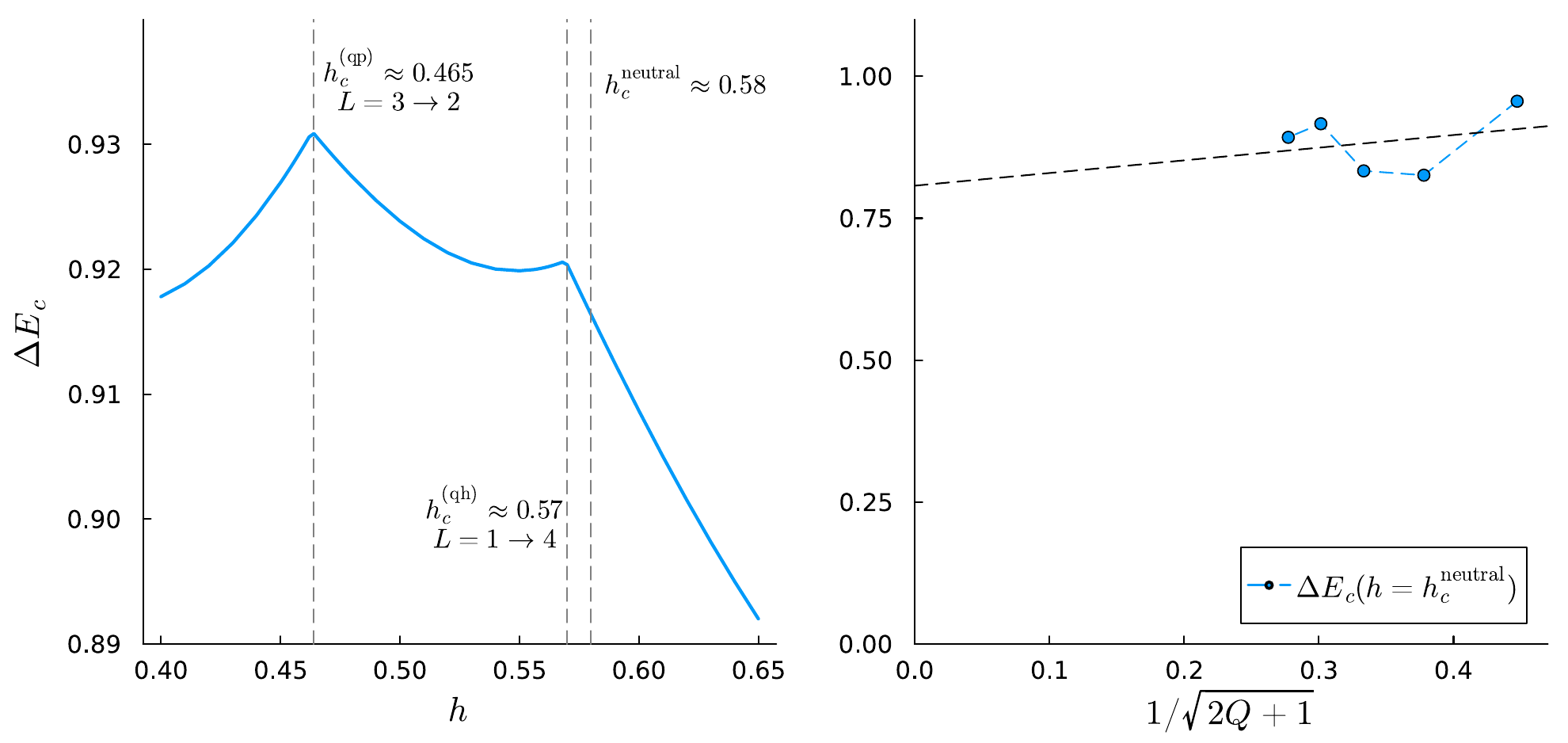}
    \caption{(a) Evolution for the charge gap $\Delta E_c$ as a function of tunneling, for a system of $N=12$ bosons. Two apparent transitions are encountered. The transition in the 2-quasiparticle state at flux $2Q-1$ happens at $h_c^{\text{(qp)}} \approx 0.465$, and the angular momentum sector changes from $L=3$ to $L=2$, and one corresponding to the quasihole flux at $2Q+1$; similarly, the quasihole transition happens at $h_c^{\text{(qh)}} \approx 0.57$, closer to the critical field identified in the neutral sector, with the angular momentum changing from $L=1$ to $L=4$. (b) Extrapolation of the charge gap for $N=6-14$ particles. While there are strong finite-size effects, the charge sector appears to have a non-zero gap in the thermodynamic limit. }
    \label{fig:charge_gap}
\end{figure}

We consider the charge gap defined as 
\begin{equation}
    \Delta E_c = f\big[ \tilde{E}_0(N, 2Q + 1) + \tilde{E}_0(N, 2Q-1) - 2 \tilde{E}_0(N,2Q) \big]/n_q \, .
\end{equation} 
In this system, the charged excitations can only be created in pairs (each one contains half of a flux quantum), and therefore to calculate the energy of a single excitation we set $n_q = 2$. The energies are also rescaled by a factor $f=\sqrt{2Q\nu/N}$ to account for the deviation from the filling factor $\nu$ in the thermodynamic limit. \cite{Morf02}. The corrected energies $\tilde{E}_0$ are calculated from the ground state energies at a given flux $E_0$ by including the correction from the neutralizing background charge, in terms of the quasiparticle number $n_q$, the quasiparticle charge $e_q$, and the average charging energy per particle pair $C_{2Q}$~\cite{Jain97, Balram20b}:
\begin{equation}\label{eq:charge gap}
    \tilde{E}_0(N, 2Q) = E_0(N,2Q) - C_{2Q}(N^2 - n_q^2 e_q^2)/2 \, , \quad \mathrm{with} \quad C_{2Q} = V_0 \frac{4Q+1}{(2Q+1)^2} \, .
\end{equation}
Here, the excitation charge is $\pm e/4$, and we use the averaged pseudopotential $V_0 = (V_0^\text{inter}+ V_0^\text{intra})/2$. 

\cref{fig:charge_gap} shows the behavior of the charged excitations across the phase transition. We observe two upward cusps, corresponding to transitions in the quasiparticle sector (i.e. $\tilde{E}_0(N, 2Q-1) - \tilde{E}_0(N,2Q)$) and the quasihole sector (i.e. $\tilde{E}_0(N, 2Q+1) - \tilde{E}_0(N,2Q)$). In finite-size systems, the quasihole, the quasiparticle and the neutral sector critical points do not coincide, which we attribute to finite-size effects; however, since the charged excitations of the Halperin 220 and Pfaffian states have specific countings in different angular momentum sectors~\cite{Read96}, the ground state sectors at the first order transitions provide an additional signature of the Abelian to non-Abelian phase transition. For example, the 2-quasihole state of the Pffafian phase can only take angular momentum values $L \in \{ N/2, N/2-2, \dots \}$ in a particular system size $N$, which our model is in agreement with.

\section{S3. Operator content at the transition}

To confirm the intuition that the phase transition is driven by interlayer tunneling, we numerically verify the overlap of $n_x(\Omega)$ with the Majorana bilinear $\bar{\psi}\psi$, the only relevant singlet of the theory. Near the critical point, the local tunneling term can be expressed in terms of the symmetry-allowed CFT operators:
\begin{equation}\label{eq:tunneling expansion}
    n_x = c_I I + c_{\bar{\psi}\psi}\bar{\psi}\psi + c_{\partial\bar{\psi}\psi}\partial\bar{\psi}\psi + c_{T} T_{\mu\nu} + \dots \, ,
\end{equation}
where the dots stand for operators with $\Delta > 3$. Note that we will focus on CFT operators with small angular momenta $L$, such that the low-energy spectrum does not feature gapped modes such as the magnetoroton. 

Operators on the fuzzy sphere can be related back to flat space via the Weyl transformation $\tau = R \ln r$, where $R$ is the fuzzy sphere radius, $(\tau, \Omega)$ are the $\mathbb{R} \times S^2$ coordinates and $(r, \Omega)$ are the original coordinates in $\mathbb{R}^3$. With the half-integer spin operators of the theory being inaccessible via local microscopic operators, let us focus on bosonic operators. The corresponding transformation for a scalar with scaling dimension $\Delta$ is 
\begin{equation}\label{eq:weyl transf}
    \phi(\tau, \Omega) = \bigg( \frac{r}{R} \bigg) ^\Delta \phi(r, \Omega) \, .
\end{equation}

\begin{figure}[tbh]
    \centering
    \includegraphics[width=0.55\textwidth]{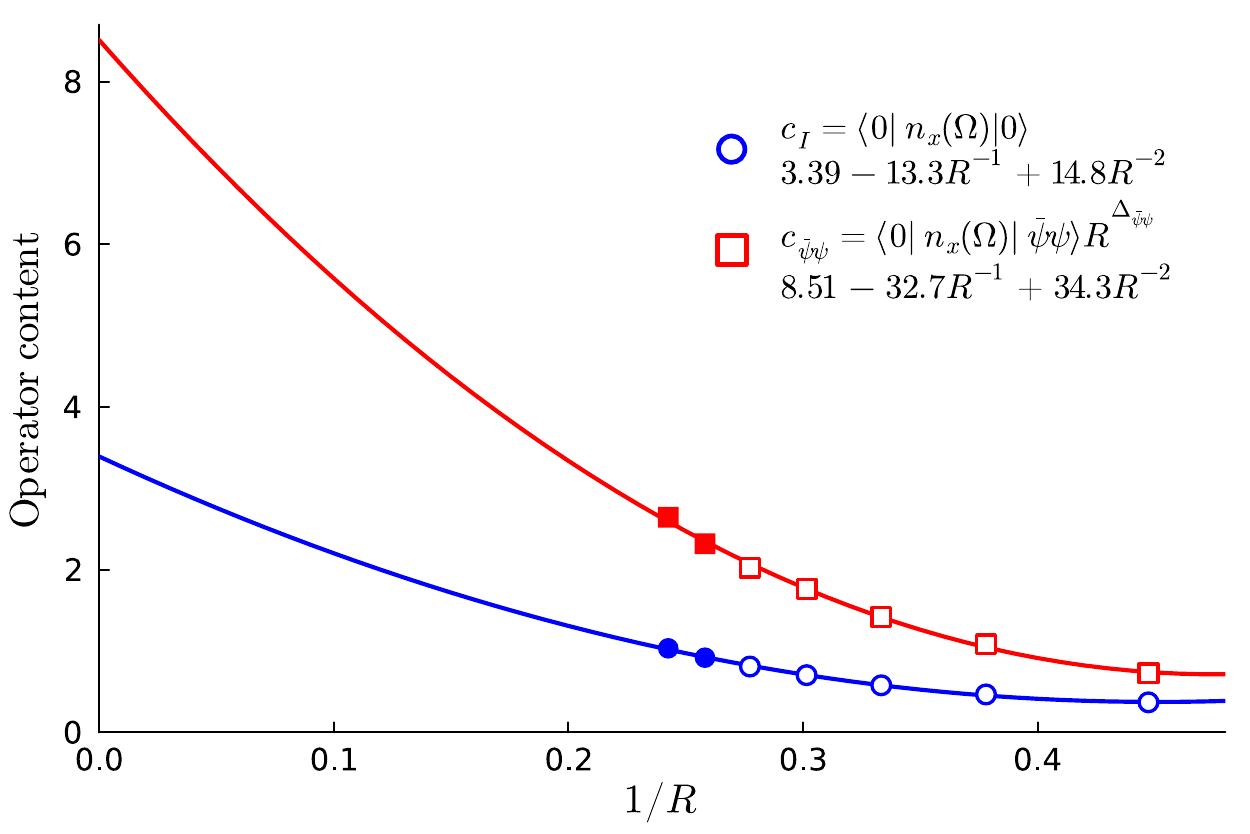}
    \caption{Operator content of the interlayer tunneling field $n_x(\Omega)$. Empty markers are system sizes accessible in exact diagonalization, while solid markers are calculated using DMRG ($N=16,18$ are converged using bond dimensions of up to 16000 and 20000, respectively). The large overlap with the relevant singlet $\bar{\psi}\psi$ confirms that the critical point can be reached by tuning the tunneling.}
    \label{fig:nx_operator_content}
\end{figure}

The operator content of $n_x$ can be extracted using the properties of one- and two-point correlations in a CFT, $\langle \phi(r,\Omega)\rangle = 0$ and $\langle \phi_i(r,\Omega)\phi_j(0)\rangle = \delta_{ij} r^{-2\Delta}$. Therefore, the identity and Majorana bilinear components take the form:
\begin{equation}\label{eq:tunneling content}
    c_I = \langle 0 | n_x(\Omega) | 0 \rangle \, \quad \text{and} \quad c_{\bar{\psi}\psi} = R^{\Delta_{\bar{\psi}\psi}} \langle 0 | n_x(\Omega) | \bar{\psi}\psi \rangle + \mathcal{O} \left( \frac{1}{R} \right) \, .
\end{equation}
We tested these relations at the critical point $(V_0^\text{inter} = 0.48, h = 0.58)$, see \cref{fig:nx_operator_content}. The tunneling field indeed has significant overlap with the fermion bilinear, confirming our intuition. In principle, correlators of this field can be used to further probe the CFT description of the transition, such as OPE coefficients. However, the absence of space-time parity as a microscopic UV symmetry, as well as the presence of gapped collective excitations in the accessible low energy spectra, render their extraction beyond the scope of the current work.

\end{document}